\documentclass[twocolumn,superscriptaddress]{revtex4-1}
\usepackage{gnuplottex}
\usepackage{mathtools}
\usepackage{float}
\usepackage{array}
\usepackage[english]{babel}
\usepackage[utf8]{inputenc}
\usepackage{amsmath}
\usepackage[table]{xcolor}
\usepackage{amsfonts}
\usepackage{graphicx}
\usepackage{dsfont}
\usepackage[colorlinks=true, linkcolor=blue, citecolor=dred]{hyperref}
\usepackage{algorithm}
\usepackage[noend]{algpseudocode}
\usepackage{bbold}
\usepackage{mathtools}
\usepackage{amssymb}
\usepackage{bm}
\usepackage{bbm}
\usepackage{tabularx, booktabs}
\usepackage{xspace}
\usepackage{listings}
\usepackage{algorithm}
\usepackage[noend]{algpseudocode}
\newcolumntype{Y}{>{\centering\arraybackslash}X}
\usepackage{physics}
\definecolor{dgreen}{rgb}{0,.5,0}
\definecolor{dblue}{rgb}{0,0,.5}
\definecolor{dred}{rgb}{0.5,0,.5}

\usepackage{scalerel,stackengine}
\stackMath
\newcommand\reallywidehat[1]{%
\savestack{\tmpbox}{\stretchto{%
  \scaleto{%
    \scalerel*[\widthof{\ensuremath{#1}}]{\kern-.6pt\bigwedge\kern-.6pt}%
    {\rule[-\textheight/2]{1ex}{\textheight}}%WIDTH-LIMITED BIG WEDGE
  }{\textheight}% 
}{0.5ex}}%
\stackon[1pt]{#1}{\tmpbox}%
}
\newcommand{\bfr}{\mathbf{r}}
\newcommand{\bfR}{\mathbf{R}}
\newcommand{\psh}[2]{\ensuremath{\langle #1|#2\rangle}\xspace}
\def\ddroit{{\rm d}}
\begin{document}

\title{A state-averaged orbital-optimized hybrid quantum-classical algorithm for a democratic description of ground and excited states
%State-Averaged Orbital-Optimized VQE: a chemistry inspired quantum algorithm to describe conical intersections
} % Article title
\author{Saad Yalouz}
\email{s.yalouz@vu.nl}
\affiliation{Theoretical Chemistry, Vrije Universiteit, De Boelelaan 1083, NL-1081 HV, Amsterdam, The Netherlands}
\affiliation{Instituut-Lorentz, Universiteit Leiden, P.O. Box 9506, 2300 RA Leiden, The Netherlands}
\author{Bruno Senjean}
\email{bsenjean@gmail.com}
\affiliation{Instituut-Lorentz, Universiteit Leiden, P.O. Box 9506, 2300 RA Leiden, The Netherlands}
\affiliation{Theoretical Chemistry, Vrije Universiteit, De Boelelaan 1083, NL-1081 HV, Amsterdam, The Netherlands}
\author{Jakob G\"unther} 
\affiliation{Theoretical Chemistry, Vrije Universiteit, De Boelelaan 1083, NL-1081 HV, Amsterdam, The Netherlands}
\author{Francesco Buda}
\affiliation{Leiden Institute of Chemistry,
Leiden University, Einsteinweg 55, P.O. Box 9502, 2300 RA Leiden, The Netherlands.}
\author{Thomas E. O'Brien}
\affiliation{Google Research, 80636 Munich, Germany}
\affiliation{Instituut-Lorentz, Universiteit Leiden, P.O. Box 9506, 2300 RA Leiden, The Netherlands}
\author{Lucas Visscher} 
\affiliation{Theoretical Chemistry, Vrije Universiteit, De Boelelaan 1083, NL-1081 HV, Amsterdam, The Netherlands}

\begin{abstract}
In the Noisy Intermediate-Scale Quantum (NISQ) era, solving the electronic structure problem from chemistry is considered as the ``killer application'' for near-term quantum devices. 
In spite of the success of variational hybrid quantum/classical algorithms in providing accurate energy profiles for small molecules, careful considerations are still required for the description of complicated features of potential energy surfaces.
Because the current quantum resources are very limited, it is common to focus on a restricted part of the Hilbert space (determined by the set of active orbitals).
While physically motivated, this approximation can severely impact the description of these complicated features.
A perfect example is that of conical intersections (\textit{i.e.} a singular point of degeneracy between electronic states), which are of primary importance to understand many prominent reactions.
Designing active spaces so that the improved accuracy from a quantum computer is not rendered useless is key to finding useful applications of these promising devices within the field of chemistry.
To answer this issue, we introduce a NISQ-friendly method called ``State-Averaged Orbital-Optimized  Variational Quantum Eigensolver'' (SA-OO-VQE)  which  combines  two  algorithms: (1) a  state-averaged  orbital-optimizer, and (2) a state-averaged VQE. 
To demonstrate the success of the method, we classically simulate it on a minimal Schiff base model (namely the formaldimine molecule CH$_\text{2}$NH) relevant also for the photoisomerization in rhodopsin --- a crucial step in the process of vision mediated by the presence of a conical intersection.
We show that merging both algorithms fulfil the necessary condition to describe the molecule's conical intersection, \textit{i.e.} the ability to treat degenerate (or quasi-degenerate) states on the same footing.
\end{abstract}

\maketitle

\section{Introduction}

Quantum computing promises to solve problems that are beyond the capacities of classical devices.
Technological advances in quantum computers occur rapidly~\cite{omalley2016scalable,kandala2017hardware,colless2018computation,hempel2018quantum,bruzewicz2019trapped,arute2019quantum,arute2020hartree,nam2020ground}, and it is now of primary importance to develop quantum algorithms dedicated to specific tasks of high industrial and societal impact.
The electronic structure problem in chemistry is considered as one of the killer applications of quantum computers~\cite{reiher2017elucidating,li2019electronic,cao2019quantum,mcardle2020quantum,bauer2020quantum,von2020quantum}, and it remains an open question whether these applications will be achieved in the noisy intermediate-scale quantum (NISQ) era~\cite{preskill2018quantum}.
While quantum phase estimation algorithms~\cite{abrams1999quantum,aspuru2005simulated,obrien2019quantum} can in principle solve the electronic structure problem in its entirety, they have received limited focus in the NISQ era due to their apparent longer circuit depth requirements.
Instead, variational hybrid quantum/classical algorithms require much shallower circuits at the cost of additional measurements, with as a spearhead the variational quantum eigensolver (VQE)~\cite{peruzzo2014variational,mcclean2016theory,arute2020hartree}. 
Those NISQ-friendly algorithms are hoped to solve quantum chemical problems of moderate-large size in the near future.

While computing the ground state of a molecule gives information about the equilibrium geometries and transition states (essential to determine the energy barrier of a chemical reaction), several chemical reactions also depend on excited states (see for instance Ref.~\onlinecite{mai2020molecular} and references therein).
To date, most variational quantum algorithms have been specifically designed to prepare ground states~\cite{omalley2016scalable,romero2018strategies,lee2018generalized,ryabinkin2018qubit,mcardle2019variational,mizukami2019orbital,kuhn2019accuracy,rattew2019domain,sokolov2020quantum,ryabinkin2020iterative,lang2020iterative,matsuzawa2020jastrow,huggins2020non,gomes2020efficient,meitei2020gate,wang2020resourceoptimized,zhang2020mutual} while excited state calculations have received less attention until only recently~\cite{mcclean2017hybrid,ollitrault2019quantum,nakanishi2019subspace,ibe2020calculating,lee2018generalized,higgott2019variational,jones2019variational,jouzdani2019method,parrish2019quantum,parrish2019hybrid,bauman2019quantum,motta2020determining,zhang2020variational}.
It should be noted that some excited states can also be obtained as the lowest energy solutions of a given set of symmetries~\cite{ryabinkin2018constrained,ryabinkin2018symmetry,greene2019generalized,zhang2020shallow,gard2020efficient,seki2020symmetry}.

The landscape of ground- and excited-state potential energy surfaces (PES) can be quite complicated, making even qualitative numerical description challenging.
As atomic nuclei move, ground- and excited-state energies may tend towards each other, leading to either avoided crossings or true inversions in the ordering of the electronic states.
These inversions are referred to as conical intersections and are ubiquitous in photochemistry and photobiology~\cite{domcke2011conical,yarkony2012nonadiabatic}.
Their presence mediates important reactions, for instance the photoisomerisation of the chromophore of the visual photoreceptor rhodopsin, ensuing absorption of light.
Photoisomerisation is a crucial step in the process of vision~\cite{garavelli1997c5h6nh2+,gonzalez2000computational,polli2010conical,valsson2013rhodopsin,manathunga2016probing}.
Conical intersections also find their importance in photosynthesis~\cite{olaso2006ultrafast}, photostabilization of DNA~\cite{kang2002intrinsic,groenhof2007ultrafast,barbatti2010relaxation,frances2018dynamics} and excitation-energy transfer~\cite{may2008charge} such as in supramolecular light-harvesting antennae~\cite{ho2019diabatic}.
Despite their indisputable importance, their description remains a challenge in the chemistry community~\cite{domcke2011conical,yarkony2012nonadiabatic,gozem2014shape}.
Diagonalization of the electronic Hamiltonian matrix being intractable on classical devices, it is of common use to reduce the Hilbert space size by considering only a subspace of (so-called active) orbitals of the full problem.
However, such a truncated Hilbert space is not guaranteed to be accurate enough to represent a conical intersection anymore.
A possible solution is provided by state-averaged orbital-optimization, which consists in modifying the truncated Hilbert space (without modifying its size) to hopefully recover the conical intersection.
On classical computers, this orbital-optimization is part of the so-called state-averaged complete active space self-consistent field (SA-CASSCF) method~\cite{siegbahn1981complete,helgaker2014molecular}, known for its democratic treatment of multiple (possibly degenerate) eigenstates.

For quantum computers in the NISQ era, the concept of active space also represents an important issue.
Indeed, the limitations of quantum resources (e.g. maximum number of qubits) constrain algorithms to focus on small-sized active spaces, thus preventing any calculation on large systems.
Designing Hilbert spaces so that the improved accuracy from a quantum computer is not rendered useless by the active space approximation itself is key to finding useful applications of these promising devices within the field of chemistry.
In this context, getting access to complicated chemical features such as avoided crossing or conical intersections can appear as a real challenge for near-term quantum computers.

Motivated by this issue, we introduce in this paper a NISQ-friendly variational hybrid  quantum/classical algorithm inspired by the SA-CASSCF method, so-called state-averaged orbital-optimized VQE (SA-OO-VQE).
The ability of SA-OO-VQE to capture a conical intersection is demonstrated on a minimal Schiff base model of the rhodopsin chromophore, i.e. the formaldimine molecule.

The paper is organized as follows.
After a brief description of the electronic structure problem in Sec.~\ref{sec:ESP}, the concept of conical intersection is introduced in Sec.~\ref{sec:CI}.
A particular attention is paid to the efficiency of classical algorithms to represent this characteristic in chemistry (Sec.~\ref{sec:CI_num}), as shown on the formaldimine molecule in Sec.~\ref{sec:formaldimine}.
Then, the SA-OO-VQE hybrid quantum/classical algorithm is detailed in Sec.~\ref{sec:method}, where each building blocks of the algorithm is described separately.
Following a brief introduction on quantum variational algorithms in Sec.~\ref{sec:VQE}, we detail how SA-OO-VQE mimics its classical analog SA-CASSCF by combining a state-averaged VQE (Sec.~\ref{sec:SA-VQE}) with a state-averaged orbital-optimization procedure (Sec.~\ref{sec:SA-OO}). 
The importance of choosing a flexible wavefunction ansatz such as the generalized unitary coupled cluster (with double excitations only) is discussed in Sec.~\ref{sec:ansatz}.
Following the computational details in Sec.~\ref{sec:comp}, we demonstrate the efficiency of our method on the formaldimine molecule in Sec.~\ref{sec:results}.
Finally, conclusions and perspectives are given in Sec.~\ref{sec:conclusion}.

\section{Theory}\label{sec:theory}

\subsection{Electronic structure problem}\label{sec:ESP}

In quantum chemistry, the Born-Oppenheimer approximation assumes that molecular structure and dynamics can be treated in two parts: considering first the atomic nuclei, and then the motion of the electrons around them.
Atomic nuclei, being many times heavier than their electronic counterparts, are often treated as classical point particles with definite position $\mathbf{R}_A$ and momentum $\mathbf{P}_A$ (with $A$ the index of the nucleus).
As classical particles have no zero-point energy, the ground state of the nuclei is motionless, \textit{i.e.} $\mathbf{P}_A=0$.
By contrast, the light electrons behave as fundamentally quantum particles, and form a collective many-body wavefunction $\Psi(\mathbf{r})$, where $\mathbf{r}=(\mathbf{r}_1,\mathbf{r}_2,\ldots,\mathbf{r}_{N_e})$ is the $3N_e$-dimensional global position vector of the $N_e$ electrons.
This wavefunction evolves by the Schr\"odinger equation with the fixed nuclei providing a background potential field.
The quantum operator describing this dynamics is the so-called electronic structure Hamiltonian (in atomic units)
\begin{eqnarray}
    \hat{\mathcal{H}}  =
    - \dfrac{1}{2}\sum_{i=1}^{N_e} \grad_{\bfr_i} +  \dfrac{1}{2}\sum_{i\neq j}^{N_e} \dfrac{1}{\bfr_{ij}}
    - \sum_{i=1}^{N_e}\sum_{A=1}^{N_a}\dfrac{Z_A}{\bfr_{iA}},
\end{eqnarray}
where $\bfr_{ij} = | \bfr_i - \bfr_j |$, $\bfr_{iA} = | \bfr_i - \bfR_A |$ and $Z_A$ are respectively the distance between electrons $i$ and $j$, the distance between electron $i$ and nuclei $A$, and the atomic number of atom $A$.
In this context, solving the electronic structure problem means solving the time-independent Schr\"{o}dinger equation
\begin{equation}
\hat{\mathcal{H}} \ket{\Psi_\ell} = E_\ell \ket{\Psi_\ell}
\end{equation}
which yields a set of electronic eigenstates $|\Psi_\ell\rangle\equiv \Psi_\ell(\mathbf{r})$ with corresponding energies $E_\ell$.
These eigenstates and energies depend on the atomic positions themselves --- $E_\ell=E_\ell(\mathbf{R})$ and $|\Psi_\ell\rangle = |\Psi_\ell(\mathbf{R})\rangle$, where $\mathbf{R}$ is the $3N_a$-dimensional position vector of the $N_a$ nuclei.
(Note the change of notation: $\Psi_\ell(\mathbf{r})$ is a single number from evaluating the function $\Psi_\ell$ at the fixed electronic co-ordinates $\mathbf{r}$, while $|\Psi_\ell(\mathbf{R})\rangle$ is an entire function defined by a fixed set of nuclear co-ordinates $\mathbf{R}$).
In practice, it is convenient to work in a finite basis of $N_o$ orthonormal molecular orbitals (MO) $\lbrace \phi_p (\bfr) \rbrace$. 
The latter are usually solutions to the mean-field single-particle problem, determined by the Hartree--Fock (HF) method \cite{helgaker2014molecular}.
In this basis, the spin-free Hamiltonian reads
\begin{eqnarray}\label{eq:Ham_elec}
\hat{\mathcal{H}}(\bfR) = \sum_{pq}^{N_o} h_{pq}(\bfR) \hat{E}_{pq} + \dfrac{1}{2} \sum_{pqrs}^{N_o} g_{pqrs}(\bfR) \hat{e}_{pqrs},
\label{eq:el_Ham}
\end{eqnarray}
where (we drop the dependence on $\bfR$ for convenience),
\begin{equation}
h_{pq}  = \int \phi_p^*(\bfr_1 ) \left(- \dfrac{1}{2} \grad_{\bfr_1} - \sum_{A=1}^{N_a}\dfrac{Z_A}{\bfr_{1A}} \right) \phi_q(\bfr_1 ) \ddroit \bfr_1
\end{equation}
and 
\begin{equation}
g_{pqrs}  =  \iint  
\phi_p^*(\bfr_1 ) \phi_r^*(\bfr_2 )
\dfrac{1}{\bfr_{12}}\phi_q(\bfr_1 ) \phi_s(\bfr_2 )  \ddroit \bfr_1 \ddroit \bfr_2 
\end{equation}
are the one- and two-electron integrals, and $\hat{E}_{pq} = \sum_{\sigma} \hat{a}_{p\sigma}^\dagger \hat{a}_{q\sigma}$ and $ \hat{e}_{pqrs} = \sum_{\sigma,\tau} \hat{a}_{p\sigma}^\dagger \hat{a}_{r\tau}^\dagger \hat{a}_{s\tau}\hat{a}_{q\sigma} $ are the one- and two-body spin-free operators. 
%Note that in chemistry notation we have $g_{pqrs} = (pq\vert rs) = \psh{pr}{qs}$.
Inverting the problem, the set $\{E_\ell(\mathbf{R})\}$ (now basis-dependent) may be considered as a set of potential energy manifolds for the classical motion of the nuclei.

% At room temperature, and in the absence of photoexcitation, stable molecular geometries are found at local minima of the electronic ground-state PES $E_0(\mathbf{R})$.
% When photoexcitations are considered, different types of transitions can occur vertically between the different PESs. In this case, a molecule can absorb a photon to get promoted to an excited state. The molecule can then return to a lower energy state by emitting a photon (via a photo-induced emission or simple photon relaxation). Contrasting with these inter-PES transitions, it is also important to note the existence of non-radiative intra-PES transitions, i.e. by smooth relaxation mediated by vibrational motions of the molecule (phonons).
% These transitions may lead to significant nuclear motions and rearrangements, as there is no \emph{a priori} reason to expect that local minima of $E_0(\mathbf{R})$ are also local minima of $E_1(\mathbf{R})$. In the presence of a conical intersection, such a non-radiative relaxation makes it possible for a molecule to smoothly transits from an excited state to a lower state via the singular point of degeneracy connecting the two PESs \cite{}.

\subsection{Conical intersections in nature}\label{sec:CI}

A naive study of quantum mechanics would suggest that the two lowest PESs of a molecule $E_0(\mathbf{R})$ and $E_1(\mathbf{R})$  would never cross.
Perturbation theory suggests that such a crossing point or degeneracy would be prone to any disturbance: if $E_0(\mathbf{R})=E_1(\mathbf{R})$, we would expect a perturbation $V$ to induce a gap $E_1(\mathbf{R})-E_0(\mathbf{R})\sim |V|$.
However, some degeneracies are more robust than others.
Consider the effective block Hamiltonian obtained by projecting the full Hamiltonian $\hat{\cal{H}}(\bf{R})$ in Eq.~(\ref{eq:el_Ham}) onto the two lowest states decoupled from other states,
\begin{eqnarray}
\hat{H} = \begin{bmatrix}
H_{00} & H_{01} \\
H_{10} & H_{11} 
\end{bmatrix},
\end{eqnarray}
where $H_{\rm IJ} = \bra{\Psi_{\rm I}}\hat{H}\ket{\Psi_{\rm J}}$.
An extensive discussion about the necessary conditions for this effective Hamiltonian to depict a conical intersection is provided in Ref.~\onlinecite{gozem2014shape}.
For simplicity, suppose that, in the absence of a perturbation, this low-energy effective Hamiltonian around a degeneracy at $\mathbf{R}_0$ takes the form
\begin{align}
    \hat{H} = h_0(\mathbf{R})I &+ h_X(\mathbf{R}-\mathbf{R}_0)\cdot\mathbf{R}_XX\nonumber \\&+ h_Z(\mathbf{R}-\mathbf{R}_0)\cdot\mathbf{R}_ZZ,\label{eq:low_energy_effective_theory}
\end{align}
where $I$ is the $2\times 2$ identity matrix, $X$ and $Z$ are the two real Pauli matrices, and $\mathbf{R}_X$ and $\mathbf{R}_Z$ are two arbitrary $3N_a$-dimensional vectors.
This amounts to $H_{00} = h_0(\bfR) + h_Z(\bfR - \bfR_0)\cdot \bfR_Z$, $H_{01} = H_{10} = h_X(\bfR - \bfR_0)\cdot \bfR_X$ and $H_{11} = h_0(\bfR) - h_Z(\bfR - \bfR_0)\cdot \bfR_Z$.
Such a degeneracy is known as a conical intersection, as the energy surfaces form a cone in the $\mathbf{R}_X$ and $\mathbf{R}_Z$ directions (modulo the background shift $h_0(\mathbf{R})$):
\begin{align}
    &E_\pm(\mathbf{R}) = h_0(\mathbf{R}) \pm\\
    & \sqrt{h_X|(\mathbf{R}-\mathbf{R}_0)\cdot\mathbf{R}_X|^2 + h_Z|(\mathbf{R}-\mathbf{R}_0)\cdot\mathbf{R}_Z|^2}.\nonumber
\end{align}
This conical intersection necessarily connects the two surfaces $E_+(\mathbf{R})$ and $E_-(\mathbf{R})$; a continuous path along one surface passing through $\mathbf{R}_0$ will invert the bands.
Note that a (real-valued) perturbation will shift the degeneracy to some $\mathbf{R}'_0$, rather than breaking it (although a gap at $\mathbf{R}=\mathbf{R}_0$ is still induced)~\cite{gozem2014shape}.
A proof is provided in Appendix~\ref{app:proof}.

In photochemistry, this degeneracy allows for non-radiative relaxation from an electronic excited state.
Consider a photoinduced chemical reaction, triggered by the absorption of a photon that promotes the electrons of the molecule from the ground state to an excited state.
As there is \emph{a priori} no reason to expect that local minima of the ground state are also local minima of the excited state, the wavepacket will evolve in time, driven by atomic forces.
Through its evolution, it may decay to a lower state thanks to different phenomena~\cite{mai2020molecular}, finally reaching a new ground-state minima which may be in a drastically different nuclear geometry than the initial one.
One such phenomena is called internal conversion, i.e. a non-radiative transition induced by the coupling between electronic and nuclear (vibrational) degrees of freedom (so-called vibronic coupling). 
Internal conversion is mediated by a degenerate point with double cone topography, so-called conical intersection~\cite{klessinger1995excited,robb1995conical,bernardi1996potential,domcke2004conical}, as described in the example above.
Such an internal nuclear rearrangement is critical e.g. for sight, where the (\textit{cis} to \textit{trans}) isomerization of the retinal protonated Schiff base (RPSB) chromophore --- in the visual receptor rhodopsin in the retina --- is the key photochemical event following the absorption of a photon~\cite{birge1990nature}.
Conical intersections are also of fundamental interest, as the (fundamentally quantum) behaviour of the nuclear motion along the degrees of freedom ($\mathbf{R}_X$ and $\mathbf{R}_Z$ in the above example) is no longer well-described by the Born--Oppenheimer approximation.
In other words, vibronic couplings diverge at the conical intersection.
These quantum dynamics are critically important to consider when modelling e.g. photoinduced electron transfer processes relevant in solar energy conversion~\cite{menzel2019photoinduced}. Also for the reverse process~\cite{Baldo98}, emission rather than harvesting of light, (intersystem) crossings~\cite{Marian2012} of potential energy surfaces play a crucial role and require balanced treatments of all involved surfaces and their conical intersections. 

\subsection{Predicting conical intersections numerically}\label{sec:CI_num}

The accurate description of the PESs of a molecular system is essential to understand complex phenomena in excited-state dynamics~\cite{gatti2017applications,gonzalez2020quantum}.
While vertical transitions and excited-state PES around equilibrium geometries are usually well described by standard chemistry methods such as linear response time-dependent density-functional theory (TDDFT)~\cite{casida1995time,marques2004time,ullrich2011time,casida2012progress}, this is much more challenging when states are degenerate or closely degenerate.
Indeed, conical intersections between the ground and first-excited state are ill-defined in linear response TDDFT~\cite{gozem2014shape}, in addition to other deficiencies regarding doubly excited~\cite{maitra2004double} and charge transfer states~\cite{fuks2014challenging}.
In order to correctly describe a conical intersection, the crossing states have to be treated on the same footing, i.e. they have to result from a same high level theory.
In this context, the full configuration interaction (FCI) approach, which consists in diagonalizing the entire electronic structure Hamiltonian in the full MO basis, would be the ideal method to describe the electronic spectrum.
Unfortunately, FCI scales exponentially with the system size.
Instead, it is common to employ the frozen core approximation, \textit{i.e.} to assume that some core orbitals are always occupied by electrons, and some virtual orbitals are never occupied.
These orbitals may be `frozen', leaving us to solve the problem using only the remaining `active' spin-orbitals.
This method is typically known as complete active space configuration interaction (CASCI) approximation.
In contrast to FCI, the active space selection is user-dependent and is not invariant with respect to orbital rotations.
As a consequence, it is usual to variationally optimize the orbitals with respect to the ground-state CASCI energy, thus leading to the so-called state-specific complete active space self-consistent field method (CASSCF).
These three methods are ordered both in terms of their complexity cost $T$ and their accuracy in finding the ground-state energy $E_0$ --- we have $T_{\mathrm{FCI}}\geq T_{\mathrm{CASSCF}}\geq T_{\mathrm{CASCI}}$, and $E_0\leq E_0^{\mathrm{FCI}}\leq E_0^{\mathrm{CASSCF}}\leq E_0^{\mathrm{CASCI}}$.
Finding an appropriate level of theory for a given problem requires finding the best tradeoff in this cost-accuracy balance.

Regardless of the level of theory chosen, finding the exact solution of the electronic structure problem is a difficult task.
Solving a single active space of $m$ spatial orbitals containing $n$ electrons, denoted by CASCI($m,n$), requires finding the ground state eigenvalue of a ${2m\choose n}\times {2m\choose n}$ matrix.
(The equivalent CASSCF($m,n$) problem takes a similar amount of time, multiplied by the number of iterations needed to find the optimal single-particle basis rotation.)
On a classical computer, this takes $O({2m\choose n}^3)\sim O(e^n)$ time.
This becomes impractical beyond $n\sim 20$~\cite{vogiatzis2017pushing}, though many approximations have been developed over the last century (see Refs.~\cite{williams2020direct,stair2020exploring,eriksen2020ground,loos2020note} and references therein).
Quantum computers promise a route beyond this boundary, as they can find eigenvalues of an $(m,n)$ active space to error $\epsilon$ in time polynomial in $n$ and $\epsilon$~\cite{lloyd1996universal,whitfield2011simulation}.
This has generated a huge flurry of experimental~\cite{omalley2016scalable,kandala2017hardware,colless2018computation,hempel2018quantum,bruzewicz2019trapped,arute2019quantum,arute2020hartree,nam2020ground} and theoretical~\cite{reiher2017elucidating,von2020quantum,bauer2020quantum,berry2019qubitization} activity.
As a promise, fault-tolerant quantum computer could open up areas of chemistry to accurate computational study that were previously inaccessible.
However, a quantum computer only improves the solution of a chemistry problem within the active space; it targets $E_0^{\rm CASCI}$ (or $E_0^{\rm CASSCF}$ when orbital-optimization is considered), and not the true ground state energy $E_0$.
Designing relevant active spaces is key to finding useful applications of quantum devices within the field of chemistry, and is an active field of research~\cite{sokolov2020quantum,mizukami2019orbital,bauman2019downfolding,bauman2019quantum,takeshita2020increasing,kowalski2020subsystem,motta2020quantum,metcalf2020resource,urbanek2020chemistry,mcardle2020improving,bylaska2020quantum,rossmannek2020quantum}.

Conical intersections require careful consideration in active space approximations.
Although the conical intersection might be qualitatively described by FCI, this is not guaranteed anymore when a truncated active space is considered (see Appendix~\ref{app:proof}).
Switching to state-specific CASSCF, using the orbitals optimized with respect to the ground-state energy will by construction lower the ground state, but might have the opposite effect on the excited state. Thus, the gap between the two states will increase and the description of the conical intersection can become worse.
One could think of optimizing the orbitals for the two states separately.
In this case, two separate sets of state-specific orbitals are obtained. However, the two states expressed in those two different orbital sets are not guaranteed to be orthogonal anymore.
Furthermore, when the states are closely degenerate (as around conical intersections or avoided crossings), optimizing the orbitals with respect to the excited state energy becomes impossible in practice.
Indeed, the (optimized) excited state and the ground state will inevitably (and indefinitely) swap during the optimization process, thus preventing convergence of the orbitals.
This practical problem is known as root flipping.
Finally, active spaces can also be designed differently, for instance by downfolding electronic Hamiltonians into low-dimensional active spaces. 
However, those downfolded Hamiltonians are state-specific~\cite{bauman2019downfolding}, and thus not appropriate for an equal footing treatment of ground and excited states.

In quantum chemistry, the standard method to describe conical intersections is given by the so-called state-averaged (SA)-CASSCF method. 
In this approach, the orbitals are variationally optimized with respect to the energy average of the states involved in the conical intersection~\cite{helgaker2014molecular}.
In this sense, SA-CASSCF is a {\it democratic} method (i.e. it treats on an equal footing the optimization of all the states simultaneously) ensuring a coherent representation of an  eventual degeneracy within a given active  space, which strongly contrasts with the state-specific CASSCF approach.
While the SA-CASSCF ground state might end up higher in energy than with state-specific CASSCF, results obtained from the former are most of the time more relevant as we are interested in energy difference in quantum chemistry.
Note that to go beyond the SA-CASSCF calculation, one can use second order perturbation theory (CASPT2)~\cite{andersson1992second}, or similarly the N-electron valence second order perturbation theory (NEVPT2)~\cite{angeli2001introduction} to recover the missing dynamical correlation (see also Ref.~\onlinecite{gozem2014shape} and references therein).

\subsection{Example: photoisomerization of formaldimine}\label{sec:formaldimine}

\begin{figure}
\centering
    \includegraphics[width=7.6cm]{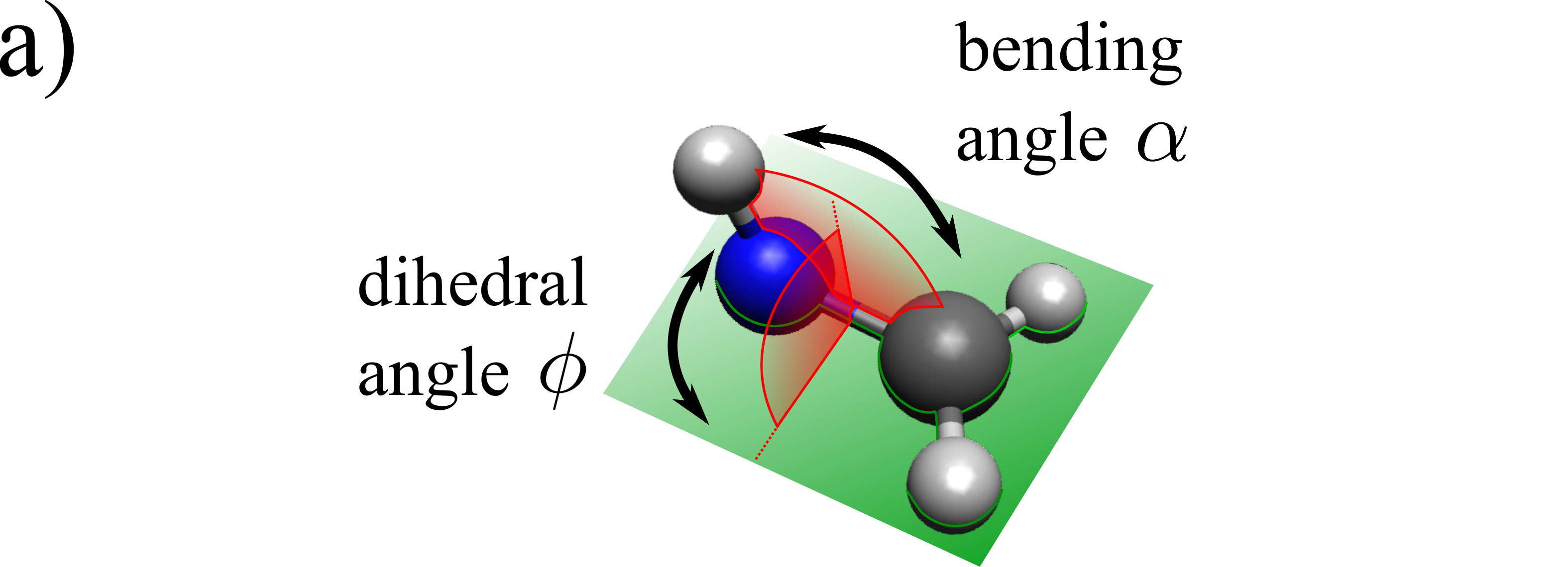}\\
    \includegraphics[width=8cm, trim= 3cm 2cm 3cm 4cm, clip=True]{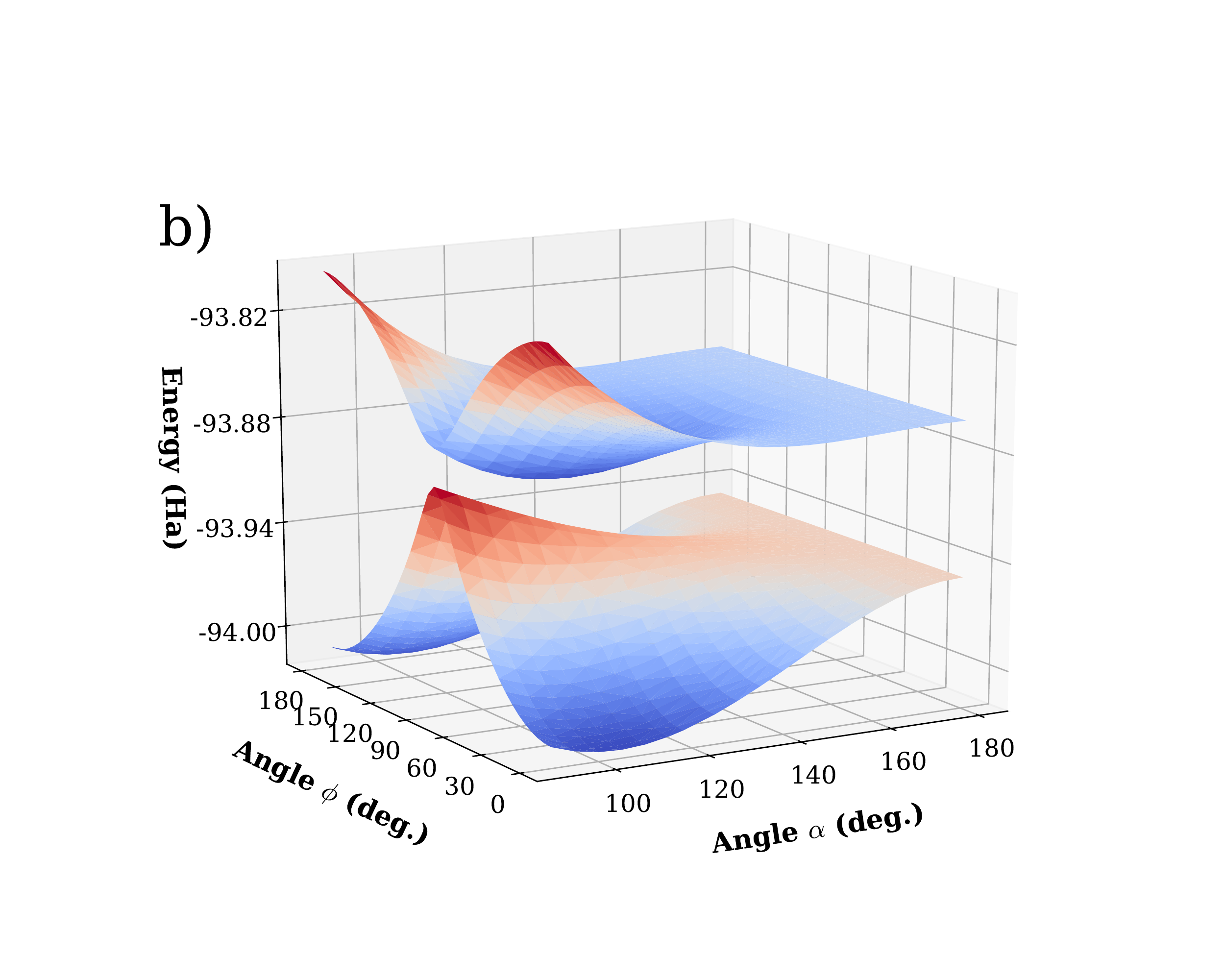}\\
    \includegraphics[width=8cm, trim= 3cm 2cm 3cm 4cm, clip=True]{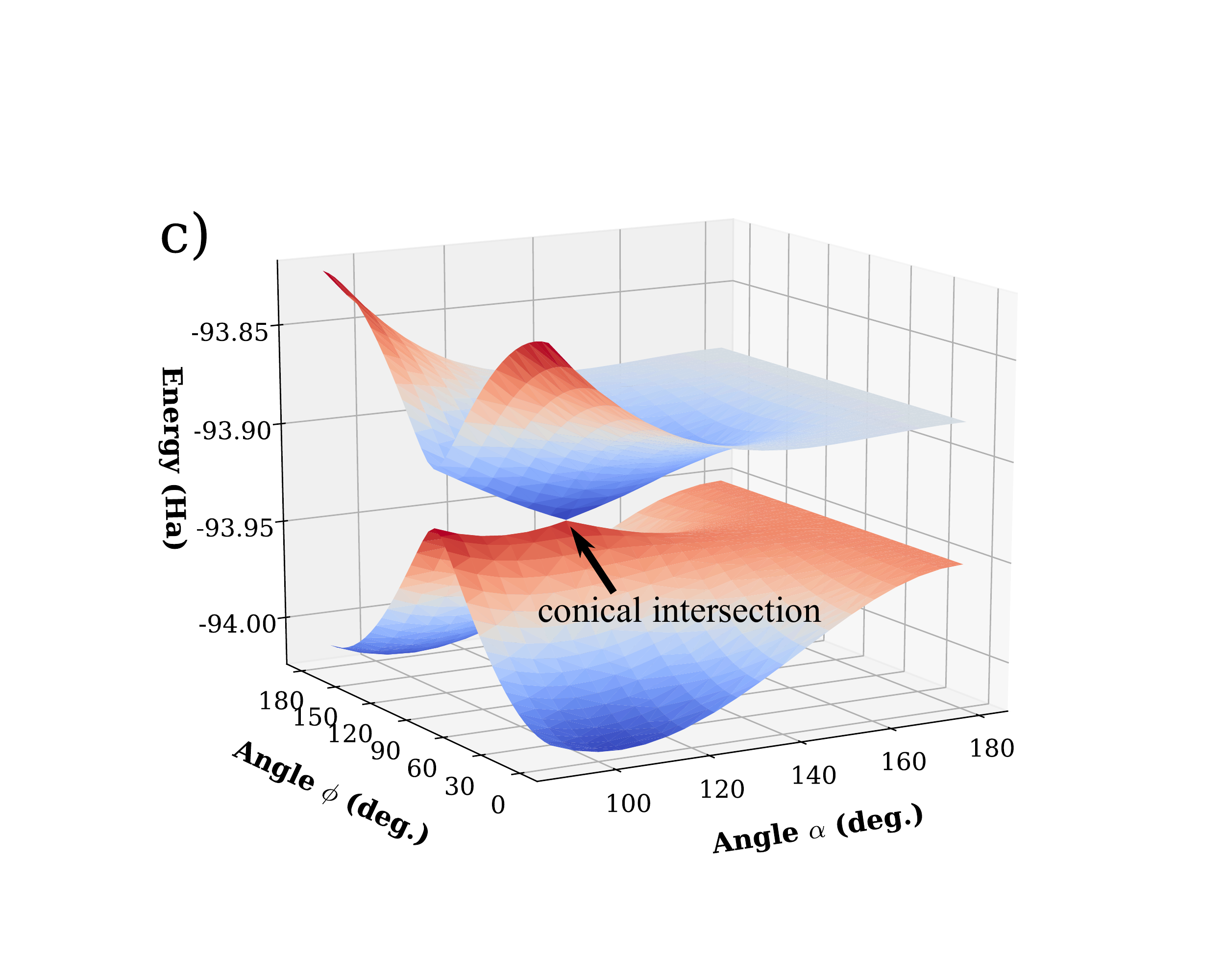}
    \caption{Two-dimensional ground- and excited-state PES of formaldimine as a function of the bending and dihedral angles $\phi$ and $\alpha$. {\bf a)} Structure of the formaldimine molecule. {\bf b)} CASCI(4,3) energies in the cc-pVDZ basis. {\bf c)} SA-CASSCF(4,3) energies in the cc-pVDZ basis (orbital optimisation realized over the first 20 MOs of the system).}
    \label{fig:3Dplot}
\end{figure}

As an illustrative example of the previous discussion, we consider the formaldimine molecule (CH$_2$NH) which exhibits a conical intersection between its ground and first-excited PESs~\cite{bonavcic1985photochemical}.
The photoisomerization of formaldimine can be considered as an illustrative  minimal model of the more complex RPSB molecule whose photoisomerisation (\textit{cis} to \textit{trans}) plays a key role in the visual cycle process~\cite{chahre1985trigger,birge1990nature}.
The description of the conical intersection can be reduced to a two-dimensional problem by varying the bending angle $\alpha \equiv \reallywidehat{\text{C--N--H}}$ and the dihedral angle $\phi \equiv \reallywidehat{\text{H--C--N--H}}$, while all other coordinates are kept frozen (see (a) in Fig.~\ref{fig:3Dplot}). It involves two states.
The first one is dominated by a closed-shell configuration composed of the $\pi$-bonding orbital between the nitrogen and carbon 2p$_x$ orbitals, and a lone pair in the nitrogen 2p$_y$ orbital denoted by $n$. 
This corresponds to the HF determinant.
The second state is dominated by an open-shell configuration called a $n\pi^*$ state, composed of one electron in the nitrogen lone pair orbital $n$, two electrons in the $\pi$-bonding orbital and one electron in the corresponding $\pi^*$-antibonding orbital~\cite{bonavcic1985photochemical}.
This corresponds to the HOMO--LUMO excited configuration
(HOMO and LUMO stand for highest occupied molecular orbital and lowest unoccupied molecular orbital, respectively).
An active space comprising those orbitals should be sufficient to qualitatively capture the conical intersection, \textit{i.e.} four electrons in three spatial-orbitals (4,3).
Note that one could even get rid of the $\pi$-bonding orbital, thus reducing the active space to two electrons in two spatial-orbitals (2,2), and still capture a conical intersection.
This could be used for an experiment on a real quantum device, as only four qubits are required.
However, we decided to keep the $\pi$-bonding orbital in the active space due to its non-negligible quantitative contribution, thus leading to a (noiseless) classical simulation of a six qubits device.

As readily seen in Fig.~\ref{fig:3Dplot}(b), realizing a CASCI(4,3) calculation in the canonical (HF) MO basis is not sufficient to capture the conical intersection.
The latter is however well described within SA-CASSCF(4,3) (Fig.~\ref{fig:3Dplot}(c)), which demonstrates the importance of working in the state-averaged orbital-optimized basis.
As a proof of the existence of the conical intersection (see also Refs.~\cite{bonavcic1985photochemical}), we considered a very large active space comprising all the electrons of the system (\textit{i.e.} $n = 16$) in 16 orbitals, in the canonical MO basis.
\begin{figure}
\centering
    \includegraphics[width=8cm]{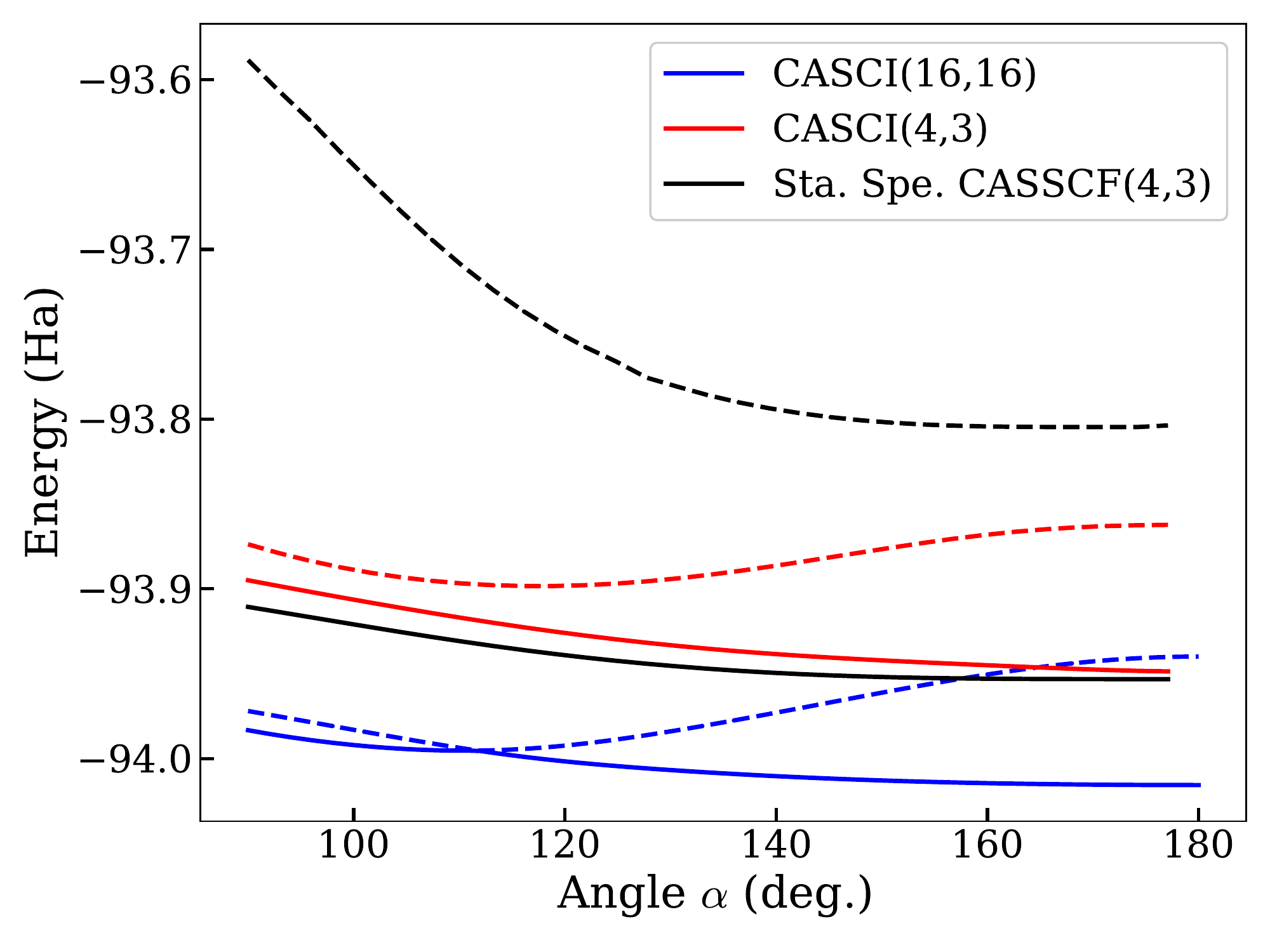}
    \caption{One-dimensional PES of formaldimine as a function of the bending angle $\alpha$ with a fixed dihedral angle $\phi=90^\circ$. 
    Ground-state (full lines) and excited-state (dashed lines) energies are computed with CASCI(16,16) (blue), CASCI(4,3) (red) and state-specific CASSCF(4,3) (black) in the cc-pVDZ basis.}
    \label{fig:SS_plot}
\end{figure}
Results are shown in Fig.~\ref{fig:SS_plot}.
For clarity, in this figure we only show the intersection plan of the 2D PES for $\phi = 90^\circ{}$ which contains the conical intersection.
By construction, the ground- and excited-state CASCI(16,16) energies are lower than the CASCI(4,3) ones, which we recall does not capture the conical intersection, as also shown in Fig.~\ref{fig:3Dplot}.
Interestingly, in contrast to CASCI(4,3), the CASCI(16,16) results feature the conical intersection around $\alpha = 112^\circ$.
This is already a numerical proof that the frozen core approximation does not guarantee the protection of the conical intersection (see Appendix~\ref{app:proof} for a mathematical proof).
To recover the conical intersection, one has to either increase the size of the active space, like in CASCI(16,16), or to switch to a more appropriate basis.
By optimizing the orbitals with respect to the ground-state energy, we end up with the state-specific CASSCF(4,3) result in Fig.~\ref{fig:SS_plot}.
As expected, the ground-state energy is lower than the CASCI(4,3) energy by construction. 
However, it remains higher than the CASCI(16,16) one.
What should be highlighted here is that while the ground-state energy is indeed getting better through the use of state-specific orbitals, this is at the expense of the description of the excited state.
As a consequence, the energy difference between the ground and first excited state deteriorates by using state-specific CASSCF, thus impacting directly the description of the conical intersection.

\section{Methods}\label{sec:method}

\subsection{Variational quantum algorithms}\label{sec:VQE}

The electronic structure problem may be solved on a quantum computer by mapping the fermionic modes onto qubits.
This may be achieved by a number of non-local transformations, most of which mapping $N$ spin-orbitals (not spatial orbitals) of an active space onto $N$ qubits (see Ref.~\onlinecite{cao2019quantum,mcardle2020quantum} for reference).
The transformed qubit Hamiltonian $\hat{\mathcal{H}}^{\mathrm{q}}$ has the same eigenvalue structure as the effective frozen core Hamiltonian $\hat{\mathcal{H}}^{\rm FC}$, constructed with respect to a given active space (see Appendix~\ref{appendix:FrozenCore} for an introduction of the frozen core approximation).
By virtue of the variational principle, all states within the $N$-qubit Hilbert space have energy $E\geq E_0^{\rm FC}$, where $E_0^{\rm FC}$ is the ground-state energy of $\hat{\mathcal{H}}^{\rm FC}$.
A variational quantum eigensolver (VQE)~\cite{peruzzo2014variational,mcclean2016theory} exploits this fact by generating a set of states $|\Psi(\vec{\theta})\rangle$ dependent on some conditional control parameters $\vec{\theta}$.
These parameters are used to tune quantum gates within a quantum circuit encoding an `ansatz'.
For a fixed choice of parameters $\vec{\theta}$, the energy of the generated state $|\Psi(\vec{\theta})\rangle$ under the Hamiltonian $\hat{\mathcal{H}}^{\mathrm{q}}$ may be estimated by partial state tomography~\cite{bonet2019nearly,huggins2020non}, which may be achieved by as simple methods as single qubit rotation and readout.
Traditionally, the energy $E(\vec{\theta})=\langle\Psi(\vec{\theta})|\hat{\mathcal{H}}^{\mathrm{q}}|\Psi(\vec{\theta})\rangle$ is used as a cost function in a (classical) optimization routine.
When $E(\vec{\theta})$ is minimized, one hopes that the resulting state $|\Psi(\vec{\theta})\rangle$ is a good approximation for the ground state of $\hat{\mathcal{H}}^{\mathrm{q}}$, and correspondingly that $E(\vec{\theta})\sim E_0^{\rm FC}$.
Various methods exist to extend this scheme to find higher-energy eigenstates~\cite{mcclean2017hybrid,ollitrault2019quantum,nakanishi2019subspace,ibe2020calculating,higgott2019variational,jones2019variational,jouzdani2019method,parrish2019quantum,parrish2019hybrid,bauman2019quantum,motta2020determining,zhang2020variational}.
However, the accuracy of the eigenenergies of $\hat{\mathcal{H}}^{\mathrm{q}}$ to the original problem is strictly limited by the active space, as we target $\hat{\mathcal{H}}^{\rm FC}$ and not the original Hamiltonian $\hat{\mathcal{H}}$.
Note that in the following, we will intentionally drop the notation $\hat{\mathcal{H}}^q$ when speaking about VQE to only use $\hat{\mathcal{H}}^{\rm FC}$ for simplification.

\subsection{State-Averaged Orbital-Optimized VQE}

\begin{figure*}
\centering
\includegraphics[width=\textwidth]{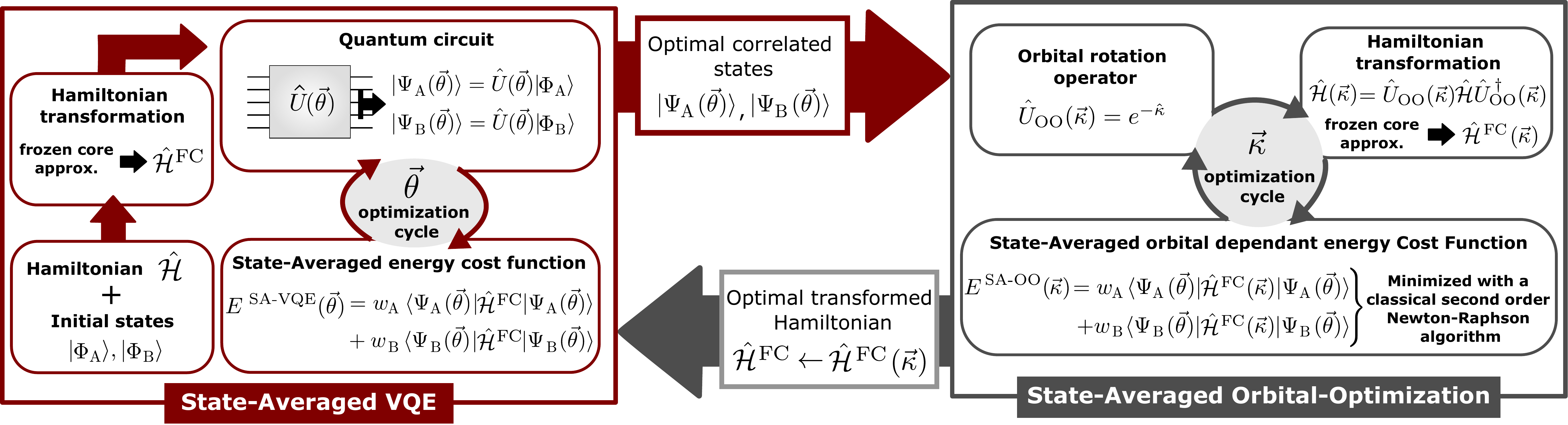}
\caption{Schematic of the SA-OO-VQE method. To produce an equal footing treatment of several states, two state-averaged algorithms are used in cycle : an hybrid quantum-classical SA-VQE algorithm (left-hand side), and a purely classical SA-Orbital-Optimization algorithm (right-hand side). In practice, the SA-VQE method employs a unique quantum circuit to determine multiple low lying eigenstates of a given Hamiltonian via a minimization of their state-averaged energy. The resulting optimal correlated states are then transmitted to the SA-OO algorithm. This second algorithm implements an optimal MO rotation over the whole Hamiltonian $\mathcal{\hat{H}}$ to pursue the minimization of the state averaged energy. The optimally rotated Hamiltonian is then sent back to the first SA-VQE algorithm, which closes the global loop of the SA-OO-VQE algorithm. }
\label{fig:scheme}
\end{figure*}

% As highlighted with the classical SA-CASSCF algorithm, a \textit{democratic} treatment of a system's eigenstates can be fundamental if we aim at capturing degeneracy with a restricted size active space. In this context, good quality results are only obtained when realizing a state-averaged orbital optimization which equitably takes into account each eigenstate during the energy optimization process. 

% To solve this problem, we can however take advantage of fundamentals concepts from classic algorithm that already demonstrated their effectiveness. As highlighted with the classical SA-CASSCF algorithm, we know that for a system such as formaldimine, a \textit{democratic} treatment of each eigenstate is fundamental if we aim at capturing degeneracy with a restricted size active space. In this context, qualitative results can only be obtained when no discrimination is done. The realization of a state-averaged orbital optimization which equitably takes into account each eigenstate during the energy optimization process.

With the current resources restriction affecting NISQ devices, VQE can only be employed to solve the electronic structure problem within very small-sized active spaces.
In this context, trying to reproduce complex chemical features such as formaldimine's conical intersection becomes a real challenge. 
Indeed, as explained previously, VQE is energetically bounded by its classical analog CASCI which is known to drastically fail when applied to a small active space (see CASCI(4,3) in Fig.\ref{fig:SS_plot}). 
If one desires to solve such a problem on near term quantum devices, it is of primary importance to develop new approaches which give coherent results while working within small-sized active spaces.

To this end, we introduce a method specially dedicated to capture (near-)degeneracies on a NISQ device: the so-called ``State-Averaged Orbital-Optimized VQE'' (SA-OO-VQE). 
Following the fundamental ideas that make the strength of the classical SA-CASSCF method, SA-OO-VQE provides a \textit{democratic} treatment of the eigenstates of a system inside and outside the active-space.
To this end, the SA-OO-VQE method is composed of two interdependent sub-algorithms working in a ``state-averaged'' manner. 
The latter are respectively called state-averaged VQE (SA-VQE) and state-averaged orbital-optimizer (SA-OO). 

In the following, we introduce how these two sub-algorithms work together in practice (as a visual support, Fig. \ref{fig:scheme} represents the general structure of the SA-OO-VQE method and provides a summary of the information exchanged between the two sub-algorithms).
Motivated by the formaldimine problem, a two-states example will be considered in the following to illustrate the different steps of the method.
SA-OO-VQE can however be straightforwardly generalized to any number of states, as well as to the particular case of a single state (thus leading to `state-specific'-OO-VQE~\cite{mizukami2019orbital,sokolov2020quantum}).

% First, a state-averaged VQE (SA-VQE) method is employed to determine the lowest energy eigenstates of the system within the active-space. Then, a state-averaged Orbital-Optimization (SA-OO) algorithm is employed to optimize the system's MOs  (inside and outside the active space) with respect to the lowest eigenstates found by SA-VQE. Looping together the SA-VQE and the SA-OO algorithms constitutes then the essence of how the SA-OO-VQE method works in practice as depicted in Fig.\ref{fig:scheme}.

\subsubsection{1st algorithm: State-Averaged VQE}\label{sec:SA-VQE}

The first algorithm acting in the SA-OO-VQE method is the SA-VQE approach, also called subspace-search VQE (SS-VQE) in Ref.~\onlinecite{nakanishi2019subspace}, which consists in finding two (or more) low-lying eigenstates of a given Hamiltonian.
To proceed, one starts by building the frozen core Hamiltonian $\mathcal{\hat{H}}^{\rm FC}$.
Then, we define a set of two (or more) orthonormal initial states $\ket{\Phi_{\rm A}}$ and $\ket{\Phi_{\rm B}}$ (with $\langle\Phi_{\rm A}| \Phi_{\rm B}\rangle = 0$) to be transformed into two correlated states via the implementation of a same unitary operator $\hat{U}(\vec{\theta})$,
\begin{eqnarray}
 |\Psi_{\rm A}(\vec{\theta})\rangle  = \hat{U}(\vec{\theta})|\Phi_{\rm A}\rangle \text{ and }
 |\Psi_{\rm B}(\vec{\theta})\rangle  = \hat{U}(\vec{\theta})|\Phi_{\rm B}\rangle.
\end{eqnarray}
The use of a same unitary $\hat{U}(\vec{\theta})$ where the parameters $\vec{\theta}$ are optimized for both states simultaneously has multiple advantages.
First, a single circuit is used to determine eigenstates which is a great asset for the NISQ era. 
Second, all target states are treated on an equal footing, thus avoiding biases that would naturally appear when using different levels of theory for each state.
In Appendix~\ref{appendix:methods}, we discuss on the ability of several methods to treat different states on an equal footing.
Finally, the orthonormality property of the initial states is transferred to the final correlated states, so that $\langle\Psi_{\rm A}(\vec{\theta}) |\Psi_{\rm B}(\vec{\theta})\rangle = \langle \Phi_{\rm A} | \Phi_{\rm B} \rangle = 0$.
The two correlated states are then used as ansatze to approximate the two low-lying eigenstates of $\hat{\mathcal{H}}^{\rm FC}$.
To this end, the parameters $\vec{\theta}$ of the VQE circuit are optimized to minimize the state-averaged energy
\begin{eqnarray}
      E^\text{SA-VQE}(\vec{\theta}) &= w_{\rm A} \langle\Psi_{\rm A} (\vec{\theta})|\hat{\mathcal{H}}^{\rm FC}|\Psi_{\rm A}(\vec{\theta})\rangle \nonumber \\ 
      &+ w_{\rm B} \langle\Psi_{\rm B} (\vec{\theta})|\hat{\mathcal{H}}^{\rm FC}|\Psi_{\rm B}(\vec{\theta})\rangle,
\label{eq:SA-VQE}
\end{eqnarray}
where the weights $w_{\rm A}$ and $w_{\rm B}$ (with $w_{\rm A}+w_{\rm B}=1$ and $w_A \geq w_B$) define the weighted contribution of each correlated state to the cost function.
This ensemble energy is governed by the Rayleigh--Ritz variational principle~\cite{gross1988rayleigh} ensuring that the lower energy bound reachable is given by the ensemble energy of the exact two low lying eigenstates of $\mathcal{\hat{H}}^{\rm FC}$ (denoted by $\ket{\Psi_0}$ and $\ket{\Psi_1}$),
\begin{eqnarray}
E^{\rm SA} &=& w_A \bra{\Psi_0} \hat{\cal{H}}^{\rm FC} \ket{\Psi_0}
+ w_B \bra{\Psi_1} \hat{\cal{H}}^{\rm FC} \ket{\Psi_1} \nonumber \\
& \leq & E^{\rm SA-VQE} (\vec{\theta}).
\end{eqnarray}
In the particular case of an equi-ensemble (\textit{i.e.} $w_A = w_B$), it is easy to show that the state-averaged energy  is invariant with respect to any rotation of the states $\ket{\Psi_0}$ and $\ket{\Psi_1}$.
Indeed, by replacing $\ket{\Psi_0}$ and $\ket{\Psi_1}$ by $\ket{\Psi_-(\varphi)} = \cos \varphi \ket{\Psi_0} - \sin \varphi \ket{\Psi_1}$ and $\ket{\Psi_+(\varphi)} = \sin \varphi \ket{\Psi_0} + \cos \varphi \ket{\Psi_1}$, one gets an ensemble energy \linebreak $E^{\rm SA}(\varphi) \geq E^{\rm SA}(0)$ where the equality is always fulfilled (for any rotation angle $\varphi$) if and only if $w_A = w_B$~\cite{gross1988rayleigh}.
Hence, taking an equi-ensemble only ensures that the subspace spanned by $\vert \Psi_A(\vec{\theta}) \rangle$ and $\vert \Psi_B(\vec{\theta}) \rangle$ best approximates the subspace spanned by $\ket{\Psi_0}$ and $\ket{\Psi_1}$, instead of approximating each individual states directly.
To extract the expected individual states and energies from the equi-ensemble, additional treatments are required such as  additional maximizations~\cite{nakanishi2019subspace} or classical diagonalization~\cite{parrish2019quantum}.
Alternatively, one can just take $w_A > w_B$ to impose a constraint in the ordering of the states~\cite{nakanishi2019subspace}, which may complicate the optimization considerably~\cite{zhang2020variational}.
We refer to this alternative as the fully-weighted SS-VQE, described in Sec.~IIC of Ref.~\onlinecite{nakanishi2019subspace}.

In this work, we considered the equi-ensemble formalism.
While there is no guarantee to capture the correct individual states after a single minimization of the state-average energy [Eq.~(\ref{eq:SA-VQE})], this remains possible by choosing appropriate initial states and wavefunction ansatz.
Indeed, in Sec.~\ref{sec:results} we show that the correct individual energies are obtained with errors even below chemical accuracy.
Furthermore, as no ordering is specified, the PES of the final states ($\ket{\Psi_{\rm A}}$ and $\ket{\Psi_{\rm B}}$) can cross: $\ket{\Psi_{\rm A}}$ and $\ket{\Psi_{\rm B}}$ can respectively be the ground and excited state for a given set of nuclear coordinates $\bfR$, and reverse order for another set $\bfR'$.
This interesting feature is not without significance and will be discussed further in Sec.~\ref{sec:results}.

Nevertheless, in general one has to ensure the correct capture of the eigenstates. 
For this, one possibility is to complement the cost function in Eq.~(\ref{eq:SA-VQE}) by the state-averaged variance~\cite{zhang2020variational}
\begin{eqnarray}
\Delta^{\rm SA}(\vec{\theta}) = w_{\rm A} \Delta_{\rm A}(\vec{\theta}) + w_{\rm B} \Delta_{\rm B}(\vec{\theta}),
\label{eq:SA-variance}
\end{eqnarray}
where
\begin{equation}
\begin{split}
    \Delta_k (\vec{\theta}) &= \langle \Psi_k(\vec{\theta}) \vert (\hat{\mathcal{H}}^\text{FC})^2 \vert \Psi_k(\vec{\theta})\rangle \\
&- \langle \Psi_k(\vec{\theta})\vert\hat{\cal{H}}^\text{FC} \vert\Psi_k(\vec{\theta})\rangle^2
\end{split}
\end{equation}
is the energy variance of Hamiltonian $\hat{\cal{H}}^\text{FC}$ with wavefunction $|\Psi_k(\vec{\theta})\rangle$. In practice, the measurement of the energy variance on a quantum computer can be realized via the averaging of a quantum covariance matrix depending on the output states and the elements of the Hamiltonian $\hat{H}$ (expressed in qubit form). For more information about this method, we refer the interested reader to Ref.~\onlinecite{zhang2020variational} (and references within).

\subsubsection{2nd algorithm: State-Averaged Orbital-Optimization}\label{sec:SA-OO}

Once the SA-VQE sub-algorithm reaches convergence, the next step consists in using a classical SA-OO procedure.
Directly inspired by the SA-CASSCF method, the goal of the SA-OO algorithm is to pursue the minimization of the state-averaged energy of $| \Psi_{\rm A}(\vec{\theta}) \rangle $ and $| \Psi_{\rm B}(\vec{\theta}) \rangle$ via the implementation of an optimal rotation of the system's MOs. 
To do so, throughout the whole SA-OO process, one considers that the shape of the correlated states never changes (\textit{i.e.} the $\vec{\theta}$ parameters are frozen). 
Then, one introduces an unitary operator $U_\text{OO}(\vec{\kappa})$ such as
\begin{equation}
    \hat{U}_\text{OO}(\vec{\kappa}) = e^{-\hat{\kappa}} \text{ with } \hat{\kappa} = \sum_{p>q}^\text{MOs} \kappa_{pq}(\hat{E}_{pq}-\hat{E}_{qp}),
\label{eq:U-OO}
\end{equation}
where $p,q$ denote any arbitrary spatial orbitals.
This operator is used to transform the system's MOs.
Such a transformation acts simultaneously inside and outside the active space~\cite{helgaker2014molecular}. 
In practice, a restricted set of MOs is considered, including all doubly occupied MOs, all active space MOs and a few low virtuals. 
Using the $\hat{U}_\text{OO}(\vec{\kappa})$ operator, the MO basis transformation is applied to the full second quantized Hamiltonian of the system,
\begin{equation}
    \hat{\mathcal{H}}(\vec{\kappa}) = \hat{U}_\text{OO}^\dagger(\vec{\kappa})\, \hat{\mathcal{H}}\, \hat{U}_\text{OO}(\vec{\kappa}).
\label{eq:ham_OO}
\end{equation}
This transformation parametrizes the system's Hamiltonian which becomes explicitly dependent on the MO basis, through rotation $\vec{\kappa}$
 (see Appendix.~\ref{appendix:HamTransfo} for details about this transformation).
The resulting transformed Hamiltonian $\hat{\mathcal{H}}(\vec{\kappa})$ still shares the same spectrum as the original operator $\hat{\mathcal{H}}$ since we use a unitary transformation. 
However, this is not the case for the associated frozen core Hamiltonian $\hat{\mathcal{H}}^{\rm FC}(\vec{\kappa})$ which has a different spectrum compared to the original frozen core Hamiltonian $\hat{\mathcal{H}}^{\rm FC}$ (in the non-optimized MO basis).
The energy of the correlated states computed in the active space will therefore change with the parameters $\vec{\kappa}$.
Using this new definition of the frozen core Hamiltonian, the state-averaged energy of the two correlated states now reads (for fixed parameters $\vec{\theta}$)
\begin{eqnarray}
      E^\text{SA-OO}(\vec{\kappa})  &=&
    w_{\rm A} \langle\Psi_{\rm A} (\vec{\theta})|\hat{\mathcal{H}}^{\rm FC}(\vec{\kappa})|\Psi_{ \rm A} (\vec{\theta})\rangle \nonumber \\
      &&+ w_{\rm B} \langle\Psi_{\rm B} (\vec{\theta})|\hat{\mathcal{H}}^{\rm FC}(\vec{\kappa})|\Psi_{ \rm B} (\vec{\theta})\rangle .
\label{eq:OO-CF}
\end{eqnarray} 
By minimizing Eq.~(\ref{eq:OO-CF}) with respect to the parameters $\vec{\kappa}$, one obtains a  variationally optimized MO basis designed to treat both correlated states on an equal footing.

Throughout this work, this optimization is carried out by using a state-averaged version of the classical Newton-Raphson method \cite{fletcher2013practical} (see Appendix~\ref{appendix:NewtonRaphson} for technical details). 
In practice, this algorithm requires the knowledge of the two converged correlated states $|\Psi_A(\vec{\theta})\rangle$ and $|\Psi_B(\vec{\theta})\rangle$ obtained from the SA-VQE algorithm. 
More precisely, it requires their respective one- and two-electron reduced density matrices, which are measured from the VQE circuit anyhow to compute the energy.
Hence, no additional measurements are required.
These matrices are used in the analytical formula for the so-called orbital gradient and Hessian which represent fundamental ingredients of the orbital-optimisation process~ \cite{siegbahn1981complete,helgaker2014molecular,yarkony1995modern}. 
Note that this approach based on reduced density matrices contrasts with regular VQE gradients which require additional measurements \cite{grimsley2019adaptive,tang2019qubit}.

To close the loop of the SA-OO-VQE algorithm, the (now orbital-optimized) frozen core Hamiltonian $\hat{\mathcal{H}}^{\rm FC}(\vec{\kappa})$ is sent back to the SA-VQE algorithm which will determine the two low-lying states again. 
Such a loop between the SA-VQE and the SA-OO algorithms is carried out until reaching a global convergence of the state-averaged energy (in addition to the state-averaged variance if used).

\subsubsection{Generalized unitary coupled cluster with double-excitation operators}\label{sec:ansatz}

In VQE-type algorithms, like SA-VQE, one has to choose a specific (and usually approximate) wavefunction ansatz.
One promising (and now traditional) ansatz considered for quantum chemistry applications within VQE is the unitary coupled cluster ansatz with single- and double-excitation operators (UCCSD)~\cite{mcclean2016theory}:
\begin{eqnarray}
    &&U(\vec{\theta})=e^{\hat{T}(\vec{\theta})-\hat{T}^{\dag}(\vec{\theta})},\\
    &&\hat{T}\equiv \hat{T}_{\rm SD} =  \sum_{a}^\text{virt.} \sum_{i}^\text{occ.} \theta_{ai} \, \hat{a}^\dagger_{a}  \hat{a}_{i}   + \sum_{a>b}^\text{virt.} \sum_{i>j}^\text{occ.} \theta_{abij}  \, \hat{a}^\dagger_{a} \hat{a}^\dagger_{b} \hat{a}_{i} \hat{a}_{j},\nonumber\\
    \label{eq:SD}
\end{eqnarray}
where $i,j$ and $a,b$ denote occupied and virtual spin-orbitals (in reference to the HF Slater determinant) \textit{within the active space}, respectively.
However, it might be of interest to consider generalizations of Eq.~(\ref{eq:SD})~\cite{lee2018generalized,greene2019generalized,higgott2019variational}.
In this work, we use a UCC ansatz based on a generalized spin-free form of double-excitation operator
\begin{align}
    \hat{T} &= \sum_{t, v, w, u}^\text{active} \theta_{tuvw} \sum_{\sigma,\tau=\uparrow,\downarrow} \hat{a}^\dagger_{t\sigma} \hat{a}^\dagger_{v\tau} \hat{a}_{w\tau} \hat{a}_{u\sigma},
\label{eq:GCC}
\end{align}
where $t,u,v,w$ denote any active spatial orbital. Comparing Eq~(\ref{eq:GCC}) to Eq.~(\ref{eq:SD}), there are three key alterations using a generalized spin-free double-excitation operator.
First, all single excitations can be removed in Eq.~(\ref{eq:GCC}) as these are already included in the unitary operator $\hat{U}_\text{OO}(\vec{\kappa})$ [Eq.~(\ref{eq:U-OO})] as part of the SA-OO algorithm, which effectively encodes all single excitations in the active space.
This allows savings of quantum resources by performing the single-excitations classically instead of within the quantum circuit.
Second, the traditional UCCSD ansatz only considers excitations from occupied to unoccupied spin-orbitals, and are usually defined according to one specific Slater determinant (the HF determinant).
However, different initial states are considered in the SA-OO-VQE method.
As a direct consequence, operators such as in Eq.~(\ref{eq:SD}) are not flexible enough to simultaneously generate multiple correlated states.
For this task, generalized excitation operators are more suitable as the latter remain intrinsically state-agnostic.
Finally, Eq.~(\ref{eq:GCC}) is chosen to match the spin-free form of the two-body operator present in the electronic Hamiltonian in Eq.~(\ref{eq:el_Ham}).
This further reduces the number of independent cluster parameters $\theta$ to optimize (four chains of creation-annihilation operators share the same parameters and the symmetry $\theta_{tuvw}=\theta_{vwtu}$ can be exploited).

In Appendix~\ref{appendix:gen_ansatz}, we discuss how we implemented this ansatz in practice to produce the results presented in our study.
Notably, we introduce step-by-step every manipulations realized to make the ansatz tractable in our simulations (\textit{i.e.} using restrictions to reduce the number of parameters to optimize and invoking symmetries). 
We also provide additional informations about the resulting quantum circuit (such as the ordering of the unitaries and the total number of quantum gates).

% the computational cost of optimisation 

% Note that in practice, due to the large computational cost arising when using such a generalized ansatz, we chose to simplify its shape by discarding some excitations terms (and invoking some symmetry properties). In appendix \ref{appendix:gen_ansatz}, we discuss these features and introduce every technical details necessary to reproduce the results presented in the following sections.

% Note that in practice, due to the large computational cost arising when using such a generalized ansatz, we chose to simplify its shape by discarding some excitations terms (and invoking some symmetry properties). In appendix \ref{appendix:gen_ansatz}, we discuss these features and introduce every technical details necessary to reproduce the results presented in the following sections.

\section{Computational details}\label{sec:comp}

The simulation of the SA-OO-VQE algorithm is realized using the python quantum computing packages \textbf{OpenFermion}~\cite{mcclean2020openfermion} and \textbf{Cirq}~\cite{cirq}. 
In practice, SA-VQE parameters $\vec{\theta}$ are always initialized to zero and optimized using the gradient-free Sequential Least Squares Programming method (SLSQP) from the python \textbf{Scipy} package.
For each call of SA-VQE, the SLSQP method is run with a maximum number of 400 iterations and a precision threshold of $10^{-4}$ Ha (\textit{i.e} we set the optimizer parameter ``ftol'' to $10^{-4}$ as implemented in the scipy package).
This low expectation on the performance of SA-VQE is rendered possible in SA-OO-VQE thanks to alternating repetitions of SA-VQE and SA-OO algorithms.
The threshold for the global convergence of SA-OO-VQE is also set to $10^{-4}$ Ha. 
The two initial orthonormal states considered are the HF Slater determinant $| \Phi_{\rm A} \rangle = | \text{HF}\rangle$, and its HOMO--LUMO excited version $| \Phi_{\rm B} \rangle = \hat{E}_\text{LH}| \text{HF} \rangle = (1/\sqrt{2}) \sum_\sigma \hat{a}^\dagger_{\text{L},\sigma} \hat{a}_{\text{H},\sigma}\ket{\rm HF}$ (with `L' and `H' referring to the LUMO and HOMO spatial orbitals, respectively).
Concerning the SA-OO algorithm, a homemade state-averaged Netwon-Raphson code has been developed based on Refs.~\cite{helgaker2014molecular,siegbahn1981complete,fletcher2013practical}.

Turning to the geometry of formaldimine, the N--CH2 part of the molecule is frozen and constrained in a same plane (see green plane in Fig.\ref{fig:3Dplot}(a)). 
The interatomic distances are $d_\text{N--C}=1.498~\text{\AA}$ and $d_\text{C--H}=1.067~\text{\AA}$ and the internuclear angles are $\reallywidehat{\text{N--C--H}} = 118.36^\circ$. 
The second H atom is symmetric to the first one with respect to the N--C axis.
The two remaining degrees of freedom characterize the out-of-plane bending angle $\alpha \equiv \reallywidehat{\text{H--N--C}}$ and the dihedral angle $\phi \equiv \reallywidehat{\text{H--N--C--H}}$.

In the following, the cc-pVDZ basis is used and an active space of four electrons in three orbitals (4,3) is considered, unless stated otherwise.
The SA-OO procedure is always realized over the first 20 spatial-orbitals of the system (for SA-OO-VQE and SA-CASSCF) composed of the six frozen occupied, the three active and the first eleven virtual orbitals.
CASCI and SA-CASSCF calculations are realized with the quantum chemistry code Psi4~\cite{smith2020psi4}.

\section{Results}\label{sec:results}
We now present the results obtained with SA-OO-VQE for formaldimine's ground- and excited-state spectra around the conical intersection.

\subsection{Convergence of the SA-OO-VQE method}

\begin{figure}
\centering
    \includegraphics[width=8cm]{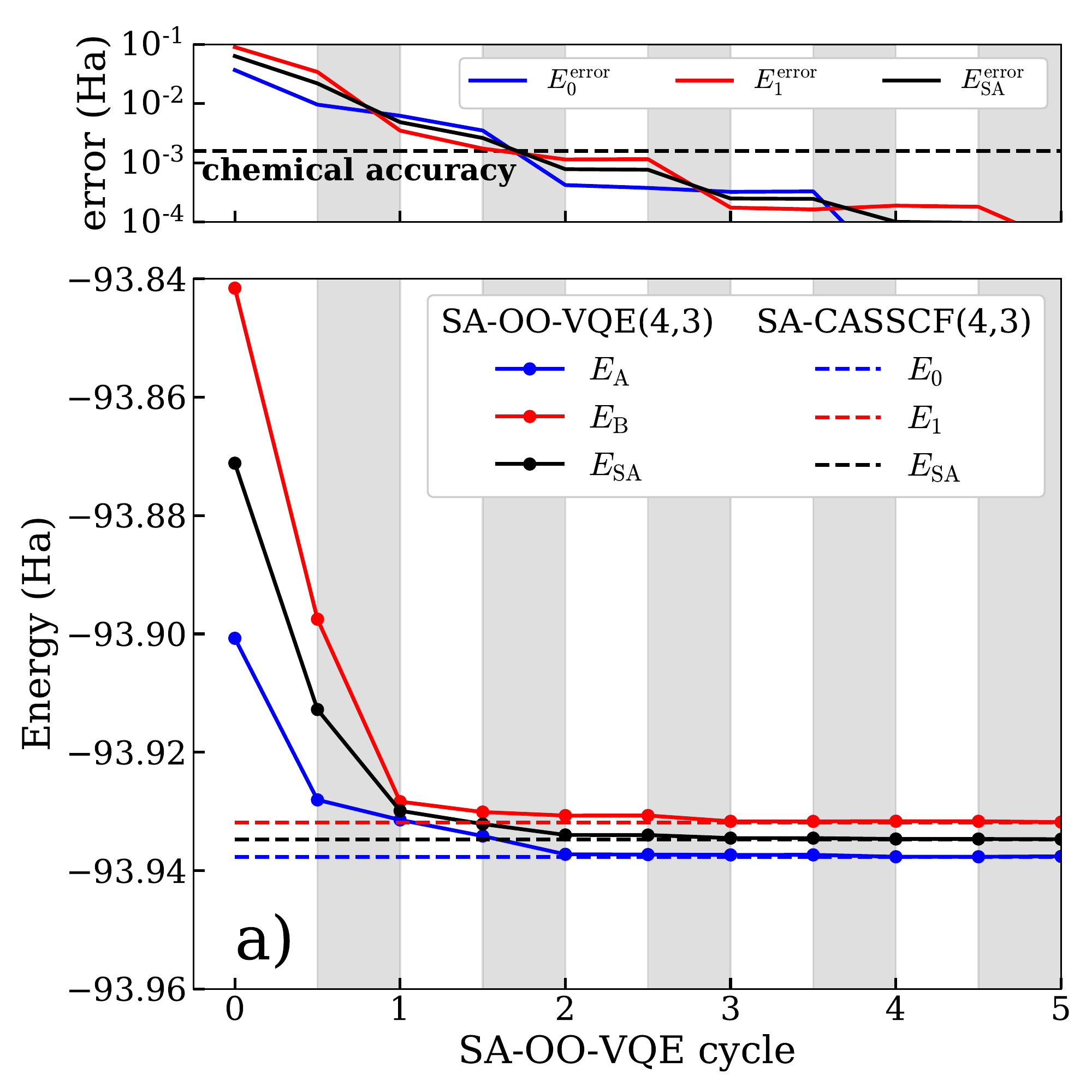}
    \includegraphics[width=8cm]{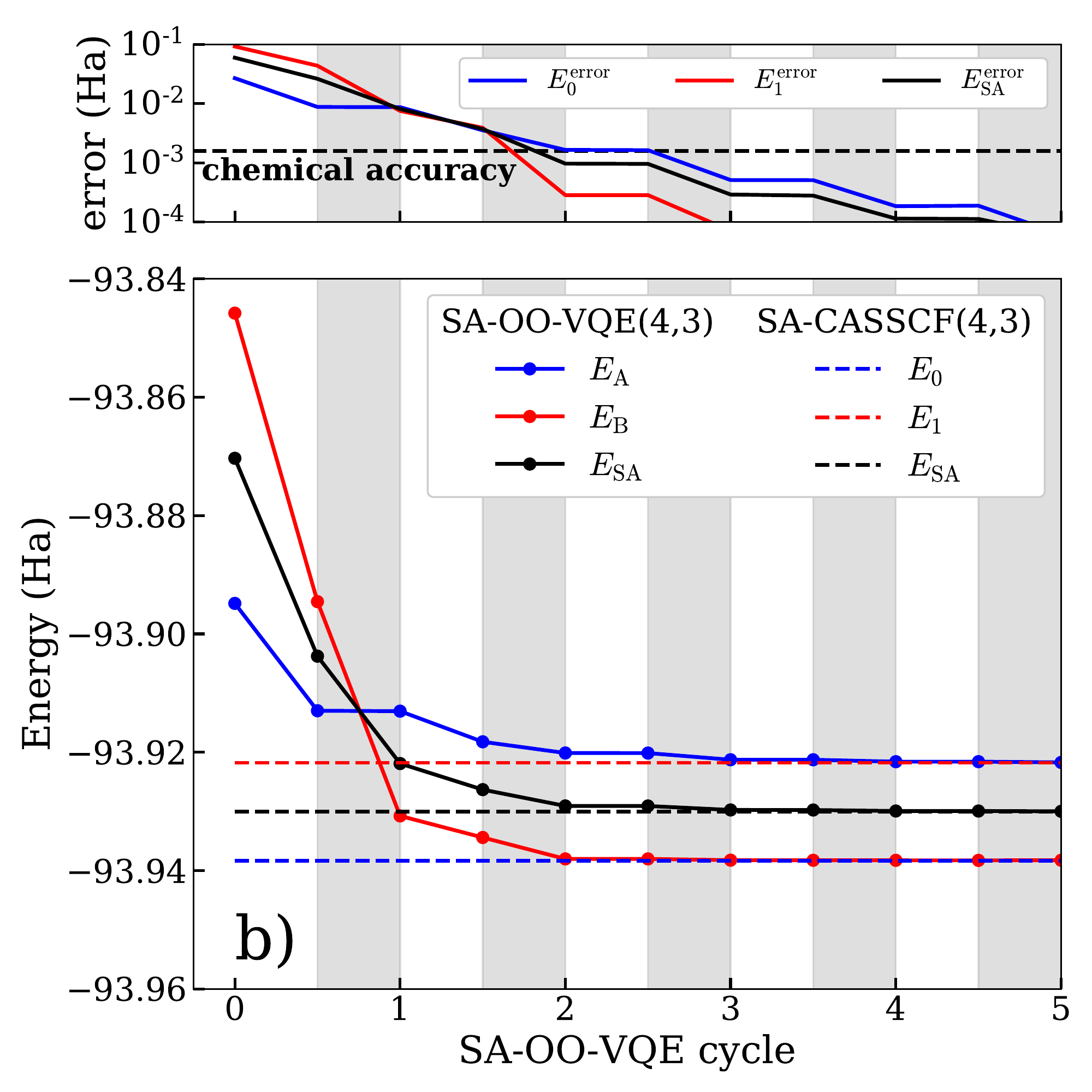}
    \caption{Evolution of the state-averaged energy optimization during the different steps of the SA-OO-VQE algorithm.
    {\bf a)} For $\phi=90^\circ$ and $\alpha=122.7^\circ$, and {\bf b)} for $\phi=90^\circ$ and $\alpha=106^\circ$.
    For both cases, white strips delimit SA-VQE phases whereas grey stripes delimit SA-OO phases. A global cycle of SA-OO-VQE is then given by two consecutive stripes (white then grey). The upper panels represent the energy error comparing SA-OO-VQE to reference SA-CASSCF. }
    \label{fig:optimization_process}
\end{figure}

We start by illustrating how SA-OO-VQE converges towards the minimal state-averaged energy for a given geometry of the molecule ($\phi=90^\circ$ and $\alpha=122.7^\circ$).
The different energy values produced by the algorithm after each step of optimization are given in Fig.~\ref{fig:optimization_process}a (where $E_0$ and $E_1$ stand for the ground- and first excited-state energies, $E_{\rm A}$ and $E_{\rm B}$ are the associated energies of $\ket{\Psi_{\rm A}}$ and $\ket{\Psi_{\rm B}}$, and $E_{\rm SA}$ is the state-averaged energy).
SA-VQE phases are represented by white stripes while SA-OO phases are represented with grey stripes.
One SA-OO-VQE cycle is then given by two consecutive white and grey stripes, repeated until global convergence of the state-averaged energy is reached.
We compare the convergence of SA-OO-VQE to the SA-CASSCF reference energies (horizontal dashed lines), which form a natural lower bound on the performance of the method.\linebreak
As readily seen in Fig.~\ref{fig:optimization_process}(a), alternating between the SA-VQE and the SA-OO algorithms progressively lowers the energies, requiring five SA-OO-VQE cycles to reach global convergence, and only two cycles to reach chemical accuracy compared to the SA-CASSCF reference.
Similar convergence was also observed across all simulated formaldimine geometries, with all SA-OO-VQE simulations converging within a maximum of 10 cycles to within chemical accuracy (1.6 mHa) of the SA-CASSCF reference.
This fast convergence on the whole one-dimension PES is a consequence of the equi-ensemble theory.
Indeed, each state can naturally evolve towards the final state which they have the strongest overlap with, without having to obey any \textit{a priori} ordering (see Sec.~\ref{sec:SA-VQE}).
This is depicted in Fig.~\ref{fig:optimization_process}(b) where the energies $E_{\rm A}$ and $E_{\rm B}$ cross each other in the first cycle of SA-OO-VQE, which is by construction not favored by setting $w_{\rm A} > w_{\rm B}$.
The importance of such a feature will be discussed in the following.

\begin{figure}
\centering
    \includegraphics[width=8cm]{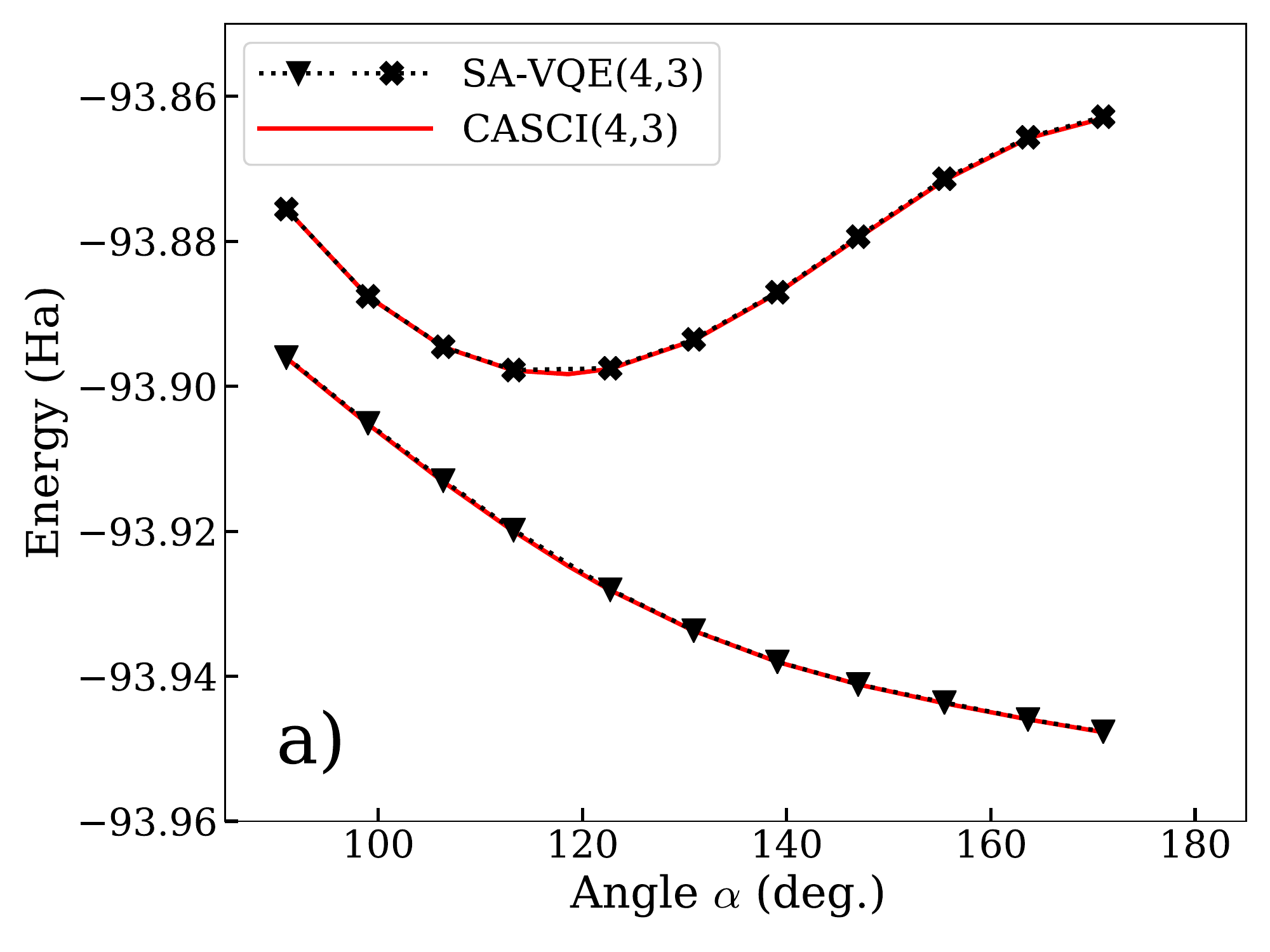}\\
    \includegraphics[width=8cm]{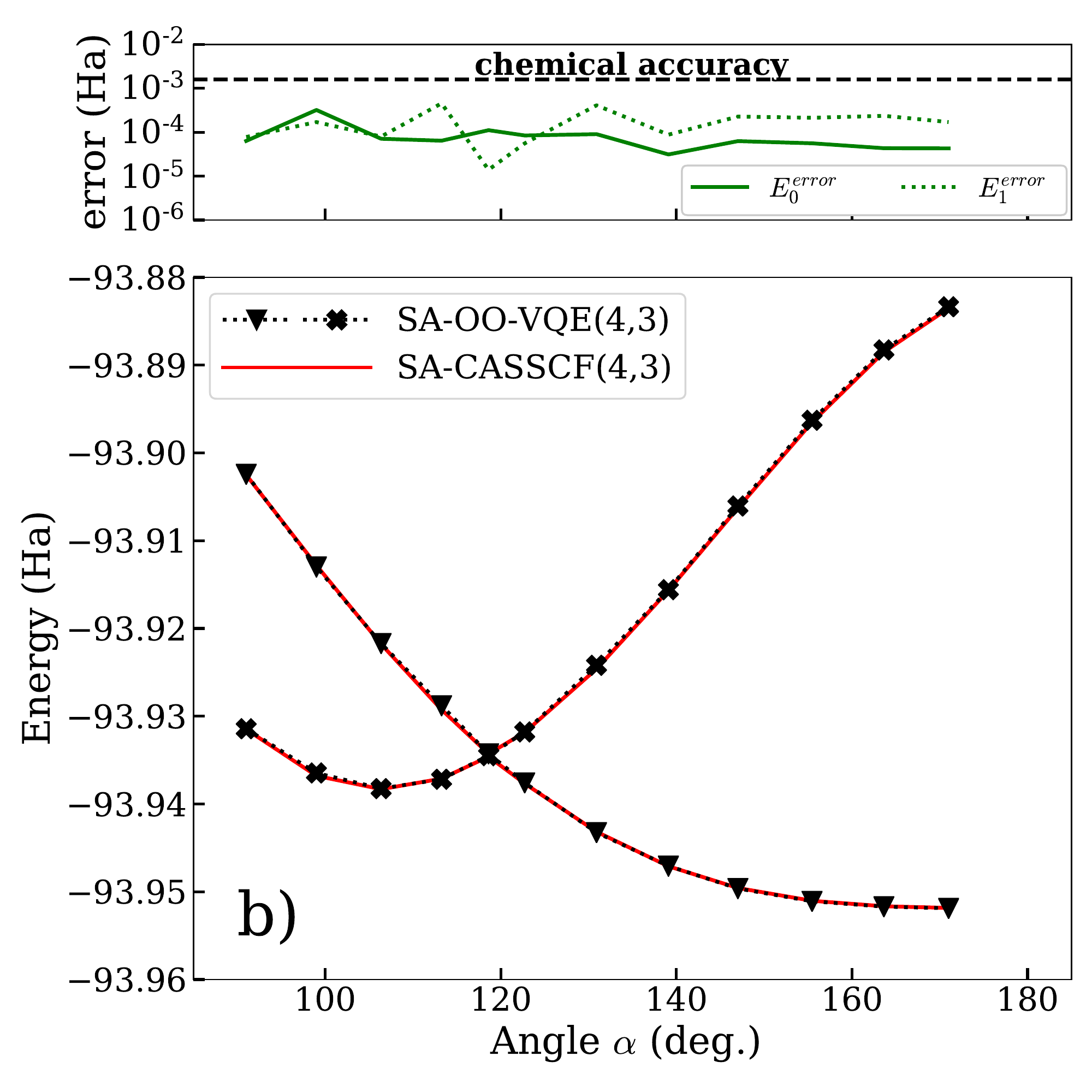}
    \caption{One-dimensional PES of formaldimine as a function of the bending angle $\alpha$ with a fixed dihedral angle $\phi=90^\circ$. Triangles and crosses markers represent the energy of the final states $\ket{\Psi_{\rm A}}$ and $\ket{\Psi_{\rm B}}$ associated to the initial states  $\ket{\Phi_{\rm A}}=\ket{\text{HF}}$ and $\ket{\Phi_{\rm B}}=\hat{E}_\text{LH} \ket{\text{HF}}$, respectively.
    {\bf a)} Comparison of CASCI vs. SA-VQE. {\bf b)} Lower panel: Comparison of SA-CASSCF vs. SA-OO-VQE. Upper panel: energy error of SA-OO-VQE compared to SA-CASSCF. }
    \label{fig:2D-PES}
\end{figure}

\subsection{Description of the conical intersection}

Let us now investigate the ability of SA-OO-VQE to reproduce the conical intersection of the formaldimine molecule.
We focus on the calculation of the one-dimensional PES of the system as function of the bending angle $\alpha$ with a fixed dihedral angle $\phi=90^\circ$ (Fig.~\ref{fig:2D-PES}).
In the absence of orbital optimization (i.e. when the SA-VQE algorithm acts alone), the energies are in very good agreement with the classical CASCI calculation [Fig.~\ref{fig:2D-PES}(a)].
However, no conical intersection is present in the energy profiles thus demonstrating a qualitative failure of the frozen core approximation in the canonical MO basis (see Secs.~\ref{sec:CI_num} and \ref{sec:formaldimine}).
To reproduce the conical intersection, the SA-OO algorithm is required.
In Fig.~\ref{fig:2D-PES}(b), we plot the energies obtained following global convergence of the full SA-OO-VQE method.
We see excellent qualitative and quantitative agreements between SA-OO-VQE and its classical analog SA-CASSCF.
Crucially, both methods are able to capture the presence of a conical intersection around $\alpha =118.5^\circ$. Furthermore, we measure all over the PES an individual energy error between SA-OO-VQE and SA-CASSCF that is always below chemical accuracy (see upper panel of Fig.~\ref{fig:2D-PES}(b)).

\begin{figure}
\centering
    \includegraphics[width=8cm]{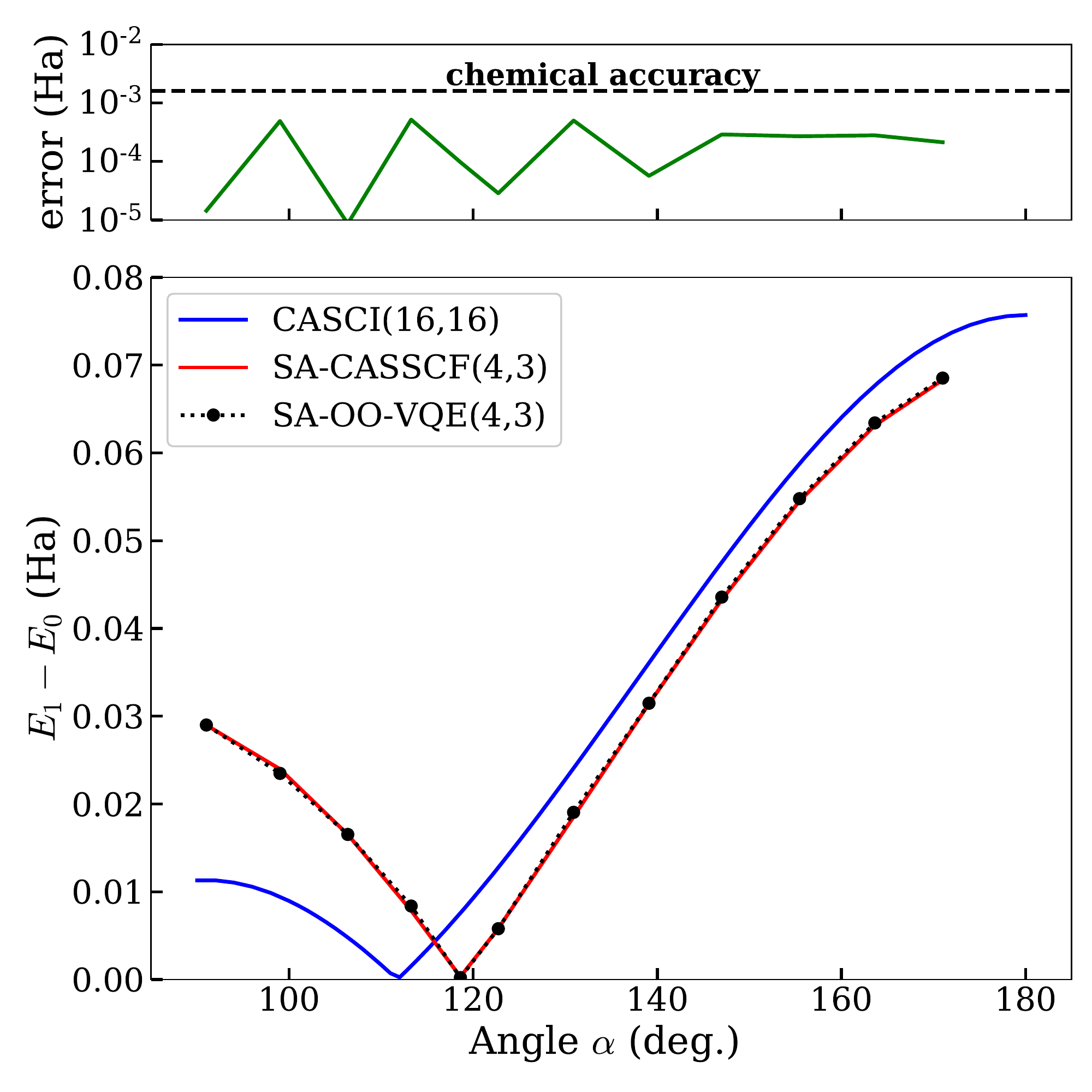}
    \caption{Lower panel : Comparison of the energy difference obtained with SA-CASSCF, SA-OO-VQE and a reference CI calculation with a very large active space CASCI(16,16). Upper panel : energy error comparing SA-OO-VQE excitation energies to reference SA-CASSCF. }
    \label{fig:energy_diff}
\end{figure}
Another important quantity in photochemistry is the excitation energy --- the energy difference between the ground and the first excited states.
We compare the estimated excitation energies from SA-CASSCF and SA-OO-VQE in Fig.~\ref{fig:energy_diff}.
Here again, SA-OO-VQE matches SA-CASSCF perfectly.
Nevertheless, even though the SA-CASSCF reference is reproduced with errors always below chemical accuracy, we recall that the latter will still deviate from the exact energy as defined by the FCI value.
As FCI is intractable for the formaldimine in the cc-pVDZ basis, we instead consider CASCI(16,16), expected to be very close to the FCI value.
We see that SA-CASSCF (or SA-OO-VQE) and CASCI(16,16) agree qualitatively on the existence of a conical intersection, but differ quantitatively: the location of the conical intersection is shifted by $6.5^\circ$.
This is not a failure of the SA-OO-VQE algorithm itself, but rather that the small active space is insufficient to capture the full dynamic correlation present in the FCI space.
This can be solved by enlarging the active space, or potentially by perturbative approaches such as considered in Ref.~\onlinecite{takeshita2020increasing}.
However, in many cases the qualitative agreement is sufficient, and the correct reproduction of the conical intersection at all is vindication of the SA-OO-VQE method.

\subsection{Diabatic versus adiabatic representation}

One interesting feature of our SA-OO-VQE implementation is that the converged states $\ket{\Psi_{\rm A}}$ and $\ket{\Psi_{\rm B}}$ cross each other at the conical intersection point, as shown by the triangles and crosses markers in Fig.~\ref{fig:2D-PES}(b).
As discussed in Sec.~\ref{sec:SA-VQE} (see also Ref.~\onlinecite{zhang2020variational}), this is only possible when considering the equi-weighted version of SA-VQE, as the fully-weighted version of SS-VQE with condition $w_A > w_B$ will force an \textit{a-priori} (energy based) ordering of the states $\vert \Psi_{\rm A}(\vec{\theta})\rangle$ and $\vert \Psi_{\rm B}(\vec{\theta})\rangle$~\cite{nakanishi2019subspace}.

In contrast to the fully-weighted SS-VQE (or SA-CASSCF) method which generate adiabatic states, our states $\vert \Psi_{\rm A}(\vec{\theta}) \rangle$ and $\vert \Psi_{\rm B} (\vec{\theta}) \rangle$ keep their nature along the whole PES.
This can be rationalized by looking at the decomposition of the wavefunction in the basis of Slater determinants (\textit{i.e.} the configuration interaction coefficients).
Consider the converged SA-OO-VQE wavefunctions $\vert \Psi_{\rm A}(\vec{\tilde{\kappa}}) \rangle$ and $\vert \Psi_{\rm B}(\vec{\tilde{\kappa}}) \rangle$,
where $\vec{\tilde{\kappa}}$ denote the converged SA-OO basis.
The dominant configuration in $\vert \Psi_{\rm A}(\vec{\tilde{\kappa}}) \rangle$ is $\vert \Phi_{\rm A}(\vec{\tilde{\kappa}}) \rangle$ --- the HF Slater determinant expressed in the SA-OO basis.
The dominant configuration in $\vert \Psi_{\rm B}(\vec{\tilde{\kappa}})\rangle$ is $\vert \Phi_{\rm B}(\vec{\tilde{\kappa}})\rangle$) --- the singly-excited HF Slater determinant expressed in the SA-OO basis.
These respective dominating configurations contribute to more than 93\% for $\vert \Psi_{\rm A}(\vec{\tilde{\kappa}})\rangle$ ($|\psh{\Phi_{\rm A}(\vec{\tilde{\kappa}})}{\Psi_{\rm A}(\vec{\tilde{\kappa}})}|^2 > 0.93$) and 88\% for $\vert \Psi_{\rm B}(\vec{\tilde{\kappa}})\rangle$ ($|\psh{\Phi_{\rm B}(\vec{\tilde{\kappa}})}{\Psi_{\rm B}(\vec{\tilde{\kappa}})}|^2 > 0.88$) along the whole PES, while their contribution on the other state is negligible ($\psh{\Phi_{\rm A}(\vec{\tilde{\kappa}})}{\Psi_{\rm B}(\vec{\tilde{\kappa}})}$ and $\psh{\Phi_{\rm B}(\vec{\tilde{\kappa}})}{\Psi_{\rm A}(\vec{\tilde{\kappa}})} \sim 0$).\linebreak
Hence, it is expected that the SA-OO-VQE states $\vert \Psi_{\rm A}(\vec{\kappa})\rangle$ and $\vert \Psi_{\rm B}(\vec{\kappa})\rangle$ will always be associated to $\vert \Phi_{\rm A}(\vec{\kappa})\rangle$ and $\vert \Phi_{\rm B}(\vec{\kappa})\rangle$, respectively, which are the initial states of the SA-VQE algorithm.
As the ground and first-excited state of the frozen-core Hamiltonian in the SA-OO basis exhibit a conical intersection (in contrast to the canonical MO basis), $\vert \Psi_{\rm A}(\vec{\kappa})\rangle$ and $\vert \Psi_{\rm B}(\vec{\kappa})\rangle$ correspond to the ground and first-excited states for $\alpha > 118.5^\circ$ and conversely for $\alpha < 118.5^\circ$, respectively.
This change of ordering in the energies of $\vert \Psi_{\rm A}(\vec{\kappa})\rangle$ and $\vert \Psi_{\rm B}(\vec{\kappa})\rangle$ in Fig.~\ref{fig:2D-PES}(b) is reflected by the crossing in the energy curves in Fig.~\ref{fig:optimization_process}(b) and is made possible by using an equi-ensemble.

These observations are not without significance, and suggest that SA-OO-VQE can lead to a diabatic representation of the states, in contrast to SA-CASSCF or fully-weighted SS-VQE.
In multistate quantum dynamics simulations, considering diabatic states is crucial to enforce smoothness of physical properties and to reduce the magnitude of non-adiabatic couplings, which explode around conical intersections or avoided crossings, signaling a break down of the Born--Oppenheimer approximation~\cite{baer2006beyond,gatti2017applications}.
However, switching to a diabatic representation is not straightforward~\cite{zhang2020novel}.
Hence, representing diabatic states like in Fig.~\ref{fig:2D-PES}(b) with SA-OO-VQE will allow the calculation of, for instance, smooth ground- and excited-state energy gradient to perform quantum dynamics around challenging PES featuring conical intersections or avoided crossings.

This diabatic representation in SA-OO-VQE needs to be assessed for more challenging systems, where more states could be required together with more complicated initial states.
A rigorous way of capturing the expected individual states in an equi-ensemble formalism is provided by adding the state-averaged variance to the cost function in Eq.~(\ref{eq:SA-variance}).
We implemented and tested this augmented cost-function on the formaldimine, but did not see any noticeable improvement compared to using the state-averaged energy alone, as in our case the latter was sufficient to capture the correct eigenstates.
As an alternative approach (not employed in this work), one can also use the fully-weighted SS-VQE method (starting with $w_{\rm A} > w_{\rm B}$) together with the equi-ensemble SA-OO.
As already mentioned in this work and in Ref.~\onlinecite{zhang2020variational}, taking $w_{\rm A} > w_{\rm B}$ imposes an ordering of the states based on their energy (\textit{i.e.} $E_{\rm A} < E_{\rm B}$), which might not be ideal and may complicate the optimization process.
When combining this method with the equi-ensemble SA-OO, the ordering constraint during the SA-OO cycle is relaxed and the energy ordering is allowed to swap freely (just as in Fig.~\ref{fig:optimization_process}(b)).
If this swapping leads to $E_{\rm A} > E_{\rm B}$, one can simply change the ordering of the weights for the next fully-weighted SS-VQE cycle (\textit{i.e.} with $w_{\rm A} < w_{\rm B}$), potentially leading to a much easier optimization.

\subsection{State fidelity}

As a final study, we estimate the similarity of the converged correlated states generated with SA-OO-VQE ($\ket{\Psi_A}$ or $\ket{\Psi_B}$) compared to SA-CASSCF reference wavefunctions
($\ket{\Psi_0}$ and $\ket{\Psi_1}$) by measuring the fidelity
\begin{equation}
    \mathcal{F}( \Psi_\text{ref}|\Psi ) = \left| \psh{ \Psi_\text{ref} }{\Psi } \right|^2.
    \label{eq:fidelity}
\end{equation}
The estimation of the state fidelity involves the evaluation of the overlap between two wavefunctions expressed in two different optimized MO basis (the one from SA-OO-VQE and the other one from SA-CASSCF).
In practice, this calculation is non-trivial and requires manipulating projections of the MOs from the two different basis~\cite{plasser2016efficient, lowdin1955quantum,malmqvist2002restricted} (generally non-orthogonal to each other, as shown in Appendix~\ref{appendix:overlap}).

\begin{figure}
\centering
    \includegraphics[width=8cm]{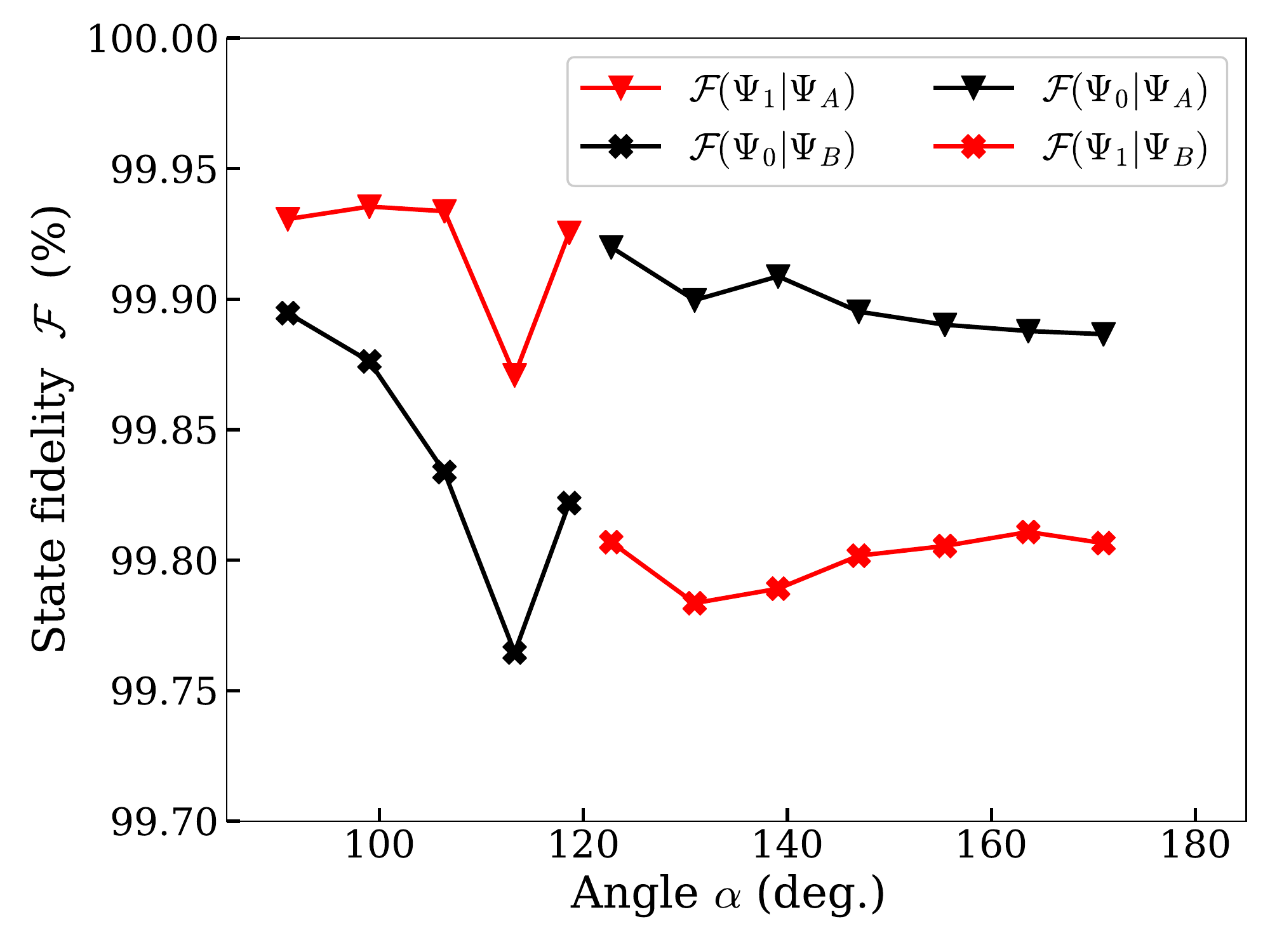}
    \caption{Estimation of the state-fidelity $\mathcal{F}$ between SA-OO-VQE and SA-CASSCF wavefunctions. Depending on the convergence of the SA-OO-VQE states, different lines and symbols are used. Colors are used for a same SA-CASSCF reference state : black for the ground-state and red for the first excited state. Symbols are used for a same converged SA-OO-VQE state : crosses for $\ket{\Psi_{\rm A}}$ and triangles for $\ket{\Psi_{\rm B}}$.}
    \label{fig:fidelity}
\end{figure}

The state-fidelity over the one dimensional PES of formaldimine is shown in Fig. \ref{fig:fidelity}.
The correlated states obtained with the SA-OO-VQE method exhibit a very high fidelity with an average of $\sim 99.85\%$ and a lower bound of  $\sim 99.75\%$, thus demonstrating an excellent correspondence to SA-CASSCF wavefunctions.
Interestingly, $\ket{\Psi_{\rm B}}$ shows a slightly lower fidelity compared to $\ket{\Psi_{\rm A}}$.
This could be attributed to the difference in the overlap between the initial states and the final one  ($|\psh{\Phi_{\rm A}(\vec{\tilde{\kappa}})}{\Psi_{\rm A}(\vec{\tilde{\kappa}})}|^2 > 0.93$ and $|\psh{\Phi_{\rm B}(\vec{\tilde{\kappa}})}{\Psi_{\rm B}(\vec{\tilde{\kappa}})}|^2 > 0.88$),
suggesting that the exact eigenstate $\ket{\Psi_{\rm B}}$ is more difficult to prepare.
This high fidelity together with the accurate energy representation and the diabatic representation of the states clearly encourages the use of SA-OO-VQE wavefunctions as a good starting point for realizing more complicated studies such as, for example, excited state dynamics.

\section{Conclusions}\label{sec:conclusion}

In this work, we implemented a variational hybrid quantum/classical algorithm able to capture non-trivial spectral features such as conical intersections.
We showed the importance of treating several states on an equal footing, as this is a necessary condition to correctly describe closely degenerate states such as in avoided crossings or conical intersections, which are prominent in photochemistry.
Our method remains appropriate in the NISQ era, where quantum resources are limited, as it doesn't increase the circuit depth of each call of VQE.
Instead, it only requires a careful choice of orbital-optimization, i.e. the state-averaged orbital-optimization, together with a quantum solver able to treat all states on an equal footing, such as SA-VQE~\cite{nakanishi2019subspace}.
We called this method the SA-OO-VQE.
As this orbital-optimization is performed classically, the only additional cost on a quantum computer is given by the number of updates of the frozen core Hamiltonian to be solved by SA-VQE, which happens to be smaller than 8 in our simulation (for a convergence criteria of $10^{-4}$ hartree on the state-averaged energy).
As a proof of concept, the performance of SA-OO-VQE was simulated classically and successfully applied to the description of the conical intersection in the formaldimine molecule, a minimal model for a photoisomerization process similar to that occurring in the  rhodopsine chromophore. 
The latter is known to participate in the process of vision, through a photoisomerization reaction mediated by the presence of a conical intersection.
Therefore, we expect our algorithm to provide accurate potential energy surfaces for both the ground and excited states, especially for challenging cases in photochemistry applications such as presented in this work.
It should be noted that the orbital-optimization procedure (requiring repeated two-electron integral transformations to obtain an updated Hamiltonian) can be costly when applied to large active space and basis sets.
As quantum computers promise to increase in size and in accuracy, this could be a potential bottleneck of the algorithm.
However, state-averaged orbital-optimization can also be performed differently, as demonstrated for instance in Ref.~\onlinecite{shepard2019all} on large active spaces. 
The procedure described in Ref.~\onlinecite{shepard2019all} alleviates the need of updating the frozen core Hamiltonian at each step, such that a single SA-VQE step would be performed on a quantum computer.

This work highlights important perspectives for near-term quantum computers.
Combined with our approach, extending the calculation of atomic forces~\cite{kassal2009quantum,obrien2019calculating,mitarai2020theory,parrish2019hybrid,sokolov2020microcanonical} to excited states will pave the way towards excited-state quantum dynamics beyond what is currently tractable on a classical device. 
Having a control on the excited state dynamics can have important implications in energy conversion and catalysis~\cite{paulus2020leveraging}.
This is a clear direction for future research, as it can be used to unravel photochemical processes that are hard to describe with classical methods for which the size of the active spaces becomes prohibitive. 
This encompasses industrially relevant applications such as the development as OLED devices with improved phosphorescent emitters~\cite{Lighthart2018} as well as fundamental questions on how nature has evolved protection mechanisms to prevent damages caused by absorption of too many~\cite{Ostroumov.Holzwarth.2020} or too energetic~\cite{Markovitsi2016} photons.

%------------------------------------------------
%\phantomsection
\section*{Acknowledgments} % The \section*{} command stops section numbering
SY and BS sincerely thank Benjamin Lasorne and Emmanuel Fromager for fruitful discussions.
SY and BS acknowledge support from the Netherlands Organization for Scientific Research (NWO/OCW).
BS acknowledges support from Shell Global Solutions BV.
%\addcontentsline{toc}{section}{Acknowledgments} % Adds this section to the table of contents

% =======================================================
% =======================================================
%                   A P P E N D I X
% =======================================================
% =======================================================
\appendix

\section{Protection of conical intersections}\label{app:proof}

Adding a real-valued perturbation $V=V(\mathbf{R})$ to a system with a conical intersection will not break the degeneracy, but will instead shift it to a new point $\mathbf{R}$ (as long as Eq.~(\ref{eq:low_energy_effective_theory}) remains valid).
We may write $V(\mathbf{R})$ in a general form
\begin{equation}
    V(\mathbf{R}) = V_0(\mathbf{R})I + V_X(\mathbf{R})X + V_Z(\mathbf{R})Z,
\end{equation}
and then the conical intersection will re-appear around any points $\mathbf{R}$ where the following two conditions are satisfied
\begin{align}
    V_X(\mathbf{R}) + h_X(\mathbf{R}-\mathbf{R}_0)\cdot\mathbf{R}_X = 0\label{eq:app_pert_vx}\\
    V_Z(\mathbf{R}) + h_Z(\mathbf{R}-\mathbf{R}_0)\cdot\mathbf{R}_Z = 0.\label{eq:app_pert_vz}
\end{align}
A solution to these equations exists as long as $\mathbf{R}_X$ and $\mathbf{R}_Z$ are linearly independent and $V_X$ and $V_Z$ are continuous.
To lowest order in $\mathbf{R}$, we may write
\begin{align}
    V_X(\mathbf{R}) &\sim V_X(\mathbf{R}_0) + \nabla_{\mathbf{R}}V_X(\mathbf{R}_0)\cdot (\mathbf{R}-\mathbf{R}_0)\\
    V_Z(\mathbf{R}) &\sim V_Z(\mathbf{R}_0) + \nabla_{\mathbf{R}}V_Z(\mathbf{R}_0)\cdot (\mathbf{R}-\mathbf{R}_0),
\end{align}
which may be substituted into Eqs.~(\ref{eq:app_pert_vx}) and (\ref{eq:app_pert_vz}) to obtain
\begin{align}
    (\mathbf{R}-\mathbf{R}_0)\cdot(h_X\mathbf{R}_X + \nabla_{\mathbf{R}}V_X(\mathbf{R}_0)) = - V_X(\mathbf{R}_0)\\
    (\mathbf{R}-\mathbf{R}_0)\cdot(h_Z\mathbf{R}_Z + \nabla_{\mathbf{R}}V_Z(\mathbf{R}_0)) = - V_Z(\mathbf{R}_0).
\end{align}
Then, if we define
\begin{align}
    h'_X\mathbf{R}'_X = h_X\mathbf{R}_X + \nabla_{\mathbf{R}}V_X(\mathbf{R}_0)\\
    h'_Z\mathbf{R}'_Z = h_Z\mathbf{R}_Z + \nabla_{\mathbf{R}}V_Z(\mathbf{R}_0),
\end{align}
we recover the form of Eq.~(\ref{eq:low_energy_effective_theory}) with $h_X\rightarrow h'_X$, $h_Z\rightarrow h'_Z$, $\mathbf{R}_X\rightarrow\mathbf{R}'_X$, $\mathbf{R}_Z\rightarrow\mathbf{R}'_Z$, and
\begin{align}
    \mathbf{R}_0& \rightarrow \mathbf{R}_0 - \frac{V_X(\mathbf{R}_0)}{h'_X|\mathbf{R}'_X|}\mathbf{R}'_X \nonumber\\
    &-\left(\frac{V_Z(\mathbf{R}_0)}{h'_Z|\mathbf{R}'_Z|} + \frac{V_X(\mathbf{R}_0)}{h'_X|\mathbf{R}'_X|}\mathbf{R}'_X\cdot\mathbf{R}'_Z\right)\nonumber\\&\times\frac{|\mathbf{R}'_Z|^2|\mathbf{R}'_X|^2}{|\mathbf{R}'_Z|^2|\mathbf{R}'_X|^2 - |\mathbf{R}_Z\cdot\mathbf{R}_X|^2}\nonumber\\&\times\left(\mathbf{R}'_Z-\frac{\mathbf{R}'_Z\cdot\mathbf{R}'_X}{|\mathbf{R}'_X|^2}\mathbf{R}'_X\right)
\end{align}

Although conical intersections cannot be lifted by a perturbation, they may be lifted by projection.
Consider a three-level system spanned by eigenstates $|\Psi_0(\mathbf{R})\rangle$, $|\Psi_1(\mathbf{R})\rangle$, and $|\Psi_2(\mathbf{R})\rangle$, with eigenenergies $E_0(\mathbf{R})$, $E_1(\mathbf{R})$, and $E_2(\mathbf{R})$.
Suppose now that there exists a conical intersection between $E_0(\mathbf{R})$ and $E_1(\mathbf{R})$ at $\mathbf{R}=\mathbf{R}_0$, but $E_2(\mathbf{R})>E_1(\mathbf{R})\geq E_0(\mathbf{R})$ everywhere.
The following linear combination of eigenstates 
\begin{equation}
 |\Phi(\mathbf{R})\rangle =a_1|\Psi_1(\mathbf{R})\rangle + a_2|\Psi_2(\mathbf{R})\rangle
\end{equation}
will have energy 
\begin{equation}
    E_{\Phi}(\mathbf{R})=|a_1|^2E_1(\mathbf{R})+|a_2|^2E_2(\mathbf{R}).
\end{equation}
If we project our three-level system into the subspace spanned only by $|\Phi(\mathbf{R})\rangle$ and $|\Psi_0\rangle$, the resulting Hamiltonian has eigenenergies $E_0(\mathbf{R})$ and $E_{\Phi}(\mathbf{R})$, which are by our assumption never equal for any $\mathbf{R}$.
As the active space Hamiltonian $\hat{\mathcal{H}}^{\rm FC}$ is generated by projecting the full Hamiltonian $\mathcal{\hat{H}}$ onto the active space, this gives a mechanism for such a process breaking the conical intersection.
Unless carefully treated (e.g. by the methods in this text), this represents a qualitative failure of the frozen core approximation.

\section{Frozen core Hamiltonian}\label{appendix:FrozenCore}

We describe how to build the frozen core Hamiltonian in practice. 
We start from the original shape of the second quantized electronic Hamiltonian
\begin{eqnarray}\label{eq:Ham_elec_}
\mathcal{\hat{H}}  = \sum_{pq}^{N_{o}} h_{pq}  \hat{E}_{pq} + \dfrac{1}{2} \sum_{pqrs}^{N_{o}} g_{pqrs}  \hat{e}_{pqrs}.
\end{eqnarray}
Applying the frozen core approximation to this Hamiltonian consists in assuming the existence of a set of frozen orbitals (always occupied), another set of active orbitals (belonging to an active space), and a set of virtual orbitals (always unoccupied). 
Based on this partitioning, every Slater determinant $|\Phi\rangle$ used to describe properties of the system will always take the form 
\begin{equation}
    |\Phi\rangle = | \Phi_\text{frozen}\Phi_\text{active} \rangle,
    \label{eq:state}
\end{equation}
where the left contribution $\Phi_{\rm frozen}$ represents a part of the determinant encoding the frozen orbitals of the system (always occupied) whereas $\Phi_{\rm AS}$ is a part encoding the occupancy of the remaining electrons in the active orbitals of the system. 
In this context, if one considers that every correlated electronic wavefunction is always expanding in terms of Slater determinants following Eq.~(\ref{eq:state}), one can demonstrate by projections onto Eq.~(\ref{eq:Ham_elec_}) that the system Hamiltonian takes an effective form 
\begin{eqnarray}\label{eq:Ham_elec}
\bra{\Phi} \mathcal{\hat{H}} \ket{\Phi} \equiv \bra{\Phi_{\rm active}} \mathcal{\hat{H}}^{\rm FC} \ket{\Phi_{\rm active}},
\end{eqnarray}
with $\mathcal{\hat{H}}^{\rm FC}$ the so-called ``frozen core Hamiltonian'' defined as follows,
\begin{eqnarray}
\label{eq:Ham_elec}
\mathcal{\hat{H}}^{\rm FC} = \mathcal{\hat{H}}_\text{active} + E_\text{frozen}^\text{MF} + \mathcal{\hat{V}}.
\end{eqnarray}
Here, $\mathcal{\hat{H}}_\text{AS}$ is the Hamiltonian encoding the one- and two- body terms only acting in the active space,
\begin{equation}
\label{eq:AS_HAM}
\mathcal{\hat{H}}_\text{active} = \sum_{tu}^\text{active} h_{tu} \hat{E}_{tu} + \sum_{tuvw}^\text{active} g_{tuvw} \hat{e}_{tuvw},
\end{equation}
where $t,u,v,w$ denote active space orbitals.
The second term $E_{\rm frozen}^{\rm MF}$ is a scalar representing the mean-field-like energy obtained from the frozen orbitals,
\begin{equation}
    E_\text{frozen}^\text{MF} = 2\sum_i^\text{frozen} h_{ii} + \sum_{ij}^\text{frozen} (2g_{iijj}- g_{ijji}),
    \label{eq:shift}
\end{equation}
and the third term
\begin{equation}
\label{eq:emb}
\mathcal{\hat{V}} = \sum_{tu}^\text{active} \mathcal{V}_{tu} \hat{E}_{tu} \text{, with } \mathcal{V}_{tu} =  \sum_i^\text{frozen} (2g_{tuii}- g_{tiiu} )
\end{equation}
represents an effective one body potential which encodes the interaction of the frozen electrons with the active space electrons. To summarize, the main effect of the frozen core approximation [Eq.~(\ref{eq:Ham_elec})] is first to introduce an energetic shift [Eq.~(\ref{eq:shift})], and second to augment the one body term of the Hamiltonian operator [Eq.~(\ref{eq:AS_HAM})] (that only lives in the active space) with an additional effective one body operator [Eq.~(\ref{eq:emb})].

\section{Orbital optimization with Newton-Raphson}\label{appendix:NewtonRaphson}

We give here some additional details about the SA-OO algorithm in Sec.~\ref{sec:SA-OO}, based on the classical Netwon-Raphson approach. 
For simplicity, let us first focus on the case of a single correlated state noted $| \Psi \rangle $ for which we want to optimize the orbitals. 
The parametrized energy of the state thus reads
\begin{equation}
\begin{split}
    E_\Psi(\vec{\kappa}) &= \bra{\Psi} e^{\hat{\kappa}} \hat{\mathcal{H}} e^{-\hat{\kappa}} \ket{\Psi}.
\end{split}
\label{eq:param_energ}
\end{equation}
Using a second-order Baker-Campbell-Hausdorff development of $e^{-\hat{\kappa}} \hat{\mathcal{H}} e^{\hat{\kappa}} $, Eq.~(\ref{eq:param_energ}) becomes
\begin{equation}
\begin{split}
    E_\Psi(\vec{\kappa}) &\simeq \bra{\Psi} \hat{\mathcal{H}} \ket{\Psi} +\bra{\Psi} [\hat{\kappa}, \hat{\mathcal{H}}] \ket{\Psi}\\
                &+\frac{1}{2} \bra{\Psi} \big[ \hat{\kappa},[\hat{\kappa} ,\hat{\mathcal{H}} ]\big]\ket{\Psi}.
\end{split}
\label{dev1}
\end{equation}
In parallel, we also develop $E_\Psi(\vec{\kappa})$ using a second-order Taylor expansion with respect to the $\vec{\kappa}$ parameters. In matrix form, this expansion reads
\begin{equation}
\begin{split}
    E_\Psi(\vec{\kappa}) &\simeq E_\Psi(0) +\bold{G}^\dagger \boldsymbol{\kappa} + \frac{1}{2} \boldsymbol{\kappa}^\dagger \mathbf{H}\boldsymbol{\kappa}
\end{split}
\label{dev2}
\end{equation}
where $\bold{G}$ and $\bold{H}$ are the column MO-gradient vector and the MO-hessian matrix of the energy, and $\boldsymbol{\kappa}$ is a vector encoding the parameters $\vec{\kappa}$. 
From Eqs.~(\ref{dev1}) and (\ref{dev2}), we obtain the MO-gradient and MO-Hessian elements
\begin{equation}
    \begin{split}
           \bold{G}_{pq} &= \bra{\Psi} [\hat{E}_{pq}^-,\hat{\mathcal{H}}] \ket{\Psi} \\ 
           \bold{H}_{pq,rs} &= \frac{1}{2}(1+\mathcal{S}_{(pq)}^{(rs)})\bra{ \Psi } \big[\hat{E}_{pq}^-,  [\hat{E}_{rs}^-,\hat{\mathcal{H}}]\big] \ket{\Psi}  
    \end{split}
\end{equation}
where $\hat{E}_{pq}^- = \hat{E}_{pq} - \hat{E}_{qp}$ and $\mathcal{S}_{(pq)}^{(rs)}$ is an operator that permutes the two couples of indices $(pq)$ and $(rs)$. 
In practice, $\bold{G}$ and $\bold{H}$ can be expressed analytically in terms of the one- and two-electron reduced density matrices, and one- and two-electron integrals (see Refs.~\cite{siegbahn1981complete,helgaker2014molecular,yarkony1995modern}).
Once $\bold{G}$ and $\bold{H}$ are built, a Newton-Raphson step is computed as follows:
\begin{equation}
    \Delta\boldsymbol{\kappa} = -\bold{H}^{-1} \bold{G},
    \label{eq:step}
\end{equation}
and used to update the value of the rotation parameters
\begin{equation}
    \boldsymbol{\kappa} \leftarrow \boldsymbol{\kappa} + \Delta\boldsymbol{\kappa}.
\end{equation}
The new parameters are used to transform the MO coefficient matrix $\bold{C}$ [Eq.~(\ref{eq:C_OO})] and by extension the system Hamiltonian in Eq.~(\ref{eq:Hnew}) (via a redefinition of the one- and two-electron integrals of the system, see Eqs.~(\ref{eq:one_elec_new}) and (\ref{eq:two_elec_new})].
In practice, the initial conditions considered in the optimisation process are usually located far away from the minimum targeted.
Hence, the convergence of the algorithm is not necessarily guaranteed, mainly because the Hessian matrix used to drive the step in Eq.~(\ref{eq:step}) can be either non-positive or singular. 
To solve the issue, a more complicated definition of the Newton-Raphson step can be employed, using for example the so-called ``augmented Hessian'' approach:
\begin{equation}
\bold{H} \leftarrow \bold{H} + \nu \bold{1} ,
\end{equation}
with $\nu$ a positive number used to ensure the positive-definiteness of the matrix, and $\bold{1}$ the identity matrix. 
Naturally, the value of $\nu$ has to be wisely chosen and many different techniques can be employed to this end (see for instance Refs.~\cite{fletcher2013practical,helgaker2014molecular}). 
Using this definition to compute the step in Eq.~(\ref{eq:step}) ensures convergence of the Newton-Raphson algorithm to a local minimum.

The above description was for a single state. 
Let us discuss how to optimize molecular orbitals for a group of $N_{\rm S}$ correlated states $\lbrace \Psi_k \rbrace_{k=1,\ldots,N_{\rm S}}$. 
In this case, a same number of individual Hessians $\bold{H}_{\Psi_k}$ and gradients $\bold{G}_{\Psi_k}$ have to be computed. 
One then builds a state-averaged version of these matrices as follows:
\begin{equation}
    \bold{H}_{\rm SA} = \sum^{N_{\rm S}}_{k} w_k\bold{H}_{\Psi_k} \text{, and } \bold{G}_{\rm SA} = \sum^{N_{\rm S}}_{k} w_k\bold{G}_{\Psi_k}, 
\end{equation}
where $\bold{H}_{\rm SA}$  and $\bold{G}_{\rm SA}$ are respectively the state-averaged Hessian and gradient of the ensemble $\lbrace \Psi_k \rbrace_{k=1,\ldots,N_{\rm S}}$. 
The weights $w_k$ (with $\sum_k^{N_{\rm S}} w_k = 1$) encode the contribution of each state into the orbital-optimization process. Using these state-averaged matrices, a Newton-Raphson step like in Eq.~(\ref{eq:step}) is built and the system's MOs can be optimized in \textit{a democratic way} with respect to all states considered in the ensemble.

To close this discussion, we mention two potential numerical bottlenecks that can arise when using orbital optimization. The first one is the repetitive transformations used to build two-electron integrals in new optimised MO basis. 
Such a transformation is known to scale as $N^5$ (see Ref. \onlinecite{yoshimine1973}), when not employing approximations such as density fitting (see Ref. \onlinecite{feyereisen1993use}). 
The second bottleneck is related to the explicit construction of the orbital Hessian during each Newton-Raphson step. 
This is a very active research problem in quantum chemistry for which many new approaches and methods are being developed currently (see for example Refs.~ \onlinecite{vogiatzis2017pushing,kreplin2020mcscf,kreplin2019second}).

\section{Hamiltonian transformation}
\label{appendix:HamTransfo}

We describe how the full second-quantized Hamiltonian transformation is realized in practice during the orbital-optimization process. 
Let us start with the $\bold{C}$ matrix encoding the MO coefficients of the system,
\begin{equation}
    \phi_p(\bold{r}) = \sum_\mu \bold{C}_{lp} \chi_\mu(\bold{r}).
\end{equation}
where $\phi_p(\bold{r}) $ is a generic MO expressed as a linear combination of atomic orbitals $\chi_\nu(\bold{r})$. 
Considering the initial MO basis $\lbrace \phi_p \rbrace$, we build a unitary operator
\begin{equation}
    \bold{U}_\text{OO} = e^{-\bold{K}} \text{ with } \bold{K} = \text{skew}(\boldsymbol{\kappa}),
\end{equation}
where $\bold{K}$ is a skew-symmetric matrix shaped by the column vector $\boldsymbol{\kappa}$ containing the rotational parameters. 
This operator is used to change the original MO basis into a new one $\lbrace \phi'_q \rbrace$ with
\begin{equation}
    \phi'_q(\bold{r}) = \sum_p  \phi_p(\bold{r}) (\bold{U}_\text{OO})_{pq}.
\end{equation}
To realize this transformation and obtain a $\bold{C}$ matrix encoding the new MO basis, one applies $\bold{U}_\text{OO}$ on the right of $\bold{C}$ and then update the latter such that
\begin{equation}
    \bold{C} \leftarrow \bold{C} \bold{U}_\text{OO}.\label{eq:C_OO}
\end{equation}
The new MO coefficients matrix $\bold{C}$ is then used to transform the one- and two-electron integrals in the new MO basis. To do so, one starts from the one- and two-electron integrals expressed in the atomic orbital basis, respectively $h_{\mu\nu}$ and $g_{\mu\nu\gamma\delta}$, and implements the following two- and four-indexes transformations:
\begin{equation}
 h_{pq}^\text{new} =  \sum_{\mu,\nu} h_{\mu\nu} \bold{C}_{\mu p}\bold{C}_{\nu q},\label{eq:one_elec_new}
\end{equation}
and 
\begin{equation}
 g_{pqrs}^\text{new} =  \sum_{\mu,\nu,\gamma,\delta} g_{\mu\nu\gamma\delta} \bold{C}_{\mu p}\bold{C}_{\nu q}\bold{C}_{\gamma r}\bold{C}_{\delta s}.\label{eq:two_elec_new}
\end{equation}
The Hamiltonian of the system expressed in the new MO basis then reads
\begin{eqnarray} 
\mathcal{\hat{H}}  = \sum_{pq}^{N_{\rm MO}} h_{pq}^\text{new}  \hat{E}_{pq}^\text{new} + \dfrac{1}{2} \sum_{pqrs}^{N_{\rm MO}} g_{pqrs}^\text{new} \hat{e}_{pqrs}^\text{new}.\label{eq:Hnew}
\end{eqnarray}

\section{Numerical implementation of the generalized spin-free double-excitation ansatz}
\label{appendix:gen_ansatz}

We detail here our practical implementation of the generalized spin-free double excitation ansatz. 
Starting from the original definition of the excitation operator
\begin{eqnarray}
    && \hat{T}(\vec{\theta}) = \sum_{t,  v, w, u }^\text{active} \theta_{tuvw}  \sum_{\sigma,\tau=\uparrow,\downarrow} \hat{a}^\dagger_{t\sigma} \hat{a}^\dagger_{v\tau} \hat{a}_{w\tau} \hat{a}_{u\sigma},
    \label{eq:origin_GCC}
\end{eqnarray}
the scaling of the total number of independent variational parameters $\theta$ to optimize with SA-VQE is $N^4$ (with $N$ the number of active space MOs). 
In practice, this scaling could represent a bottleneck for the energy optimization realized with the SA-VQE sub-algorithm (especially when multiple calls of SA-VQE are required to reach global convergence of the SA-OO-VQE method). 
To reduce the number of parameters and favor a faster convergence, we simplify Eq.~(\ref{eq:origin_GCC}) by imposing two conditions.
First, we considered the indices restriction $t \geq  v \geq w \geq u $ over the active space MOs (in addition to discarding terms where $t =  v = w = u $). 
In practice, this reduces the total number of parameters from 81 to 12 for the three MO active-space considered in our simulations. 
Second, we invoke the symmetry property $\theta_{tuvw}=\theta_{vwtu}$ to recover part of the variational parameters that were discarded with the previous restriction on the indices. 
This is inspired by the properties $\hat{e}_{pqrs} = \hat{e}_{rspq}$ of the spin-free two-body term of the Hamiltonian [Eq.~(\ref{eq:el_Ham})] (see Ref.  \onlinecite{helgaker2014molecular}). 
Within the above conditions, the excitation operator simplifies as
\begin{equation}
     \hat{T}(\vec{\theta}) = \sum_{\substack{t \geq v  \geq w  \geq u \\ \text{not}(t = v  = w = u)} }^\text{active} \theta_{tuvw} (1+\mathcal{S}_{(tu)}^{(vw)}) \sum_{\sigma,\tau=\uparrow,\downarrow} \hat{a}^\dagger_{t\sigma} \hat{a}^\dagger_{v\tau} \hat{a}_{w\tau} \hat{a}_{u\sigma}
\end{equation}
where $\mathcal{S}_{(tu)}^{(vw)}$ is an operator that permutes the couples of indices $(tu)$ and $(vw)$.
Based on this simplified version of the excitation operator, we built up a quantum circuit implementing the unitary $\hat{U}(\vec{\theta}) = e^{\hat{T}(\vec{\theta})-\hat{T}(\vec{\theta})^\dagger}$. 
To proceed, we chose to apply a single Trotter step to split up the original unitary into products of exponentials as follows:
\begin{equation}
    \begin{split}   
       \hat{U}(\vec{\theta}) &= e^{\hat{T}(\vec{\theta})-\hat{T}(\vec{\theta})^\dagger} \\
       &\approx \prod_{\substack{t \geq  v \geq w \geq u \\ \text{not}(t = v  = w = u) } } \prod_{\sigma,\tau=\uparrow,\downarrow} e^{  \theta_{tuvw} ( \hat{a}^\dagger_{t\sigma} \hat{a}^\dagger_{v\tau} \hat{a}_{w\tau} \hat{a}_{u\sigma} - \text{h.c.})}  \\
   & \prod_{\sigma,\tau=\uparrow,\downarrow} e^{ \theta_{tuvw} ( \hat{a}^\dagger_{v\sigma} \hat{a}^\dagger_{t\tau} \hat{a}_{u\tau} \hat{a}_{w\sigma} - \text{h.c.} ) } ,
    \end{split} 
    \label{eq:encoding_ansatz}
\end{equation}
where ``h.c.'' stands for ``hermitian conjugate''. Then, we encode each of the operators into a quantum circuit with an order described by the following pseudo-code:
\begin{algorithm}[H]
\caption{ansatz encoding}
\begin{algorithmic}[1]

\State $N = 3$ \Comment{Number of active MOs}

\For{$u=1:N$}
    \For{$t=1:N$}
        \For{$w=1:N$}
            \For{$v=1:N$}
                \If{ $\text{not}(t=u=v=w)$}
                    \If{ ($ t \geq v \geq w \geq u $)}

                        \State \textbf{Set} $\sigma = \uparrow, \tau = \uparrow$
                        
                        \State \textbf{Encode} $e^{  \theta_{tuvw} ( \hat{a}^\dagger_{t\sigma} \hat{a}^\dagger_{v\tau} \hat{a}_{w\tau} \hat{a}_{u\sigma} - \text{h.c.})}$

                        \State \textbf{Set} $\sigma = \downarrow, \tau = \uparrow$
                        
                        \State \textbf{Encode} $e^{  \theta_{tuvw} ( \hat{a}^\dagger_{t\sigma} \hat{a}^\dagger_{v\tau} \hat{a}_{w\tau} \hat{a}_{u\sigma} - \text{h.c.})}$

                        \State \textbf{Set} $\sigma = \uparrow, \tau = \downarrow$
                        
                        \State \textbf{Encode} $e^{  \theta_{tuvw} ( \hat{a}^\dagger_{t\sigma} \hat{a}^\dagger_{v\tau} \hat{a}_{w\tau} \hat{a}_{u\sigma} - \text{h.c.})}$
                        
                        \State \textbf{Set} $\sigma = \downarrow, \tau = \downarrow$
                        
                        \State \textbf{Encode} $e^{  \theta_{tuvw} ( \hat{a}^\dagger_{t\sigma} \hat{a}^\dagger_{v\tau} \hat{a}_{w\tau} \hat{a}_{u\sigma} - \text{h.c.})}$
                        
                        \State \textbf{Swap the indices :} $t \leftrightarrow  v$,  $u \leftrightarrow  w$
                         
                        \State \textbf{Encode} the same exponentials 
                        \State (with the same parameter $\theta_{tuvw}$)
                    \EndIf
                 \EndIf
              \EndFor
           \EndFor
        \EndFor
     \EndFor
\end{algorithmic}
\end{algorithm}
In practice, every time we have to encode a fermionic exponential in the quantum circuit, we apply the Jordan--Wigner transformation to the associated difference of chain of fermionic operators. We then obtain a series of exponentiated Pauli-strings we translate into circuit form using pretty common implementation techniques (namely, chains of CNOT gates and single gate rotations, as described in Ref.~\onlinecite{seeley2012bravyi}). Within this scheme, the total number of quantum gates in our six-qubits quantum circuit is estimated at 3904 gates (2080 single-qubit gates and 1824 two-qubit gates).

% Then, we apply the Jordan--Wigner transformation to each chain of fermionic operator, thus leading to exponentiated Pauli-strings that are encoded into the quantum circuit using pretty common implementation techniques (namely, chains of CNOT gates and single gate rotations, as described in Ref.~\onlinecite{seeley2012bravyi}).
% The order of the unitary operations follows Eq.~(\ref{eq:encoding_ansatz}), where we gather in a row all unitaries sharing the same parameter.
% Within this scheme, the total number of quantum gates in our six-qubits quantum circuit is estimated at 3904 gates (2080 single-qubit gates and 1824 two-qubit gates).

\section{Variational quantum algorithms for excited states}
\label{appendix:methods}

While variational quantum algorithms were originally designed to extract the ground state of a given Hamiltonian, several extensions to excited states have been recently developed.
We provide here a brief discussion on these methods and their capability to describe several states on an equal footing.

In Sec.~\ref{sec:theory}, we insisted on the importance of a democratic description of the states involved in the conical intersection.
This democratisation can be achieved by performing a single minimization for all states sharing the same ansatz, as in multistate-contracted VQE (MC-VQE)~\cite{parrish2019quantum,parrish2019hybrid}, fully-weighted subspace-search VQE (SS-VQE)~\cite{nakanishi2019subspace} and variance-VQE~\cite{zhang2020variational}.
The other versions described in the SS-VQE paper require additional minimizations or maximizations to extract the excited states one by one, thus potentially breaking the democratic treatment of the states as the optimization landscape might be more complex from one state to another.

Other methods to extract excited states can be grouped in three types.\\
First, methods which consists in expanding about the reference ground state to form a linear (or beyond) subspace.
The optimal solutions within this subspace are approximations to the excited states and are obtained by solving a generalized eigenvalue equation on a classical computer.
These methods include quantum subspace expansion ~\cite{mcclean2017hybrid,colless2018computation} and
the quantum equation-of-motion coupled-cluster~\cite{ollitrault2019quantum}.
Although the excited states are treated on the same footing, the ground state is favored by construction.
Hence, an accurate description of a conical intersection between the ground and first excited state is unlikely, as in linear response TDDFT~\cite{gozem2014shape}.
Similar conclusions also hold for the Quantum Lanczos algorithm described in Ref.~\onlinecite{motta2020determining}.
Note that other methods imply to solve generalized eigenvalue equations on a classical computer but without favoring any state and are therefore democratic, for instance in the quantum filter diagonalization method~\cite{parrish2019quantumfilter,bespalova2020hamiltonian}.\\
Second, variational algorithms based on penalization of the Hamiltonian. 
In order to get the $k$-th excited state, these methods penalize the Hamiltonian by the (orthogonal) ground and ($k-1$)-th excited states~\cite{higgott2019variational,jones2019variational,jouzdani2019method,ibe2020calculating}.
This procedure is sequential, which may lead to error accumulation~\cite{higgott2019variational}.
As each state is determined by a separate minimization, these methods are by construction ``state-specific''.
Hence, although in principle exact, the latter are not expected to coherently treat a conical intersection, except maybe for relatively simple cases.\\
Third, methods based on phase estimation (like for instance the witness-assisted variational eigenspectra solver~\cite{santagati2018witnessing}) are democratic, as phase estimation is the quantum algorithm equivalent to matrix diagonalization. However, for NISQ devices these approaches remain ill-adapted due to their very high computational cost.

\section{Overlap of wavefunctions expressed in different MO basis}
\label{appendix:overlap}

We describe how to compute the non-trivial overlap between two many-electron wavefunctions expressed in two different MO basis.
let us consider two distinct correlated wavefunctions $\ket{\Psi}$ and $|\tilde{\Psi}\rangle$ which form a linear combination of Slater determinants
\begin{equation}
    \ket{\Psi} = \sum_I d_{I} \ket{\Phi_I} \text{ and } |\tilde{\Psi}\rangle = \sum_J \tilde{d}_{J} |\tilde{\Phi}_J\rangle,
\end{equation}
where $\lbrace \ket{\Phi_I} \rbrace$ and $\lbrace |\tilde{\Phi}_J\rangle \rbrace$ represent two (sometimes different) sets of Slater determinant expressed in two different MO basis $\lbrace \ket{\phi_p} \rbrace $ and  $\lbrace |\tilde{\phi}_q\rangle \rbrace $. Our goal here is to compute 
\begin{equation}
    \psh{\Psi}{\tilde{\Psi}} = \sum_I\sum_J d_{I}^* \tilde{d}_{J} \psh{\Phi_I}{\tilde{\Phi}_J}.
\end{equation}
To proceed, the overlap $\psh{\Phi_I}{\tilde{\Phi}_J}$ between two Slater determinants needs to be estimated. 
This calculation is non-trivial as the two determinants considered are expressed in different (and usually non-orthogonal) MO basis which generally leads to $\psh{\Phi_I}{\tilde{\Phi}_J} \neq \delta_{IJ}$. 
Based on the spin-orbital occupancy defining the Slater determinants,
\begin{equation}
\ket{\Phi_I} = | \phi_1 \ldots \phi_{N_e} | \text{ and }  |\tilde{\Phi}_J\rangle = | \tilde{\phi}_1 \ldots \tilde{\phi}_{N_e} |,
\end{equation}
their overlap can be computed as follows  (see Refs.~\cite{plasser2016efficient,lowdin1955quantum} for a proof):
\begin{equation}
\psh{\Phi_I}{\tilde{\Phi}_J}  =
\begin{vmatrix}
\psh{\phi_1}{\tilde{\phi}_1} & \ldots & \psh{\phi_1}{\tilde{\phi}_{N_e}} \\
\vdots & \ddots & \vdots \\
\psh{\phi_{N_e}}{\tilde{\phi}_1} & \ldots & \psh{\phi_{N_e}}{\tilde{\phi}_{N_e}} \\
\end{vmatrix}.
\end{equation}
In other words, the overlap of two Slater determinants expressed in two different MO basis is given by the determinant of the matrix containing all mutual spin-orbital overlaps. 
Here, a generic spin-orbital overlap $\psh{\phi_p}{\tilde{\phi}_q} $ is defined as
\begin{equation}
    \psh{\phi_p}{\tilde{\phi}_q} = \int  \phi_p^*(\bold{x}) \tilde{\phi}_q(\bold{x})  \ \text{d}\bold{x}
\end{equation}
where $\bold{x} \equiv (\bfr,\sigma)$ denotes both spatial and spin coordinates of the spin-orbitals.
Note that when a same MO basis is used, many simplifications occur leading to $\psh{\phi_p}{\tilde{\phi}_q} = \delta_{pq}$ and to the trivial result $\psh{\Phi_I}{\tilde{\Phi}_J} = \delta_{IJ}$.

%----------------------------------------------------------------------------------------
%	REFERENCE LIST
%----------------------------------------------------------------------------------------
\phantomsection
%\bibliographystyle{apsrev4-1}
%\bibliography{biblio}

\begin{thebibliography}{140}%
\makeatletter
\providecommand \@ifxundefined [1]{%
 \@ifx{#1\undefined}
}%
\providecommand \@ifnum [1]{%
 \ifnum #1\expandafter \@firstoftwo
 \else \expandafter \@secondoftwo
 \fi
}%
\providecommand \@ifx [1]{%
 \ifx #1\expandafter \@firstoftwo
 \else \expandafter \@secondoftwo
 \fi
}%
\providecommand \natexlab [1]{#1}%
\providecommand \enquote  [1]{``#1''}%
\providecommand \bibnamefont  [1]{#1}%
\providecommand \bibfnamefont [1]{#1}%
\providecommand \citenamefont [1]{#1}%
\providecommand \href@noop [0]{\@secondoftwo}%
\providecommand \href [0]{\begingroup \@sanitize@url \@href}%
\providecommand \@href[1]{\@@startlink{#1}\@@href}%
\providecommand \@@href[1]{\endgroup#1\@@endlink}%
\providecommand \@sanitize@url [0]{\catcode `\\12\catcode `\$12\catcode
  `\&12\catcode `\#12\catcode `\^12\catcode `\_12\catcode `\%12\relax}%
\providecommand \@@startlink[1]{}%
\providecommand \@@endlink[0]{}%
\providecommand \url  [0]{\begingroup\@sanitize@url \@url }%
\providecommand \@url [1]{\endgroup\@href {#1}{\urlprefix }}%
\providecommand \urlprefix  [0]{URL }%
\providecommand \Eprint [0]{\href }%
\providecommand \doibase [0]{http://dx.doi.org/}%
\providecommand \selectlanguage [0]{\@gobble}%
\providecommand \bibinfo  [0]{\@secondoftwo}%
\providecommand \bibfield  [0]{\@secondoftwo}%
\providecommand \translation [1]{[#1]}%
\providecommand \BibitemOpen [0]{}%
\providecommand \bibitemStop [0]{}%
\providecommand \bibitemNoStop [0]{.\EOS\space}%
\providecommand \EOS [0]{\spacefactor3000\relax}%
\providecommand \BibitemShut  [1]{\csname bibitem#1\endcsname}%
\let\auto@bib@innerbib\@empty
%</preamble>
\bibitem [{\citenamefont {O'Malley}\ \emph {et~al.}(2016)\citenamefont
  {O'Malley}, \citenamefont {Babbush}, \citenamefont {Kivlichan}, \citenamefont
  {Romero}, \citenamefont {McClean}, \citenamefont {Barends}, \citenamefont
  {Kelly}, \citenamefont {Roushan}, \citenamefont {Tranter}, \citenamefont
  {Ding} \emph {et~al.}}]{omalley2016scalable}%
  \BibitemOpen
  \bibfield  {author} {\bibinfo {author} {\bibfnamefont {P.~J.}\ \bibnamefont
  {O'Malley}}, \bibinfo {author} {\bibfnamefont {R.}~\bibnamefont {Babbush}},
  \bibinfo {author} {\bibfnamefont {I.~D.}\ \bibnamefont {Kivlichan}}, \bibinfo
  {author} {\bibfnamefont {J.}~\bibnamefont {Romero}}, \bibinfo {author}
  {\bibfnamefont {J.~R.}\ \bibnamefont {McClean}}, \bibinfo {author}
  {\bibfnamefont {R.}~\bibnamefont {Barends}}, \bibinfo {author} {\bibfnamefont
  {J.}~\bibnamefont {Kelly}}, \bibinfo {author} {\bibfnamefont
  {P.}~\bibnamefont {Roushan}}, \bibinfo {author} {\bibfnamefont
  {A.}~\bibnamefont {Tranter}}, \bibinfo {author} {\bibfnamefont
  {N.}~\bibnamefont {Ding}},  \emph {et~al.},\ }\href
  {https://doi.org/10.1103/PhysRevX.6.031007} {\bibfield  {journal} {\bibinfo
  {journal} {Phys. Rev. X}\ }\textbf {\bibinfo {volume} {6}},\ \bibinfo {pages}
  {031007} (\bibinfo {year} {2016})}\BibitemShut {NoStop}%
\bibitem [{\citenamefont {Kandala}\ \emph {et~al.}(2017)\citenamefont
  {Kandala}, \citenamefont {Mezzacapo}, \citenamefont {Temme}, \citenamefont
  {Takita}, \citenamefont {Brink}, \citenamefont {Chow},\ and\ \citenamefont
  {Gambetta}}]{kandala2017hardware}%
  \BibitemOpen
  \bibfield  {author} {\bibinfo {author} {\bibfnamefont {A.}~\bibnamefont
  {Kandala}}, \bibinfo {author} {\bibfnamefont {A.}~\bibnamefont {Mezzacapo}},
  \bibinfo {author} {\bibfnamefont {K.}~\bibnamefont {Temme}}, \bibinfo
  {author} {\bibfnamefont {M.}~\bibnamefont {Takita}}, \bibinfo {author}
  {\bibfnamefont {M.}~\bibnamefont {Brink}}, \bibinfo {author} {\bibfnamefont
  {J.~M.}\ \bibnamefont {Chow}}, \ and\ \bibinfo {author} {\bibfnamefont
  {J.~M.}\ \bibnamefont {Gambetta}},\ }\href
  {https://doi.org/10.1038/nature23879} {\bibfield  {journal} {\bibinfo
  {journal} {Nature}\ }\textbf {\bibinfo {volume} {549}},\ \bibinfo {pages}
  {242} (\bibinfo {year} {2017})}\BibitemShut {NoStop}%
\bibitem [{\citenamefont {Colless}\ \emph {et~al.}(2018)\citenamefont
  {Colless}, \citenamefont {Ramasesh}, \citenamefont {Dahlen}, \citenamefont
  {Blok}, \citenamefont {Kimchi-Schwartz}, \citenamefont {McClean},
  \citenamefont {Carter}, \citenamefont {De~Jong},\ and\ \citenamefont
  {Siddiqi}}]{colless2018computation}%
  \BibitemOpen
  \bibfield  {author} {\bibinfo {author} {\bibfnamefont {J.~I.}\ \bibnamefont
  {Colless}}, \bibinfo {author} {\bibfnamefont {V.~V.}\ \bibnamefont
  {Ramasesh}}, \bibinfo {author} {\bibfnamefont {D.}~\bibnamefont {Dahlen}},
  \bibinfo {author} {\bibfnamefont {M.~S.}\ \bibnamefont {Blok}}, \bibinfo
  {author} {\bibfnamefont {M.}~\bibnamefont {Kimchi-Schwartz}}, \bibinfo
  {author} {\bibfnamefont {J.}~\bibnamefont {McClean}}, \bibinfo {author}
  {\bibfnamefont {J.}~\bibnamefont {Carter}}, \bibinfo {author} {\bibfnamefont
  {W.}~\bibnamefont {De~Jong}}, \ and\ \bibinfo {author} {\bibfnamefont
  {I.}~\bibnamefont {Siddiqi}},\ }\href
  {https://doi.org/10.1103/PhysRevX.8.011021} {\bibfield  {journal} {\bibinfo
  {journal} {Phys. Rev. X}\ }\textbf {\bibinfo {volume} {8}},\ \bibinfo {pages}
  {011021} (\bibinfo {year} {2018})}\BibitemShut {NoStop}%
\bibitem [{\citenamefont {Hempel}\ \emph {et~al.}(2018)\citenamefont {Hempel},
  \citenamefont {Maier}, \citenamefont {Romero}, \citenamefont {McClean},
  \citenamefont {Monz}, \citenamefont {Shen}, \citenamefont {Jurcevic},
  \citenamefont {Lanyon}, \citenamefont {Love}, \citenamefont {Babbush} \emph
  {et~al.}}]{hempel2018quantum}%
  \BibitemOpen
  \bibfield  {author} {\bibinfo {author} {\bibfnamefont {C.}~\bibnamefont
  {Hempel}}, \bibinfo {author} {\bibfnamefont {C.}~\bibnamefont {Maier}},
  \bibinfo {author} {\bibfnamefont {J.}~\bibnamefont {Romero}}, \bibinfo
  {author} {\bibfnamefont {J.}~\bibnamefont {McClean}}, \bibinfo {author}
  {\bibfnamefont {T.}~\bibnamefont {Monz}}, \bibinfo {author} {\bibfnamefont
  {H.}~\bibnamefont {Shen}}, \bibinfo {author} {\bibfnamefont {P.}~\bibnamefont
  {Jurcevic}}, \bibinfo {author} {\bibfnamefont {B.~P.}\ \bibnamefont
  {Lanyon}}, \bibinfo {author} {\bibfnamefont {P.}~\bibnamefont {Love}},
  \bibinfo {author} {\bibfnamefont {R.}~\bibnamefont {Babbush}},  \emph
  {et~al.},\ }\href {https://doi.org/10.1103/PhysRevX.8.031022} {\bibfield
  {journal} {\bibinfo  {journal} {Phys. Rev. X}\ }\textbf {\bibinfo {volume}
  {8}},\ \bibinfo {pages} {031022} (\bibinfo {year} {2018})}\BibitemShut
  {NoStop}%
\bibitem [{\citenamefont {Bruzewicz}\ \emph {et~al.}(2019)\citenamefont
  {Bruzewicz}, \citenamefont {Chiaverini}, \citenamefont {McConnell},\ and\
  \citenamefont {Sage}}]{bruzewicz2019trapped}%
  \BibitemOpen
  \bibfield  {author} {\bibinfo {author} {\bibfnamefont {C.~D.}\ \bibnamefont
  {Bruzewicz}}, \bibinfo {author} {\bibfnamefont {J.}~\bibnamefont
  {Chiaverini}}, \bibinfo {author} {\bibfnamefont {R.}~\bibnamefont
  {McConnell}}, \ and\ \bibinfo {author} {\bibfnamefont {J.~M.}\ \bibnamefont
  {Sage}},\ }\href {https://doi.org/10.1063/1.5088164} {\bibfield  {journal}
  {\bibinfo  {journal} {Appl. Phys. Rev.}\ }\textbf {\bibinfo {volume} {6}},\
  \bibinfo {pages} {021314} (\bibinfo {year} {2019})}\BibitemShut {NoStop}%
\bibitem [{\citenamefont {Arute}\ \emph {et~al.}(2019)\citenamefont {Arute},
  \citenamefont {Arya}, \citenamefont {Babbush}, \citenamefont {Bacon},
  \citenamefont {Bardin}, \citenamefont {Barends}, \citenamefont {Biswas},
  \citenamefont {Boixo}, \citenamefont {Brandao}, \citenamefont {Buell} \emph
  {et~al.}}]{arute2019quantum}%
  \BibitemOpen
  \bibfield  {author} {\bibinfo {author} {\bibfnamefont {F.}~\bibnamefont
  {Arute}}, \bibinfo {author} {\bibfnamefont {K.}~\bibnamefont {Arya}},
  \bibinfo {author} {\bibfnamefont {R.}~\bibnamefont {Babbush}}, \bibinfo
  {author} {\bibfnamefont {D.}~\bibnamefont {Bacon}}, \bibinfo {author}
  {\bibfnamefont {J.~C.}\ \bibnamefont {Bardin}}, \bibinfo {author}
  {\bibfnamefont {R.}~\bibnamefont {Barends}}, \bibinfo {author} {\bibfnamefont
  {R.}~\bibnamefont {Biswas}}, \bibinfo {author} {\bibfnamefont
  {S.}~\bibnamefont {Boixo}}, \bibinfo {author} {\bibfnamefont {F.~G.}\
  \bibnamefont {Brandao}}, \bibinfo {author} {\bibfnamefont {D.~A.}\
  \bibnamefont {Buell}},  \emph {et~al.},\ }\href
  {https://doi.org/10.1038/s41586-019-1666-5} {\bibfield  {journal} {\bibinfo
  {journal} {Nature}\ }\textbf {\bibinfo {volume} {574}},\ \bibinfo {pages}
  {505} (\bibinfo {year} {2019})}\BibitemShut {NoStop}%
\bibitem [{\citenamefont {Arute}\ \emph {et~al.}(2020)\citenamefont {Arute},
  \citenamefont {Arya}, \citenamefont {Babbush}, \citenamefont {Bacon},
  \citenamefont {Bardin}, \citenamefont {Barends}, \citenamefont {Boixo},
  \citenamefont {Broughton}, \citenamefont {Buckley}, \citenamefont {Buell},
  \citenamefont {Burkett}, \citenamefont {Bushnell}, \citenamefont {Chen},
  \citenamefont {Chen}, \citenamefont {Chiaro}, \citenamefont {Collins},
  \citenamefont {Courtney}, \citenamefont {Demura}, \citenamefont {Dunsworth},
  \citenamefont {Farhi}, \citenamefont {Fowler}, \citenamefont {Foxen},
  \citenamefont {Gidney}, \citenamefont {Giustina}, \citenamefont {Graff},
  \citenamefont {Habegger}, \citenamefont {Harrigan}, \citenamefont {Ho},
  \citenamefont {Hong}, \citenamefont {Huang}, \citenamefont {Huggins},
  \citenamefont {Ioffe}, \citenamefont {Isakov}, \citenamefont {Jeffrey},
  \citenamefont {Jiang}, \citenamefont {Jones}, \citenamefont {Kafri},
  \citenamefont {Kechedzhi}, \citenamefont {Kelly}, \citenamefont {Kim},
  \citenamefont {Klimov}, \citenamefont {Korotkov}, \citenamefont {Kostritsa},
  \citenamefont {Landhuis}, \citenamefont {Laptev}, \citenamefont {Lindmark},
  \citenamefont {Lucero}, \citenamefont {Martin}, \citenamefont {Martinis},
  \citenamefont {McClean}, \citenamefont {McEwen}, \citenamefont {Megrant},
  \citenamefont {Mi}, \citenamefont {Mohseni}, \citenamefont {Mruczkiewicz},
  \citenamefont {Mutus}, \citenamefont {Naaman}, \citenamefont {Neeley},
  \citenamefont {Neill}, \citenamefont {Neven}, \citenamefont {Niu},
  \citenamefont {O{\textquoteright}Brien}, \citenamefont {Ostby}, \citenamefont
  {Petukhov}, \citenamefont {Putterman}, \citenamefont {Quintana},
  \citenamefont {Roushan}, \citenamefont {Rubin}, \citenamefont {Sank},
  \citenamefont {Satzinger}, \citenamefont {Smelyanskiy}, \citenamefont
  {Strain}, \citenamefont {Sung}, \citenamefont {Szalay}, \citenamefont
  {Takeshita}, \citenamefont {Vainsencher}, \citenamefont {White},
  \citenamefont {Wiebe}, \citenamefont {Yao}, \citenamefont {Yeh},\ and\
  \citenamefont {Zalcman}}]{arute2020hartree}%
  \BibitemOpen
  \bibfield  {author} {\bibinfo {author} {\bibfnamefont {F.}~\bibnamefont
  {Arute}}, \bibinfo {author} {\bibfnamefont {K.}~\bibnamefont {Arya}},
  \bibinfo {author} {\bibfnamefont {R.}~\bibnamefont {Babbush}}, \bibinfo
  {author} {\bibfnamefont {D.}~\bibnamefont {Bacon}}, \bibinfo {author}
  {\bibfnamefont {J.~C.}\ \bibnamefont {Bardin}}, \bibinfo {author}
  {\bibfnamefont {R.}~\bibnamefont {Barends}}, \bibinfo {author} {\bibfnamefont
  {S.}~\bibnamefont {Boixo}}, \bibinfo {author} {\bibfnamefont
  {M.}~\bibnamefont {Broughton}}, \bibinfo {author} {\bibfnamefont {B.~B.}\
  \bibnamefont {Buckley}}, \bibinfo {author} {\bibfnamefont {D.~A.}\
  \bibnamefont {Buell}}, \bibinfo {author} {\bibfnamefont {B.}~\bibnamefont
  {Burkett}}, \bibinfo {author} {\bibfnamefont {N.}~\bibnamefont {Bushnell}},
  \bibinfo {author} {\bibfnamefont {Y.}~\bibnamefont {Chen}}, \bibinfo {author}
  {\bibfnamefont {Z.}~\bibnamefont {Chen}}, \bibinfo {author} {\bibfnamefont
  {B.}~\bibnamefont {Chiaro}}, \bibinfo {author} {\bibfnamefont
  {R.}~\bibnamefont {Collins}}, \bibinfo {author} {\bibfnamefont
  {W.}~\bibnamefont {Courtney}}, \bibinfo {author} {\bibfnamefont
  {S.}~\bibnamefont {Demura}}, \bibinfo {author} {\bibfnamefont
  {A.}~\bibnamefont {Dunsworth}}, \bibinfo {author} {\bibfnamefont
  {E.}~\bibnamefont {Farhi}}, \bibinfo {author} {\bibfnamefont
  {A.}~\bibnamefont {Fowler}}, \bibinfo {author} {\bibfnamefont
  {B.}~\bibnamefont {Foxen}}, \bibinfo {author} {\bibfnamefont
  {C.}~\bibnamefont {Gidney}}, \bibinfo {author} {\bibfnamefont
  {M.}~\bibnamefont {Giustina}}, \bibinfo {author} {\bibfnamefont
  {R.}~\bibnamefont {Graff}}, \bibinfo {author} {\bibfnamefont
  {S.}~\bibnamefont {Habegger}}, \bibinfo {author} {\bibfnamefont {M.~P.}\
  \bibnamefont {Harrigan}}, \bibinfo {author} {\bibfnamefont {A.}~\bibnamefont
  {Ho}}, \bibinfo {author} {\bibfnamefont {S.}~\bibnamefont {Hong}}, \bibinfo
  {author} {\bibfnamefont {T.}~\bibnamefont {Huang}}, \bibinfo {author}
  {\bibfnamefont {W.~J.}\ \bibnamefont {Huggins}}, \bibinfo {author}
  {\bibfnamefont {L.}~\bibnamefont {Ioffe}}, \bibinfo {author} {\bibfnamefont
  {S.~V.}\ \bibnamefont {Isakov}}, \bibinfo {author} {\bibfnamefont
  {E.}~\bibnamefont {Jeffrey}}, \bibinfo {author} {\bibfnamefont
  {Z.}~\bibnamefont {Jiang}}, \bibinfo {author} {\bibfnamefont
  {C.}~\bibnamefont {Jones}}, \bibinfo {author} {\bibfnamefont
  {D.}~\bibnamefont {Kafri}}, \bibinfo {author} {\bibfnamefont
  {K.}~\bibnamefont {Kechedzhi}}, \bibinfo {author} {\bibfnamefont
  {J.}~\bibnamefont {Kelly}}, \bibinfo {author} {\bibfnamefont
  {S.}~\bibnamefont {Kim}}, \bibinfo {author} {\bibfnamefont {P.~V.}\
  \bibnamefont {Klimov}}, \bibinfo {author} {\bibfnamefont {A.}~\bibnamefont
  {Korotkov}}, \bibinfo {author} {\bibfnamefont {F.}~\bibnamefont {Kostritsa}},
  \bibinfo {author} {\bibfnamefont {D.}~\bibnamefont {Landhuis}}, \bibinfo
  {author} {\bibfnamefont {P.}~\bibnamefont {Laptev}}, \bibinfo {author}
  {\bibfnamefont {M.}~\bibnamefont {Lindmark}}, \bibinfo {author}
  {\bibfnamefont {E.}~\bibnamefont {Lucero}}, \bibinfo {author} {\bibfnamefont
  {O.}~\bibnamefont {Martin}}, \bibinfo {author} {\bibfnamefont {J.~M.}\
  \bibnamefont {Martinis}}, \bibinfo {author} {\bibfnamefont {J.~R.}\
  \bibnamefont {McClean}}, \bibinfo {author} {\bibfnamefont {M.}~\bibnamefont
  {McEwen}}, \bibinfo {author} {\bibfnamefont {A.}~\bibnamefont {Megrant}},
  \bibinfo {author} {\bibfnamefont {X.}~\bibnamefont {Mi}}, \bibinfo {author}
  {\bibfnamefont {M.}~\bibnamefont {Mohseni}}, \bibinfo {author} {\bibfnamefont
  {W.}~\bibnamefont {Mruczkiewicz}}, \bibinfo {author} {\bibfnamefont
  {J.}~\bibnamefont {Mutus}}, \bibinfo {author} {\bibfnamefont
  {O.}~\bibnamefont {Naaman}}, \bibinfo {author} {\bibfnamefont
  {M.}~\bibnamefont {Neeley}}, \bibinfo {author} {\bibfnamefont
  {C.}~\bibnamefont {Neill}}, \bibinfo {author} {\bibfnamefont
  {H.}~\bibnamefont {Neven}}, \bibinfo {author} {\bibfnamefont {M.~Y.}\
  \bibnamefont {Niu}}, \bibinfo {author} {\bibfnamefont {T.~E.}\ \bibnamefont
  {O{\textquoteright}Brien}}, \bibinfo {author} {\bibfnamefont
  {E.}~\bibnamefont {Ostby}}, \bibinfo {author} {\bibfnamefont
  {A.}~\bibnamefont {Petukhov}}, \bibinfo {author} {\bibfnamefont
  {H.}~\bibnamefont {Putterman}}, \bibinfo {author} {\bibfnamefont
  {C.}~\bibnamefont {Quintana}}, \bibinfo {author} {\bibfnamefont
  {P.}~\bibnamefont {Roushan}}, \bibinfo {author} {\bibfnamefont {N.~C.}\
  \bibnamefont {Rubin}}, \bibinfo {author} {\bibfnamefont {D.}~\bibnamefont
  {Sank}}, \bibinfo {author} {\bibfnamefont {K.~J.}\ \bibnamefont {Satzinger}},
  \bibinfo {author} {\bibfnamefont {V.}~\bibnamefont {Smelyanskiy}}, \bibinfo
  {author} {\bibfnamefont {D.}~\bibnamefont {Strain}}, \bibinfo {author}
  {\bibfnamefont {K.~J.}\ \bibnamefont {Sung}}, \bibinfo {author}
  {\bibfnamefont {M.}~\bibnamefont {Szalay}}, \bibinfo {author} {\bibfnamefont
  {T.~Y.}\ \bibnamefont {Takeshita}}, \bibinfo {author} {\bibfnamefont
  {A.}~\bibnamefont {Vainsencher}}, \bibinfo {author} {\bibfnamefont
  {T.}~\bibnamefont {White}}, \bibinfo {author} {\bibfnamefont
  {N.}~\bibnamefont {Wiebe}}, \bibinfo {author} {\bibfnamefont {Z.~J.}\
  \bibnamefont {Yao}}, \bibinfo {author} {\bibfnamefont {P.}~\bibnamefont
  {Yeh}}, \ and\ \bibinfo {author} {\bibfnamefont {A.}~\bibnamefont
  {Zalcman}},\ }\href {https://doi.org/10.1126/science.abb9811} {\bibfield
  {journal} {\bibinfo  {journal} {Science}\ }\textbf {\bibinfo {volume}
  {369}},\ \bibinfo {pages} {1084} (\bibinfo {year} {2020})}\BibitemShut
  {NoStop}%
\bibitem [{\citenamefont {Nam}\ \emph {et~al.}(2020)\citenamefont {Nam},
  \citenamefont {Chen}, \citenamefont {Pisenti}, \citenamefont {Wright},
  \citenamefont {Delaney}, \citenamefont {Maslov}, \citenamefont {Brown},
  \citenamefont {Allen}, \citenamefont {Amini}, \citenamefont {Apisdorf} \emph
  {et~al.}}]{nam2020ground}%
  \BibitemOpen
  \bibfield  {author} {\bibinfo {author} {\bibfnamefont {Y.}~\bibnamefont
  {Nam}}, \bibinfo {author} {\bibfnamefont {J.-S.}\ \bibnamefont {Chen}},
  \bibinfo {author} {\bibfnamefont {N.~C.}\ \bibnamefont {Pisenti}}, \bibinfo
  {author} {\bibfnamefont {K.}~\bibnamefont {Wright}}, \bibinfo {author}
  {\bibfnamefont {C.}~\bibnamefont {Delaney}}, \bibinfo {author} {\bibfnamefont
  {D.}~\bibnamefont {Maslov}}, \bibinfo {author} {\bibfnamefont {K.~R.}\
  \bibnamefont {Brown}}, \bibinfo {author} {\bibfnamefont {S.}~\bibnamefont
  {Allen}}, \bibinfo {author} {\bibfnamefont {J.~M.}\ \bibnamefont {Amini}},
  \bibinfo {author} {\bibfnamefont {J.}~\bibnamefont {Apisdorf}},  \emph
  {et~al.},\ }\href {https://doi.org/10.1038/s41534-020-0259-3} {\bibfield
  {journal} {\bibinfo  {journal} {npj Quantum Inf.}\ }\textbf {\bibinfo
  {volume} {6}},\ \bibinfo {pages} {1} (\bibinfo {year} {2020})}\BibitemShut
  {NoStop}%
\bibitem [{\citenamefont {Reiher}\ \emph {et~al.}(2017)\citenamefont {Reiher},
  \citenamefont {Wiebe}, \citenamefont {Svore}, \citenamefont {Wecker},\ and\
  \citenamefont {Troyer}}]{reiher2017elucidating}%
  \BibitemOpen
  \bibfield  {author} {\bibinfo {author} {\bibfnamefont {M.}~\bibnamefont
  {Reiher}}, \bibinfo {author} {\bibfnamefont {N.}~\bibnamefont {Wiebe}},
  \bibinfo {author} {\bibfnamefont {K.~M.}\ \bibnamefont {Svore}}, \bibinfo
  {author} {\bibfnamefont {D.}~\bibnamefont {Wecker}}, \ and\ \bibinfo {author}
  {\bibfnamefont {M.}~\bibnamefont {Troyer}},\ }\href
  {https://doi.org/10.1073/pnas.1619152114} {\bibfield  {journal} {\bibinfo
  {journal} {Proc. Natl. Acad. Sci. U.S.A.}\ }\textbf {\bibinfo {volume}
  {114}},\ \bibinfo {pages} {7555} (\bibinfo {year} {2017})}\BibitemShut
  {NoStop}%
\bibitem [{\citenamefont {Li}\ \emph {et~al.}(2019)\citenamefont {Li},
  \citenamefont {Li}, \citenamefont {Dattani}, \citenamefont {Umrigar},\ and\
  \citenamefont {Chan}}]{li2019electronic}%
  \BibitemOpen
  \bibfield  {author} {\bibinfo {author} {\bibfnamefont {Z.}~\bibnamefont
  {Li}}, \bibinfo {author} {\bibfnamefont {J.}~\bibnamefont {Li}}, \bibinfo
  {author} {\bibfnamefont {N.~S.}\ \bibnamefont {Dattani}}, \bibinfo {author}
  {\bibfnamefont {C.}~\bibnamefont {Umrigar}}, \ and\ \bibinfo {author}
  {\bibfnamefont {G.~K.-L.}\ \bibnamefont {Chan}},\ }\href
  {https://doi.org/10.1063/1.5063376} {\bibfield  {journal} {\bibinfo
  {journal} {J. Chem. Phys.}\ }\textbf {\bibinfo {volume} {150}},\ \bibinfo
  {pages} {024302} (\bibinfo {year} {2019})}\BibitemShut {NoStop}%
\bibitem [{\citenamefont {Cao}\ \emph {et~al.}(2019)\citenamefont {Cao},
  \citenamefont {Romero}, \citenamefont {Olson}, \citenamefont {Degroote},
  \citenamefont {Johnson}, \citenamefont {Kieferov{\'a}}, \citenamefont
  {Kivlichan}, \citenamefont {Menke}, \citenamefont {Peropadre}, \citenamefont
  {Sawaya} \emph {et~al.}}]{cao2019quantum}%
  \BibitemOpen
  \bibfield  {author} {\bibinfo {author} {\bibfnamefont {Y.}~\bibnamefont
  {Cao}}, \bibinfo {author} {\bibfnamefont {J.}~\bibnamefont {Romero}},
  \bibinfo {author} {\bibfnamefont {J.~P.}\ \bibnamefont {Olson}}, \bibinfo
  {author} {\bibfnamefont {M.}~\bibnamefont {Degroote}}, \bibinfo {author}
  {\bibfnamefont {P.~D.}\ \bibnamefont {Johnson}}, \bibinfo {author}
  {\bibfnamefont {M.}~\bibnamefont {Kieferov{\'a}}}, \bibinfo {author}
  {\bibfnamefont {I.~D.}\ \bibnamefont {Kivlichan}}, \bibinfo {author}
  {\bibfnamefont {T.}~\bibnamefont {Menke}}, \bibinfo {author} {\bibfnamefont
  {B.}~\bibnamefont {Peropadre}}, \bibinfo {author} {\bibfnamefont {N.~P.}\
  \bibnamefont {Sawaya}},  \emph {et~al.},\ }\href
  {https://doi.org/10.1021/acs.chemrev.8b00803} {\bibfield  {journal} {\bibinfo
   {journal} {Chem. Rev.}\ }\textbf {\bibinfo {volume} {119}},\ \bibinfo
  {pages} {10856} (\bibinfo {year} {2019})}\BibitemShut {NoStop}%
\bibitem [{\citenamefont {McArdle}\ \emph {et~al.}(2020)\citenamefont
  {McArdle}, \citenamefont {Endo}, \citenamefont {Aspuru-Guzik}, \citenamefont
  {Benjamin},\ and\ \citenamefont {Yuan}}]{mcardle2020quantum}%
  \BibitemOpen
  \bibfield  {author} {\bibinfo {author} {\bibfnamefont {S.}~\bibnamefont
  {McArdle}}, \bibinfo {author} {\bibfnamefont {S.}~\bibnamefont {Endo}},
  \bibinfo {author} {\bibfnamefont {A.}~\bibnamefont {Aspuru-Guzik}}, \bibinfo
  {author} {\bibfnamefont {S.~C.}\ \bibnamefont {Benjamin}}, \ and\ \bibinfo
  {author} {\bibfnamefont {X.}~\bibnamefont {Yuan}},\ }\href
  {https://doi.org/10.1103/RevModPhys.92.015003} {\bibfield  {journal}
  {\bibinfo  {journal} {Rev. Mod. Phys.}\ }\textbf {\bibinfo {volume} {92}},\
  \bibinfo {pages} {015003} (\bibinfo {year} {2020})}\BibitemShut {NoStop}%
\bibitem [{\citenamefont {Bauer}\ \emph {et~al.}(2020)\citenamefont {Bauer},
  \citenamefont {Bravyi}, \citenamefont {Motta},\ and\ \citenamefont
  {Chan}}]{bauer2020quantum}%
  \BibitemOpen
  \bibfield  {author} {\bibinfo {author} {\bibfnamefont {B.}~\bibnamefont
  {Bauer}}, \bibinfo {author} {\bibfnamefont {S.}~\bibnamefont {Bravyi}},
  \bibinfo {author} {\bibfnamefont {M.}~\bibnamefont {Motta}}, \ and\ \bibinfo
  {author} {\bibfnamefont {G.~K.}\ \bibnamefont {Chan}},\ }\href
  {https://arxiv.org/abs/2001.03685} {\bibfield  {journal} {\bibinfo  {journal}
  {arXiv:2001.03685}\ } (\bibinfo {year} {2020})}\BibitemShut {NoStop}%
\bibitem [{\citenamefont {von Burg}\ \emph {et~al.}(2020)\citenamefont {von
  Burg}, \citenamefont {Low}, \citenamefont {H{\"a}ner}, \citenamefont
  {Steiger}, \citenamefont {Reiher}, \citenamefont {Roetteler},\ and\
  \citenamefont {Troyer}}]{von2020quantum}%
  \BibitemOpen
  \bibfield  {author} {\bibinfo {author} {\bibfnamefont {V.}~\bibnamefont {von
  Burg}}, \bibinfo {author} {\bibfnamefont {G.~H.}\ \bibnamefont {Low}},
  \bibinfo {author} {\bibfnamefont {T.}~\bibnamefont {H{\"a}ner}}, \bibinfo
  {author} {\bibfnamefont {D.~S.}\ \bibnamefont {Steiger}}, \bibinfo {author}
  {\bibfnamefont {M.}~\bibnamefont {Reiher}}, \bibinfo {author} {\bibfnamefont
  {M.}~\bibnamefont {Roetteler}}, \ and\ \bibinfo {author} {\bibfnamefont
  {M.}~\bibnamefont {Troyer}},\ }\href {https://arxiv.org/abs/2007.14460}
  {\bibfield  {journal} {\bibinfo  {journal} {arXiv:2007.14460}\ } (\bibinfo
  {year} {2020})}\BibitemShut {NoStop}%
\bibitem [{\citenamefont {Preskill}(2018)}]{preskill2018quantum}%
  \BibitemOpen
  \bibfield  {author} {\bibinfo {author} {\bibfnamefont {J.}~\bibnamefont
  {Preskill}},\ }\href {https://doi.org/10.22331/q-2018-08-06-79} {\bibfield
  {journal} {\bibinfo  {journal} {Quantum}\ }\textbf {\bibinfo {volume} {2}},\
  \bibinfo {pages} {79} (\bibinfo {year} {2018})}\BibitemShut {NoStop}%
\bibitem [{\citenamefont {Abrams}\ and\ \citenamefont
  {Lloyd}(1999)}]{abrams1999quantum}%
  \BibitemOpen
  \bibfield  {author} {\bibinfo {author} {\bibfnamefont {D.~S.}\ \bibnamefont
  {Abrams}}\ and\ \bibinfo {author} {\bibfnamefont {S.}~\bibnamefont {Lloyd}},\
  }\href {https://doi.org/10.1103/PhysRevLett.83.5162} {\bibfield  {journal}
  {\bibinfo  {journal} {Phys. Rev. Lett.}\ }\textbf {\bibinfo {volume} {83}},\
  \bibinfo {pages} {5162} (\bibinfo {year} {1999})}\BibitemShut {NoStop}%
\bibitem [{\citenamefont {Aspuru-Guzik}\ \emph {et~al.}(2005)\citenamefont
  {Aspuru-Guzik}, \citenamefont {Dutoi}, \citenamefont {Love},\ and\
  \citenamefont {Head-Gordon}}]{aspuru2005simulated}%
  \BibitemOpen
  \bibfield  {author} {\bibinfo {author} {\bibfnamefont {A.}~\bibnamefont
  {Aspuru-Guzik}}, \bibinfo {author} {\bibfnamefont {A.~D.}\ \bibnamefont
  {Dutoi}}, \bibinfo {author} {\bibfnamefont {P.~J.}\ \bibnamefont {Love}}, \
  and\ \bibinfo {author} {\bibfnamefont {M.}~\bibnamefont {Head-Gordon}},\
  }\href {https://doi.org/10.1126/science.1113479} {\bibfield  {journal}
  {\bibinfo  {journal} {Science}\ }\textbf {\bibinfo {volume} {309}},\ \bibinfo
  {pages} {1704} (\bibinfo {year} {2005})}\BibitemShut {NoStop}%
\bibitem [{\citenamefont {O'Brien}\ \emph {et~al.}(2019)\citenamefont
  {O'Brien}, \citenamefont {Tarasinski},\ and\ \citenamefont
  {Terhal}}]{obrien2019quantum}%
  \BibitemOpen
  \bibfield  {author} {\bibinfo {author} {\bibfnamefont {T.~E.}\ \bibnamefont
  {O'Brien}}, \bibinfo {author} {\bibfnamefont {B.}~\bibnamefont {Tarasinski}},
  \ and\ \bibinfo {author} {\bibfnamefont {B.~M.}\ \bibnamefont {Terhal}},\
  }\href {https://doi.org/10.1088/1367-2630/aafb8e} {\bibfield  {journal}
  {\bibinfo  {journal} {New J. Phys.}\ }\textbf {\bibinfo {volume} {21}},\
  \bibinfo {pages} {023022} (\bibinfo {year} {2019})}\BibitemShut {NoStop}%
\bibitem [{\citenamefont {Peruzzo}\ \emph {et~al.}(2014)\citenamefont
  {Peruzzo}, \citenamefont {McClean}, \citenamefont {Shadbolt}, \citenamefont
  {Yung}, \citenamefont {Zhou}, \citenamefont {Love}, \citenamefont
  {Aspuru-Guzik},\ and\ \citenamefont {O’brien}}]{peruzzo2014variational}%
  \BibitemOpen
  \bibfield  {author} {\bibinfo {author} {\bibfnamefont {A.}~\bibnamefont
  {Peruzzo}}, \bibinfo {author} {\bibfnamefont {J.}~\bibnamefont {McClean}},
  \bibinfo {author} {\bibfnamefont {P.}~\bibnamefont {Shadbolt}}, \bibinfo
  {author} {\bibfnamefont {M.-H.}\ \bibnamefont {Yung}}, \bibinfo {author}
  {\bibfnamefont {X.-Q.}\ \bibnamefont {Zhou}}, \bibinfo {author}
  {\bibfnamefont {P.~J.}\ \bibnamefont {Love}}, \bibinfo {author}
  {\bibfnamefont {A.}~\bibnamefont {Aspuru-Guzik}}, \ and\ \bibinfo {author}
  {\bibfnamefont {J.~L.}\ \bibnamefont {O’brien}},\ }\href
  {https://doi.org/10.1038/ncomms5213} {\bibfield  {journal} {\bibinfo
  {journal} {Nature Comm.}\ }\textbf {\bibinfo {volume} {5}},\ \bibinfo {pages}
  {4213} (\bibinfo {year} {2014})}\BibitemShut {NoStop}%
\bibitem [{\citenamefont {McClean}\ \emph {et~al.}(2016)\citenamefont
  {McClean}, \citenamefont {Romero}, \citenamefont {Babbush},\ and\
  \citenamefont {Aspuru-Guzik}}]{mcclean2016theory}%
  \BibitemOpen
  \bibfield  {author} {\bibinfo {author} {\bibfnamefont {J.~R.}\ \bibnamefont
  {McClean}}, \bibinfo {author} {\bibfnamefont {J.}~\bibnamefont {Romero}},
  \bibinfo {author} {\bibfnamefont {R.}~\bibnamefont {Babbush}}, \ and\
  \bibinfo {author} {\bibfnamefont {A.}~\bibnamefont {Aspuru-Guzik}},\ }\href
  {https://doi.org/10.1088/1367-2630/18/2/023023} {\bibfield  {journal}
  {\bibinfo  {journal} {New J. Phys.}\ }\textbf {\bibinfo {volume} {18}},\
  \bibinfo {pages} {023023} (\bibinfo {year} {2016})}\BibitemShut {NoStop}%
\bibitem [{\citenamefont {Mai}\ and\ \citenamefont
  {González}(2020)}]{mai2020molecular}%
  \BibitemOpen
  \bibfield  {author} {\bibinfo {author} {\bibfnamefont {S.}~\bibnamefont
  {Mai}}\ and\ \bibinfo {author} {\bibfnamefont {L.}~\bibnamefont
  {González}},\ }\href
  {https://onlinelibrary.wiley.com/doi/abs/10.1002/anie.201916381} {\bibfield
  {journal} {\bibinfo  {journal} {Angew. Chem. Int. Ed.}\ }\textbf {\bibinfo
  {volume} {59}},\ \bibinfo {pages} {16832} (\bibinfo {year}
  {2020})}\BibitemShut {NoStop}%
\bibitem [{\citenamefont {Romero}\ \emph {et~al.}(2018)\citenamefont {Romero},
  \citenamefont {Babbush}, \citenamefont {McClean}, \citenamefont {Hempel},
  \citenamefont {Love},\ and\ \citenamefont
  {Aspuru-Guzik}}]{romero2018strategies}%
  \BibitemOpen
  \bibfield  {author} {\bibinfo {author} {\bibfnamefont {J.}~\bibnamefont
  {Romero}}, \bibinfo {author} {\bibfnamefont {R.}~\bibnamefont {Babbush}},
  \bibinfo {author} {\bibfnamefont {J.~R.}\ \bibnamefont {McClean}}, \bibinfo
  {author} {\bibfnamefont {C.}~\bibnamefont {Hempel}}, \bibinfo {author}
  {\bibfnamefont {P.~J.}\ \bibnamefont {Love}}, \ and\ \bibinfo {author}
  {\bibfnamefont {A.}~\bibnamefont {Aspuru-Guzik}},\ }\href
  {https://doi.org/10.1088/2058-9565/aad3e4} {\bibfield  {journal} {\bibinfo
  {journal} {Quantum Sci. Techno.}\ }\textbf {\bibinfo {volume} {4}},\ \bibinfo
  {pages} {014008} (\bibinfo {year} {2018})}\BibitemShut {NoStop}%
\bibitem [{\citenamefont {Lee}\ \emph {et~al.}(2018)\citenamefont {Lee},
  \citenamefont {Huggins}, \citenamefont {Head-Gordon},\ and\ \citenamefont
  {Whaley}}]{lee2018generalized}%
  \BibitemOpen
  \bibfield  {author} {\bibinfo {author} {\bibfnamefont {J.}~\bibnamefont
  {Lee}}, \bibinfo {author} {\bibfnamefont {W.~J.}\ \bibnamefont {Huggins}},
  \bibinfo {author} {\bibfnamefont {M.}~\bibnamefont {Head-Gordon}}, \ and\
  \bibinfo {author} {\bibfnamefont {K.~B.}\ \bibnamefont {Whaley}},\ }\href
  {https://doi.org/10.1021/acs.jctc.8b01004} {\bibfield  {journal} {\bibinfo
  {journal} {J. Chem. Theory Comput.}\ }\textbf {\bibinfo {volume} {15}},\
  \bibinfo {pages} {311} (\bibinfo {year} {2018})}\BibitemShut {NoStop}%
\bibitem [{\citenamefont {Ryabinkin}\ \emph
  {et~al.}(2018{\natexlab{a}})\citenamefont {Ryabinkin}, \citenamefont {Yen},
  \citenamefont {Genin},\ and\ \citenamefont {Izmaylov}}]{ryabinkin2018qubit}%
  \BibitemOpen
  \bibfield  {author} {\bibinfo {author} {\bibfnamefont {I.~G.}\ \bibnamefont
  {Ryabinkin}}, \bibinfo {author} {\bibfnamefont {T.-C.}\ \bibnamefont {Yen}},
  \bibinfo {author} {\bibfnamefont {S.~N.}\ \bibnamefont {Genin}}, \ and\
  \bibinfo {author} {\bibfnamefont {A.~F.}\ \bibnamefont {Izmaylov}},\ }\href
  {https://doi.org/10.1021/acs.jctc.8b00932} {\bibfield  {journal} {\bibinfo
  {journal} {J. Chem. Theory Comput.}\ }\textbf {\bibinfo {volume} {14}},\
  \bibinfo {pages} {6317} (\bibinfo {year} {2018}{\natexlab{a}})}\BibitemShut
  {NoStop}%
\bibitem [{\citenamefont {McArdle}\ \emph {et~al.}(2019)\citenamefont
  {McArdle}, \citenamefont {Jones}, \citenamefont {Endo}, \citenamefont {Li},
  \citenamefont {Benjamin},\ and\ \citenamefont
  {Yuan}}]{mcardle2019variational}%
  \BibitemOpen
  \bibfield  {author} {\bibinfo {author} {\bibfnamefont {S.}~\bibnamefont
  {McArdle}}, \bibinfo {author} {\bibfnamefont {T.}~\bibnamefont {Jones}},
  \bibinfo {author} {\bibfnamefont {S.}~\bibnamefont {Endo}}, \bibinfo {author}
  {\bibfnamefont {Y.}~\bibnamefont {Li}}, \bibinfo {author} {\bibfnamefont
  {S.~C.}\ \bibnamefont {Benjamin}}, \ and\ \bibinfo {author} {\bibfnamefont
  {X.}~\bibnamefont {Yuan}},\ }\href
  {https://doi.org/10.1038/s41534-019-0187-2} {\bibfield  {journal} {\bibinfo
  {journal} {npj Quantum Inf.}\ }\textbf {\bibinfo {volume} {5}},\ \bibinfo
  {pages} {1} (\bibinfo {year} {2019})}\BibitemShut {NoStop}%
\bibitem [{\citenamefont {Mizukami}\ \emph {et~al.}(2019)\citenamefont
  {Mizukami}, \citenamefont {Mitarai}, \citenamefont {Nakagawa}, \citenamefont
  {Yamamoto}, \citenamefont {Yan},\ and\ \citenamefont
  {Ohnishi}}]{mizukami2019orbital}%
  \BibitemOpen
  \bibfield  {author} {\bibinfo {author} {\bibfnamefont {W.}~\bibnamefont
  {Mizukami}}, \bibinfo {author} {\bibfnamefont {K.}~\bibnamefont {Mitarai}},
  \bibinfo {author} {\bibfnamefont {Y.~O.}\ \bibnamefont {Nakagawa}}, \bibinfo
  {author} {\bibfnamefont {T.}~\bibnamefont {Yamamoto}}, \bibinfo {author}
  {\bibfnamefont {T.}~\bibnamefont {Yan}}, \ and\ \bibinfo {author}
  {\bibfnamefont {Y.-y.}\ \bibnamefont {Ohnishi}},\ }\href
  {https://arxiv.org/abs/1910.11526} {\bibfield  {journal} {\bibinfo  {journal}
  {arXiv:1910.11526}\ } (\bibinfo {year} {2019})}\BibitemShut {NoStop}%
\bibitem [{\citenamefont {K\"uhn}\ \emph {et~al.}(2019)\citenamefont {K\"uhn},
  \citenamefont {Zanker}, \citenamefont {Deglmann}, \citenamefont {Marthaler},\
  and\ \citenamefont {Wei{\ss}}}]{kuhn2019accuracy}%
  \BibitemOpen
  \bibfield  {author} {\bibinfo {author} {\bibfnamefont {M.}~\bibnamefont
  {K\"uhn}}, \bibinfo {author} {\bibfnamefont {S.}~\bibnamefont {Zanker}},
  \bibinfo {author} {\bibfnamefont {P.}~\bibnamefont {Deglmann}}, \bibinfo
  {author} {\bibfnamefont {M.}~\bibnamefont {Marthaler}}, \ and\ \bibinfo
  {author} {\bibfnamefont {H.}~\bibnamefont {Wei{\ss}}},\ }\href
  {https://doi.org/10.1021/acs.jctc.9b00236} {\bibfield  {journal} {\bibinfo
  {journal} {J. Chem. Theory Comput.}\ }\textbf {\bibinfo {volume} {15}},\
  \bibinfo {pages} {4764} (\bibinfo {year} {2019})}\BibitemShut {NoStop}%
\bibitem [{\citenamefont {Rattew}\ \emph {et~al.}(2019)\citenamefont {Rattew},
  \citenamefont {Hu}, \citenamefont {Pistoia}, \citenamefont {Chen},\ and\
  \citenamefont {Wood}}]{rattew2019domain}%
  \BibitemOpen
  \bibfield  {author} {\bibinfo {author} {\bibfnamefont {A.~G.}\ \bibnamefont
  {Rattew}}, \bibinfo {author} {\bibfnamefont {S.}~\bibnamefont {Hu}}, \bibinfo
  {author} {\bibfnamefont {M.}~\bibnamefont {Pistoia}}, \bibinfo {author}
  {\bibfnamefont {R.}~\bibnamefont {Chen}}, \ and\ \bibinfo {author}
  {\bibfnamefont {S.}~\bibnamefont {Wood}},\ }\href
  {https://arxiv.org/abs/1910.09694v4} {\bibfield  {journal} {\bibinfo
  {journal} {arXiv:1910.09694v4}\ } (\bibinfo {year} {2019})}\BibitemShut
  {NoStop}%
\bibitem [{\citenamefont {Sokolov}\ \emph
  {et~al.}(2020{\natexlab{a}})\citenamefont {Sokolov}, \citenamefont
  {Barkoutsos}, \citenamefont {Ollitrault}, \citenamefont {Greenberg},
  \citenamefont {Rice}, \citenamefont {Pistoia},\ and\ \citenamefont
  {Tavernelli}}]{sokolov2020quantum}%
  \BibitemOpen
  \bibfield  {author} {\bibinfo {author} {\bibfnamefont {I.~O.}\ \bibnamefont
  {Sokolov}}, \bibinfo {author} {\bibfnamefont {P.~K.}\ \bibnamefont
  {Barkoutsos}}, \bibinfo {author} {\bibfnamefont {P.~J.}\ \bibnamefont
  {Ollitrault}}, \bibinfo {author} {\bibfnamefont {D.}~\bibnamefont
  {Greenberg}}, \bibinfo {author} {\bibfnamefont {J.}~\bibnamefont {Rice}},
  \bibinfo {author} {\bibfnamefont {M.}~\bibnamefont {Pistoia}}, \ and\
  \bibinfo {author} {\bibfnamefont {I.}~\bibnamefont {Tavernelli}},\ }\href
  {https://doi.org/10.1063/1.5141835} {\bibfield  {journal} {\bibinfo
  {journal} {J. Chem. Theory Comput.}\ }\textbf {\bibinfo {volume} {152}},\
  \bibinfo {pages} {124107} (\bibinfo {year} {2020}{\natexlab{a}})}\BibitemShut
  {NoStop}%
\bibitem [{\citenamefont {Ryabinkin}\ \emph {et~al.}(2020)\citenamefont
  {Ryabinkin}, \citenamefont {Lang}, \citenamefont {Genin},\ and\ \citenamefont
  {Izmaylov}}]{ryabinkin2020iterative}%
  \BibitemOpen
  \bibfield  {author} {\bibinfo {author} {\bibfnamefont {I.~G.}\ \bibnamefont
  {Ryabinkin}}, \bibinfo {author} {\bibfnamefont {R.~A.}\ \bibnamefont {Lang}},
  \bibinfo {author} {\bibfnamefont {S.~N.}\ \bibnamefont {Genin}}, \ and\
  \bibinfo {author} {\bibfnamefont {A.~F.}\ \bibnamefont {Izmaylov}},\ }\href
  {https://doi.org/10.1021/acs.jctc.9b01084} {\bibfield  {journal} {\bibinfo
  {journal} {J. Chem. Theory Comput.}\ }\textbf {\bibinfo {volume} {16}},\
  \bibinfo {pages} {1055} (\bibinfo {year} {2020})}\BibitemShut {NoStop}%
\bibitem [{\citenamefont {Lang}\ \emph {et~al.}(2020)\citenamefont {Lang},
  \citenamefont {Ryabinkin},\ and\ \citenamefont
  {Izmaylov}}]{lang2020iterative}%
  \BibitemOpen
  \bibfield  {author} {\bibinfo {author} {\bibfnamefont {R.~A.}\ \bibnamefont
  {Lang}}, \bibinfo {author} {\bibfnamefont {I.~G.}\ \bibnamefont {Ryabinkin}},
  \ and\ \bibinfo {author} {\bibfnamefont {A.~F.}\ \bibnamefont {Izmaylov}},\
  }\href {https://arxiv.org/abs/2002.05701} {\bibfield  {journal} {\bibinfo
  {journal} {arXiv:2002.05701}\ } (\bibinfo {year} {2020})}\BibitemShut
  {NoStop}%
\bibitem [{\citenamefont {Matsuzawa}\ and\ \citenamefont
  {Kurashige}(2020)}]{matsuzawa2020jastrow}%
  \BibitemOpen
  \bibfield  {author} {\bibinfo {author} {\bibfnamefont {Y.}~\bibnamefont
  {Matsuzawa}}\ and\ \bibinfo {author} {\bibfnamefont {Y.}~\bibnamefont
  {Kurashige}},\ }\href {https://doi.org/10.1021/acs.jctc.9b00963} {\bibfield
  {journal} {\bibinfo  {journal} {J. Chem. Theory Comput.}\ }\textbf {\bibinfo
  {volume} {16}},\ \bibinfo {pages} {944} (\bibinfo {year} {2020})}\BibitemShut
  {NoStop}%
\bibitem [{\citenamefont {Huggins}\ \emph {et~al.}(2020)\citenamefont
  {Huggins}, \citenamefont {Lee}, \citenamefont {Baek}, \citenamefont
  {O'Gorman},\ and\ \citenamefont {Whaley}}]{huggins2020non}%
  \BibitemOpen
  \bibfield  {author} {\bibinfo {author} {\bibfnamefont {W.~J.}\ \bibnamefont
  {Huggins}}, \bibinfo {author} {\bibfnamefont {J.}~\bibnamefont {Lee}},
  \bibinfo {author} {\bibfnamefont {U.}~\bibnamefont {Baek}}, \bibinfo {author}
  {\bibfnamefont {B.}~\bibnamefont {O'Gorman}}, \ and\ \bibinfo {author}
  {\bibfnamefont {K.~B.}\ \bibnamefont {Whaley}},\ }\href
  {https://doi.org/10.1088/1367-2630/ab867b} {\bibfield  {journal} {\bibinfo
  {journal} {New J. Phys.}\ } (\bibinfo {year} {2020})}\BibitemShut {NoStop}%
\bibitem [{\citenamefont {Gomes}\ \emph {et~al.}(2020)\citenamefont {Gomes},
  \citenamefont {Zhang}, \citenamefont {Berthusen}, \citenamefont {Wang},
  \citenamefont {Ho}, \citenamefont {Orth},\ and\ \citenamefont
  {Yao}}]{gomes2020efficient}%
  \BibitemOpen
  \bibfield  {author} {\bibinfo {author} {\bibfnamefont {N.}~\bibnamefont
  {Gomes}}, \bibinfo {author} {\bibfnamefont {F.}~\bibnamefont {Zhang}},
  \bibinfo {author} {\bibfnamefont {N.~F.}\ \bibnamefont {Berthusen}}, \bibinfo
  {author} {\bibfnamefont {C.-Z.}\ \bibnamefont {Wang}}, \bibinfo {author}
  {\bibfnamefont {K.-M.}\ \bibnamefont {Ho}}, \bibinfo {author} {\bibfnamefont
  {P.~P.}\ \bibnamefont {Orth}}, \ and\ \bibinfo {author} {\bibfnamefont
  {Y.}~\bibnamefont {Yao}},\ }\href {https://arxiv.org/abs/2006.15371}
  {\bibfield  {journal} {\bibinfo  {journal} {arXiv:2006.15371}\ } (\bibinfo
  {year} {2020})}\BibitemShut {NoStop}%
\bibitem [{\citenamefont {Meitei}\ \emph {et~al.}(2020)\citenamefont {Meitei},
  \citenamefont {Gard}, \citenamefont {Barron}, \citenamefont {Pappas},
  \citenamefont {Economou}, \citenamefont {Barnes},\ and\ \citenamefont
  {Mayhall}}]{meitei2020gate}%
  \BibitemOpen
  \bibfield  {author} {\bibinfo {author} {\bibfnamefont {O.~R.}\ \bibnamefont
  {Meitei}}, \bibinfo {author} {\bibfnamefont {B.~T.}\ \bibnamefont {Gard}},
  \bibinfo {author} {\bibfnamefont {G.~S.}\ \bibnamefont {Barron}}, \bibinfo
  {author} {\bibfnamefont {D.~P.}\ \bibnamefont {Pappas}}, \bibinfo {author}
  {\bibfnamefont {S.~E.}\ \bibnamefont {Economou}}, \bibinfo {author}
  {\bibfnamefont {E.}~\bibnamefont {Barnes}}, \ and\ \bibinfo {author}
  {\bibfnamefont {N.~J.}\ \bibnamefont {Mayhall}},\ }\href
  {https://arxiv.org/abs/2008.04302} {\bibfield  {journal} {\bibinfo  {journal}
  {arXiv:2008.04302}\ } (\bibinfo {year} {2020})}\BibitemShut {NoStop}%
\bibitem [{\citenamefont {Wang}\ \emph {et~al.}(2020)\citenamefont {Wang},
  \citenamefont {Li}, \citenamefont {Monroe},\ and\ \citenamefont
  {Nam}}]{wang2020resourceoptimized}%
  \BibitemOpen
  \bibfield  {author} {\bibinfo {author} {\bibfnamefont {Q.}~\bibnamefont
  {Wang}}, \bibinfo {author} {\bibfnamefont {M.}~\bibnamefont {Li}}, \bibinfo
  {author} {\bibfnamefont {C.}~\bibnamefont {Monroe}}, \ and\ \bibinfo {author}
  {\bibfnamefont {Y.}~\bibnamefont {Nam}},\ }\href
  {https://arxiv.org/abs/2004.04151} {\bibfield  {journal} {\bibinfo  {journal}
  {arXiv:2004.04151}\ } (\bibinfo {year} {2020})}\BibitemShut {NoStop}%
\bibitem [{\citenamefont {Zhang}\ \emph
  {et~al.}(2020{\natexlab{a}})\citenamefont {Zhang}, \citenamefont {Kyaw},
  \citenamefont {Kottmann}, \citenamefont {Degroote},\ and\ \citenamefont
  {Aspuru-Guzik}}]{zhang2020mutual}%
  \BibitemOpen
  \bibfield  {author} {\bibinfo {author} {\bibfnamefont {Z.-J.}\ \bibnamefont
  {Zhang}}, \bibinfo {author} {\bibfnamefont {T.~H.}\ \bibnamefont {Kyaw}},
  \bibinfo {author} {\bibfnamefont {J.~S.}\ \bibnamefont {Kottmann}}, \bibinfo
  {author} {\bibfnamefont {M.}~\bibnamefont {Degroote}}, \ and\ \bibinfo
  {author} {\bibfnamefont {A.}~\bibnamefont {Aspuru-Guzik}},\ }\href
  {https://arxiv.org/abs/2008.07553} {\bibfield  {journal} {\bibinfo  {journal}
  {arXiv:2008.07553}\ } (\bibinfo {year} {2020}{\natexlab{a}})}\BibitemShut
  {NoStop}%
\bibitem [{\citenamefont {McClean}\ \emph {et~al.}(2017)\citenamefont
  {McClean}, \citenamefont {Kimchi-Schwartz}, \citenamefont {Carter},\ and\
  \citenamefont {de~Jong}}]{mcclean2017hybrid}%
  \BibitemOpen
  \bibfield  {author} {\bibinfo {author} {\bibfnamefont {J.~R.}\ \bibnamefont
  {McClean}}, \bibinfo {author} {\bibfnamefont {M.~E.}\ \bibnamefont
  {Kimchi-Schwartz}}, \bibinfo {author} {\bibfnamefont {J.}~\bibnamefont
  {Carter}}, \ and\ \bibinfo {author} {\bibfnamefont {W.~A.}\ \bibnamefont
  {de~Jong}},\ }\href {https://doi.org/10.1103/PhysRevA.95.042308} {\bibfield
  {journal} {\bibinfo  {journal} {Phys. Rev. A}\ }\textbf {\bibinfo {volume}
  {95}},\ \bibinfo {pages} {042308} (\bibinfo {year} {2017})}\BibitemShut
  {NoStop}%
\bibitem [{\citenamefont {Ollitrault}\ \emph {et~al.}(2019)\citenamefont
  {Ollitrault}, \citenamefont {Kandala}, \citenamefont {Chen}, \citenamefont
  {Barkoutsos}, \citenamefont {Mezzacapo}, \citenamefont {Pistoia},
  \citenamefont {Sheldon}, \citenamefont {Woerner}, \citenamefont {Gambetta},\
  and\ \citenamefont {Tavernelli}}]{ollitrault2019quantum}%
  \BibitemOpen
  \bibfield  {author} {\bibinfo {author} {\bibfnamefont {P.~J.}\ \bibnamefont
  {Ollitrault}}, \bibinfo {author} {\bibfnamefont {A.}~\bibnamefont {Kandala}},
  \bibinfo {author} {\bibfnamefont {C.-F.}\ \bibnamefont {Chen}}, \bibinfo
  {author} {\bibfnamefont {P.~K.}\ \bibnamefont {Barkoutsos}}, \bibinfo
  {author} {\bibfnamefont {A.}~\bibnamefont {Mezzacapo}}, \bibinfo {author}
  {\bibfnamefont {M.}~\bibnamefont {Pistoia}}, \bibinfo {author} {\bibfnamefont
  {S.}~\bibnamefont {Sheldon}}, \bibinfo {author} {\bibfnamefont
  {S.}~\bibnamefont {Woerner}}, \bibinfo {author} {\bibfnamefont
  {J.}~\bibnamefont {Gambetta}}, \ and\ \bibinfo {author} {\bibfnamefont
  {I.}~\bibnamefont {Tavernelli}},\ }\href {https://arxiv.org/abs/1910.12890}
  {\bibfield  {journal} {\bibinfo  {journal} {arXiv:1910.12890}\ } (\bibinfo
  {year} {2019})}\BibitemShut {NoStop}%
\bibitem [{\citenamefont {Nakanishi}\ \emph {et~al.}(2019)\citenamefont
  {Nakanishi}, \citenamefont {Mitarai},\ and\ \citenamefont
  {Fujii}}]{nakanishi2019subspace}%
  \BibitemOpen
  \bibfield  {author} {\bibinfo {author} {\bibfnamefont {K.~M.}\ \bibnamefont
  {Nakanishi}}, \bibinfo {author} {\bibfnamefont {K.}~\bibnamefont {Mitarai}},
  \ and\ \bibinfo {author} {\bibfnamefont {K.}~\bibnamefont {Fujii}},\ }\href
  {https://doi.org/10.1103/PhysRevResearch.1.033062} {\bibfield  {journal}
  {\bibinfo  {journal} {Phys. Rev. Res.}\ }\textbf {\bibinfo {volume} {1}},\
  \bibinfo {pages} {033062} (\bibinfo {year} {2019})}\BibitemShut {NoStop}%
\bibitem [{\citenamefont {Ibe}\ \emph {et~al.}(2020)\citenamefont {Ibe},
  \citenamefont {Nakagawa}, \citenamefont {Yamamoto}, \citenamefont {Mitarai},
  \citenamefont {Gao},\ and\ \citenamefont {Kobayashi}}]{ibe2020calculating}%
  \BibitemOpen
  \bibfield  {author} {\bibinfo {author} {\bibfnamefont {Y.}~\bibnamefont
  {Ibe}}, \bibinfo {author} {\bibfnamefont {Y.~O.}\ \bibnamefont {Nakagawa}},
  \bibinfo {author} {\bibfnamefont {T.}~\bibnamefont {Yamamoto}}, \bibinfo
  {author} {\bibfnamefont {K.}~\bibnamefont {Mitarai}}, \bibinfo {author}
  {\bibfnamefont {Q.}~\bibnamefont {Gao}}, \ and\ \bibinfo {author}
  {\bibfnamefont {T.}~\bibnamefont {Kobayashi}},\ }\href
  {https://arxiv.org/abs/2002.11724} {\bibfield  {journal} {\bibinfo  {journal}
  {arXiv:2002.11724}\ } (\bibinfo {year} {2020})}\BibitemShut {NoStop}%
\bibitem [{\citenamefont {Higgott}\ \emph {et~al.}(2019)\citenamefont
  {Higgott}, \citenamefont {Wang},\ and\ \citenamefont
  {Brierley}}]{higgott2019variational}%
  \BibitemOpen
  \bibfield  {author} {\bibinfo {author} {\bibfnamefont {O.}~\bibnamefont
  {Higgott}}, \bibinfo {author} {\bibfnamefont {D.}~\bibnamefont {Wang}}, \
  and\ \bibinfo {author} {\bibfnamefont {S.}~\bibnamefont {Brierley}},\ }\href
  {https://doi.org/10.22331/q-2019-07-01-156} {\bibfield  {journal} {\bibinfo
  {journal} {Quantum}\ }\textbf {\bibinfo {volume} {3}},\ \bibinfo {pages}
  {156} (\bibinfo {year} {2019})}\BibitemShut {NoStop}%
\bibitem [{\citenamefont {Jones}\ \emph {et~al.}(2019)\citenamefont {Jones},
  \citenamefont {Endo}, \citenamefont {McArdle}, \citenamefont {Yuan},\ and\
  \citenamefont {Benjamin}}]{jones2019variational}%
  \BibitemOpen
  \bibfield  {author} {\bibinfo {author} {\bibfnamefont {T.}~\bibnamefont
  {Jones}}, \bibinfo {author} {\bibfnamefont {S.}~\bibnamefont {Endo}},
  \bibinfo {author} {\bibfnamefont {S.}~\bibnamefont {McArdle}}, \bibinfo
  {author} {\bibfnamefont {X.}~\bibnamefont {Yuan}}, \ and\ \bibinfo {author}
  {\bibfnamefont {S.~C.}\ \bibnamefont {Benjamin}},\ }\href
  {https://doi.org/10.1103/PhysRevA.99.062304} {\bibfield  {journal} {\bibinfo
  {journal} {Phys. Rev. A}\ }\textbf {\bibinfo {volume} {99}},\ \bibinfo
  {pages} {062304} (\bibinfo {year} {2019})}\BibitemShut {NoStop}%
\bibitem [{\citenamefont {Jouzdani}\ \emph {et~al.}(2019)\citenamefont
  {Jouzdani}, \citenamefont {Bringuier},\ and\ \citenamefont
  {Kostuk}}]{jouzdani2019method}%
  \BibitemOpen
  \bibfield  {author} {\bibinfo {author} {\bibfnamefont {P.}~\bibnamefont
  {Jouzdani}}, \bibinfo {author} {\bibfnamefont {S.}~\bibnamefont {Bringuier}},
  \ and\ \bibinfo {author} {\bibfnamefont {M.}~\bibnamefont {Kostuk}},\ }\href
  {https://arxiv.org/abs/1908.05238} {\bibfield  {journal} {\bibinfo  {journal}
  {arXiv:1908.05238}\ } (\bibinfo {year} {2019})}\BibitemShut {NoStop}%
\bibitem [{\citenamefont {Parrish}\ \emph
  {et~al.}(2019{\natexlab{a}})\citenamefont {Parrish}, \citenamefont
  {Hohenstein}, \citenamefont {McMahon},\ and\ \citenamefont
  {Mart{\'\i}nez}}]{parrish2019quantum}%
  \BibitemOpen
  \bibfield  {author} {\bibinfo {author} {\bibfnamefont {R.~M.}\ \bibnamefont
  {Parrish}}, \bibinfo {author} {\bibfnamefont {E.~G.}\ \bibnamefont
  {Hohenstein}}, \bibinfo {author} {\bibfnamefont {P.~L.}\ \bibnamefont
  {McMahon}}, \ and\ \bibinfo {author} {\bibfnamefont {T.~J.}\ \bibnamefont
  {Mart{\'\i}nez}},\ }\href {https://doi.org/10.1103/PhysRevLett.122.230401}
  {\bibfield  {journal} {\bibinfo  {journal} {Phys. Rev. Lett.}\ }\textbf
  {\bibinfo {volume} {122}},\ \bibinfo {pages} {230401} (\bibinfo {year}
  {2019}{\natexlab{a}})}\BibitemShut {NoStop}%
\bibitem [{\citenamefont {Parrish}\ \emph
  {et~al.}(2019{\natexlab{b}})\citenamefont {Parrish}, \citenamefont
  {Hohenstein}, \citenamefont {McMahon},\ and\ \citenamefont
  {Martinez}}]{parrish2019hybrid}%
  \BibitemOpen
  \bibfield  {author} {\bibinfo {author} {\bibfnamefont {R.~M.}\ \bibnamefont
  {Parrish}}, \bibinfo {author} {\bibfnamefont {E.~G.}\ \bibnamefont
  {Hohenstein}}, \bibinfo {author} {\bibfnamefont {P.~L.}\ \bibnamefont
  {McMahon}}, \ and\ \bibinfo {author} {\bibfnamefont {T.~J.}\ \bibnamefont
  {Martinez}},\ }\href {https://arxiv.org/abs/1906.08728} {\bibfield  {journal}
  {\bibinfo  {journal} {arXiv:1906.08728}\ } (\bibinfo {year}
  {2019}{\natexlab{b}})}\BibitemShut {NoStop}%
\bibitem [{\citenamefont {Bauman}\ \emph
  {et~al.}(2019{\natexlab{a}})\citenamefont {Bauman}, \citenamefont {Low},\
  and\ \citenamefont {Kowalski}}]{bauman2019quantum}%
  \BibitemOpen
  \bibfield  {author} {\bibinfo {author} {\bibfnamefont {N.~P.}\ \bibnamefont
  {Bauman}}, \bibinfo {author} {\bibfnamefont {G.~H.}\ \bibnamefont {Low}}, \
  and\ \bibinfo {author} {\bibfnamefont {K.}~\bibnamefont {Kowalski}},\ }\href
  {https://doi.org/10.1063/1.5128103} {\bibfield  {journal} {\bibinfo
  {journal} {J. Chem. Phys.}\ }\textbf {\bibinfo {volume} {151}},\ \bibinfo
  {pages} {234114} (\bibinfo {year} {2019}{\natexlab{a}})}\BibitemShut
  {NoStop}%
\bibitem [{\citenamefont {Motta}\ \emph
  {et~al.}(2020{\natexlab{a}})\citenamefont {Motta}, \citenamefont {Sun},
  \citenamefont {Tan}, \citenamefont {O’Rourke}, \citenamefont {Ye},
  \citenamefont {Minnich}, \citenamefont {Brand{\~a}o},\ and\ \citenamefont
  {Chan}}]{motta2020determining}%
  \BibitemOpen
  \bibfield  {author} {\bibinfo {author} {\bibfnamefont {M.}~\bibnamefont
  {Motta}}, \bibinfo {author} {\bibfnamefont {C.}~\bibnamefont {Sun}}, \bibinfo
  {author} {\bibfnamefont {A.~T.}\ \bibnamefont {Tan}}, \bibinfo {author}
  {\bibfnamefont {M.~J.}\ \bibnamefont {O’Rourke}}, \bibinfo {author}
  {\bibfnamefont {E.}~\bibnamefont {Ye}}, \bibinfo {author} {\bibfnamefont
  {A.~J.}\ \bibnamefont {Minnich}}, \bibinfo {author} {\bibfnamefont {F.~G.}\
  \bibnamefont {Brand{\~a}o}}, \ and\ \bibinfo {author} {\bibfnamefont
  {G.~K.-L.}\ \bibnamefont {Chan}},\ }\href
  {https://doi.org/10.1038/s41567-019-0704-4} {\bibfield  {journal} {\bibinfo
  {journal} {Nat. Phys.}\ }\textbf {\bibinfo {volume} {16}},\ \bibinfo {pages}
  {205} (\bibinfo {year} {2020}{\natexlab{a}})}\BibitemShut {NoStop}%
\bibitem [{\citenamefont {Zhang}\ \emph
  {et~al.}(2020{\natexlab{b}})\citenamefont {Zhang}, \citenamefont {Yuan},\
  and\ \citenamefont {Yin}}]{zhang2020variational}%
  \BibitemOpen
  \bibfield  {author} {\bibinfo {author} {\bibfnamefont {D.-B.}\ \bibnamefont
  {Zhang}}, \bibinfo {author} {\bibfnamefont {Z.-H.}\ \bibnamefont {Yuan}}, \
  and\ \bibinfo {author} {\bibfnamefont {T.}~\bibnamefont {Yin}},\ }\href
  {https://arxiv.org/abs/2006.15781} {\bibfield  {journal} {\bibinfo  {journal}
  {arXiv:2006.15781}\ } (\bibinfo {year} {2020}{\natexlab{b}})}\BibitemShut
  {NoStop}%
\bibitem [{\citenamefont {Ryabinkin}\ \emph
  {et~al.}(2018{\natexlab{b}})\citenamefont {Ryabinkin}, \citenamefont
  {Genin},\ and\ \citenamefont {Izmaylov}}]{ryabinkin2018constrained}%
  \BibitemOpen
  \bibfield  {author} {\bibinfo {author} {\bibfnamefont {I.~G.}\ \bibnamefont
  {Ryabinkin}}, \bibinfo {author} {\bibfnamefont {S.~N.}\ \bibnamefont
  {Genin}}, \ and\ \bibinfo {author} {\bibfnamefont {A.~F.}\ \bibnamefont
  {Izmaylov}},\ }\href {https://doi.org/10.1021/acs.jctc.8b00943} {\bibfield
  {journal} {\bibinfo  {journal} {J. Chem. Theory Comput.}\ }\textbf {\bibinfo
  {volume} {15}},\ \bibinfo {pages} {249} (\bibinfo {year}
  {2018}{\natexlab{b}})}\BibitemShut {NoStop}%
\bibitem [{\citenamefont {Ryabinkin}\ and\ \citenamefont
  {Genin}(2018)}]{ryabinkin2018symmetry}%
  \BibitemOpen
  \bibfield  {author} {\bibinfo {author} {\bibfnamefont {I.~G.}\ \bibnamefont
  {Ryabinkin}}\ and\ \bibinfo {author} {\bibfnamefont {S.~N.}\ \bibnamefont
  {Genin}},\ }\href {https://arxiv.org/abs/1812.09812} {\bibfield  {journal}
  {\bibinfo  {journal} {arXiv:1812.09812}\ } (\bibinfo {year}
  {2018})}\BibitemShut {NoStop}%
\bibitem [{\citenamefont {Greene-Diniz}\ and\ \citenamefont
  {Ramo}(2019)}]{greene2019generalized}%
  \BibitemOpen
  \bibfield  {author} {\bibinfo {author} {\bibfnamefont {G.}~\bibnamefont
  {Greene-Diniz}}\ and\ \bibinfo {author} {\bibfnamefont {D.~M.}\ \bibnamefont
  {Ramo}},\ }\href {https://arxiv.org/abs/1910.05168} {\bibfield  {journal}
  {\bibinfo  {journal} {arXiv:1910.05168}\ } (\bibinfo {year}
  {2019})}\BibitemShut {NoStop}%
\bibitem [{\citenamefont {Zhang}\ \emph
  {et~al.}(2020{\natexlab{c}})\citenamefont {Zhang}, \citenamefont {Gomes},
  \citenamefont {Berthusen}, \citenamefont {Orth}, \citenamefont {Wang},
  \citenamefont {Ho},\ and\ \citenamefont {Yao}}]{zhang2020shallow}%
  \BibitemOpen
  \bibfield  {author} {\bibinfo {author} {\bibfnamefont {F.}~\bibnamefont
  {Zhang}}, \bibinfo {author} {\bibfnamefont {N.}~\bibnamefont {Gomes}},
  \bibinfo {author} {\bibfnamefont {N.~F.}\ \bibnamefont {Berthusen}}, \bibinfo
  {author} {\bibfnamefont {P.~P.}\ \bibnamefont {Orth}}, \bibinfo {author}
  {\bibfnamefont {C.-Z.}\ \bibnamefont {Wang}}, \bibinfo {author}
  {\bibfnamefont {K.-M.}\ \bibnamefont {Ho}}, \ and\ \bibinfo {author}
  {\bibfnamefont {Y.-X.}\ \bibnamefont {Yao}},\ }\href
  {https://arxiv.org/abs/2006.11213} {\bibfield  {journal} {\bibinfo  {journal}
  {arXiv:2006.11213}\ } (\bibinfo {year} {2020}{\natexlab{c}})}\BibitemShut
  {NoStop}%
\bibitem [{\citenamefont {Gard}\ \emph {et~al.}(2020)\citenamefont {Gard},
  \citenamefont {Zhu}, \citenamefont {Barron}, \citenamefont {Mayhall},
  \citenamefont {Economou},\ and\ \citenamefont {Barnes}}]{gard2020efficient}%
  \BibitemOpen
  \bibfield  {author} {\bibinfo {author} {\bibfnamefont {B.~T.}\ \bibnamefont
  {Gard}}, \bibinfo {author} {\bibfnamefont {L.}~\bibnamefont {Zhu}}, \bibinfo
  {author} {\bibfnamefont {G.~S.}\ \bibnamefont {Barron}}, \bibinfo {author}
  {\bibfnamefont {N.~J.}\ \bibnamefont {Mayhall}}, \bibinfo {author}
  {\bibfnamefont {S.~E.}\ \bibnamefont {Economou}}, \ and\ \bibinfo {author}
  {\bibfnamefont {E.}~\bibnamefont {Barnes}},\ }\href
  {https://doi.org/10.1038/s41534-019-0240-1} {\bibfield  {journal} {\bibinfo
  {journal} {npj Quantum Inf.}\ }\textbf {\bibinfo {volume} {6}},\ \bibinfo
  {pages} {1} (\bibinfo {year} {2020})}\BibitemShut {NoStop}%
\bibitem [{\citenamefont {Seki}\ \emph {et~al.}(2020)\citenamefont {Seki},
  \citenamefont {Shirakawa},\ and\ \citenamefont {Yunoki}}]{seki2020symmetry}%
  \BibitemOpen
  \bibfield  {author} {\bibinfo {author} {\bibfnamefont {K.}~\bibnamefont
  {Seki}}, \bibinfo {author} {\bibfnamefont {T.}~\bibnamefont {Shirakawa}}, \
  and\ \bibinfo {author} {\bibfnamefont {S.}~\bibnamefont {Yunoki}},\ }\href
  {https://doi.org/10.1103/PhysRevA.101.052340} {\bibfield  {journal} {\bibinfo
   {journal} {Phys. Rev. A}\ }\textbf {\bibinfo {volume} {101}},\ \bibinfo
  {pages} {052340} (\bibinfo {year} {2020})}\BibitemShut {NoStop}%
\bibitem [{\citenamefont {Domcke}\ \emph {et~al.}(2011)\citenamefont {Domcke},
  \citenamefont {Yarkony},\ and\ \citenamefont
  {K{\"o}ppel}}]{domcke2011conical}%
  \BibitemOpen
  \bibfield  {author} {\bibinfo {author} {\bibfnamefont {W.}~\bibnamefont
  {Domcke}}, \bibinfo {author} {\bibfnamefont {D.~R.}\ \bibnamefont {Yarkony}},
  \ and\ \bibinfo {author} {\bibfnamefont {H.}~\bibnamefont {K{\"o}ppel}},\
  }\href@noop {} {\emph {\bibinfo {title} {Conical intersections: theory,
  computation and experiment}}},\ Vol.~\bibinfo {volume} {17}\ (\bibinfo
  {publisher} {World Scientific},\ \bibinfo {year} {2011})\BibitemShut
  {NoStop}%
\bibitem [{\citenamefont {Yarkony}(2012)}]{yarkony2012nonadiabatic}%
  \BibitemOpen
  \bibfield  {author} {\bibinfo {author} {\bibfnamefont {D.~R.}\ \bibnamefont
  {Yarkony}},\ }\href {https://doi.org/10.1021/cr2001299} {\bibfield  {journal}
  {\bibinfo  {journal} {Chemical reviews}\ }\textbf {\bibinfo {volume} {112}},\
  \bibinfo {pages} {481} (\bibinfo {year} {2012})}\BibitemShut {NoStop}%
\bibitem [{\citenamefont {Garavelli}\ \emph {et~al.}(1997)\citenamefont
  {Garavelli}, \citenamefont {Celani}, \citenamefont {Bernardi}, \citenamefont
  {Robb},\ and\ \citenamefont {Olivucci}}]{garavelli1997c5h6nh2+}%
  \BibitemOpen
  \bibfield  {author} {\bibinfo {author} {\bibfnamefont {M.}~\bibnamefont
  {Garavelli}}, \bibinfo {author} {\bibfnamefont {P.}~\bibnamefont {Celani}},
  \bibinfo {author} {\bibfnamefont {F.}~\bibnamefont {Bernardi}}, \bibinfo
  {author} {\bibfnamefont {M.}~\bibnamefont {Robb}}, \ and\ \bibinfo {author}
  {\bibfnamefont {M.}~\bibnamefont {Olivucci}},\ }\href
  {https://doi.org/10.1021/ja9610895} {\bibfield  {journal} {\bibinfo
  {journal} {J. Am. Chem. Soc.}\ }\textbf {\bibinfo {volume} {119}},\ \bibinfo
  {pages} {6891} (\bibinfo {year} {1997})}\BibitemShut {NoStop}%
\bibitem [{\citenamefont {Gonz{\'a}lez-Luque}\ \emph
  {et~al.}(2000)\citenamefont {Gonz{\'a}lez-Luque}, \citenamefont {Garavelli},
  \citenamefont {Bernardi}, \citenamefont {Merch{\'a}n}, \citenamefont {Robb},\
  and\ \citenamefont {Olivucci}}]{gonzalez2000computational}%
  \BibitemOpen
  \bibfield  {author} {\bibinfo {author} {\bibfnamefont {R.}~\bibnamefont
  {Gonz{\'a}lez-Luque}}, \bibinfo {author} {\bibfnamefont {M.}~\bibnamefont
  {Garavelli}}, \bibinfo {author} {\bibfnamefont {F.}~\bibnamefont {Bernardi}},
  \bibinfo {author} {\bibfnamefont {M.}~\bibnamefont {Merch{\'a}n}}, \bibinfo
  {author} {\bibfnamefont {M.~A.}\ \bibnamefont {Robb}}, \ and\ \bibinfo
  {author} {\bibfnamefont {M.}~\bibnamefont {Olivucci}},\ }\href
  {https://doi.org/10.1073/pnas.97.17.9379} {\bibfield  {journal} {\bibinfo
  {journal} {Proc. Natl. Acad. Sci. U.S.A.}\ }\textbf {\bibinfo {volume}
  {97}},\ \bibinfo {pages} {9379} (\bibinfo {year} {2000})}\BibitemShut
  {NoStop}%
\bibitem [{\citenamefont {Polli}\ \emph {et~al.}(2010)\citenamefont {Polli},
  \citenamefont {Alto{\`e}}, \citenamefont {Weingart}, \citenamefont
  {Spillane}, \citenamefont {Manzoni}, \citenamefont {Brida}, \citenamefont
  {Tomasello}, \citenamefont {Orlandi}, \citenamefont {Kukura}, \citenamefont
  {Mathies} \emph {et~al.}}]{polli2010conical}%
  \BibitemOpen
  \bibfield  {author} {\bibinfo {author} {\bibfnamefont {D.}~\bibnamefont
  {Polli}}, \bibinfo {author} {\bibfnamefont {P.}~\bibnamefont {Alto{\`e}}},
  \bibinfo {author} {\bibfnamefont {O.}~\bibnamefont {Weingart}}, \bibinfo
  {author} {\bibfnamefont {K.~M.}\ \bibnamefont {Spillane}}, \bibinfo {author}
  {\bibfnamefont {C.}~\bibnamefont {Manzoni}}, \bibinfo {author} {\bibfnamefont
  {D.}~\bibnamefont {Brida}}, \bibinfo {author} {\bibfnamefont
  {G.}~\bibnamefont {Tomasello}}, \bibinfo {author} {\bibfnamefont
  {G.}~\bibnamefont {Orlandi}}, \bibinfo {author} {\bibfnamefont
  {P.}~\bibnamefont {Kukura}}, \bibinfo {author} {\bibfnamefont {R.~A.}\
  \bibnamefont {Mathies}},  \emph {et~al.},\ }\href
  {https://doi.org/10.1038/nature09346} {\bibfield  {journal} {\bibinfo
  {journal} {Nature}\ }\textbf {\bibinfo {volume} {467}},\ \bibinfo {pages}
  {440} (\bibinfo {year} {2010})}\BibitemShut {NoStop}%
\bibitem [{\citenamefont {Valsson}\ \emph {et~al.}(2013)\citenamefont
  {Valsson}, \citenamefont {Campomanes}, \citenamefont {Tavernelli},
  \citenamefont {Rothlisberger},\ and\ \citenamefont
  {Filippi}}]{valsson2013rhodopsin}%
  \BibitemOpen
  \bibfield  {author} {\bibinfo {author} {\bibfnamefont {O.}~\bibnamefont
  {Valsson}}, \bibinfo {author} {\bibfnamefont {P.}~\bibnamefont {Campomanes}},
  \bibinfo {author} {\bibfnamefont {I.}~\bibnamefont {Tavernelli}}, \bibinfo
  {author} {\bibfnamefont {U.}~\bibnamefont {Rothlisberger}}, \ and\ \bibinfo
  {author} {\bibfnamefont {C.}~\bibnamefont {Filippi}},\ }\href
  {https://doi.org/10.1021/ct3010408} {\bibfield  {journal} {\bibinfo
  {journal} {J. Chem. Theory Comput.}\ }\textbf {\bibinfo {volume} {9}},\
  \bibinfo {pages} {2441} (\bibinfo {year} {2013})}\BibitemShut {NoStop}%
\bibitem [{\citenamefont {Manathunga}\ \emph {et~al.}(2016)\citenamefont
  {Manathunga}, \citenamefont {Yang}, \citenamefont {Luk}, \citenamefont
  {Gozem}, \citenamefont {Frutos}, \citenamefont {Valentini}, \citenamefont
  {Ferr{\`e}},\ and\ \citenamefont {Olivucci}}]{manathunga2016probing}%
  \BibitemOpen
  \bibfield  {author} {\bibinfo {author} {\bibfnamefont {M.}~\bibnamefont
  {Manathunga}}, \bibinfo {author} {\bibfnamefont {X.}~\bibnamefont {Yang}},
  \bibinfo {author} {\bibfnamefont {H.~L.}\ \bibnamefont {Luk}}, \bibinfo
  {author} {\bibfnamefont {S.}~\bibnamefont {Gozem}}, \bibinfo {author}
  {\bibfnamefont {L.~M.}\ \bibnamefont {Frutos}}, \bibinfo {author}
  {\bibfnamefont {A.}~\bibnamefont {Valentini}}, \bibinfo {author}
  {\bibfnamefont {N.}~\bibnamefont {Ferr{\`e}}}, \ and\ \bibinfo {author}
  {\bibfnamefont {M.}~\bibnamefont {Olivucci}},\ }\href
  {https://doi.org/10.1021/acs.jctc.5b00945} {\bibfield  {journal} {\bibinfo
  {journal} {J. Chem. Theory Comput.}\ }\textbf {\bibinfo {volume} {12}},\
  \bibinfo {pages} {839} (\bibinfo {year} {2016})}\BibitemShut {NoStop}%
\bibitem [{\citenamefont {Olaso-Gonz{\'a}lez}\ \emph
  {et~al.}(2006)\citenamefont {Olaso-Gonz{\'a}lez}, \citenamefont
  {Merch{\'a}n},\ and\ \citenamefont
  {Serrano-Andr{\'e}s}}]{olaso2006ultrafast}%
  \BibitemOpen
  \bibfield  {author} {\bibinfo {author} {\bibfnamefont {G.}~\bibnamefont
  {Olaso-Gonz{\'a}lez}}, \bibinfo {author} {\bibfnamefont {M.}~\bibnamefont
  {Merch{\'a}n}}, \ and\ \bibinfo {author} {\bibfnamefont {L.}~\bibnamefont
  {Serrano-Andr{\'e}s}},\ }\href {https://doi.org/10.1021/jp063915u} {\bibfield
   {journal} {\bibinfo  {journal} {J. Phys. Chem. B}\ }\textbf {\bibinfo
  {volume} {110}},\ \bibinfo {pages} {24734} (\bibinfo {year}
  {2006})}\BibitemShut {NoStop}%
\bibitem [{\citenamefont {Kang}\ \emph {et~al.}(2002)\citenamefont {Kang},
  \citenamefont {Lee}, \citenamefont {Jung}, \citenamefont {Ko},\ and\
  \citenamefont {Kim}}]{kang2002intrinsic}%
  \BibitemOpen
  \bibfield  {author} {\bibinfo {author} {\bibfnamefont {H.}~\bibnamefont
  {Kang}}, \bibinfo {author} {\bibfnamefont {K.~T.}\ \bibnamefont {Lee}},
  \bibinfo {author} {\bibfnamefont {B.}~\bibnamefont {Jung}}, \bibinfo {author}
  {\bibfnamefont {Y.~J.}\ \bibnamefont {Ko}}, \ and\ \bibinfo {author}
  {\bibfnamefont {S.~K.}\ \bibnamefont {Kim}},\ }\href
  {https://doi.org/10.1021/ja027627x} {\bibfield  {journal} {\bibinfo
  {journal} {J. Am. Chem. Soc.}\ }\textbf {\bibinfo {volume} {124}},\ \bibinfo
  {pages} {12958} (\bibinfo {year} {2002})}\BibitemShut {NoStop}%
\bibitem [{\citenamefont {Groenhof}\ \emph {et~al.}(2007)\citenamefont
  {Groenhof}, \citenamefont {Sch{\"a}fer}, \citenamefont {Boggio-Pasqua},
  \citenamefont {Goette}, \citenamefont {Grubm{\"u}ller},\ and\ \citenamefont
  {Robb}}]{groenhof2007ultrafast}%
  \BibitemOpen
  \bibfield  {author} {\bibinfo {author} {\bibfnamefont {G.}~\bibnamefont
  {Groenhof}}, \bibinfo {author} {\bibfnamefont {L.~V.}\ \bibnamefont
  {Sch{\"a}fer}}, \bibinfo {author} {\bibfnamefont {M.}~\bibnamefont
  {Boggio-Pasqua}}, \bibinfo {author} {\bibfnamefont {M.}~\bibnamefont
  {Goette}}, \bibinfo {author} {\bibfnamefont {H.}~\bibnamefont
  {Grubm{\"u}ller}}, \ and\ \bibinfo {author} {\bibfnamefont {M.~A.}\
  \bibnamefont {Robb}},\ }\href {https://doi.org/10.1021/ja069176c} {\bibfield
  {journal} {\bibinfo  {journal} {J. Am. Chem. Soc.}\ }\textbf {\bibinfo
  {volume} {129}},\ \bibinfo {pages} {6812} (\bibinfo {year}
  {2007})}\BibitemShut {NoStop}%
\bibitem [{\citenamefont {Barbatti}\ \emph {et~al.}(2010)\citenamefont
  {Barbatti}, \citenamefont {Aquino}, \citenamefont {Szymczak}, \citenamefont
  {Nachtigallov{\'a}}, \citenamefont {Hobza},\ and\ \citenamefont
  {Lischka}}]{barbatti2010relaxation}%
  \BibitemOpen
  \bibfield  {author} {\bibinfo {author} {\bibfnamefont {M.}~\bibnamefont
  {Barbatti}}, \bibinfo {author} {\bibfnamefont {A.~J.}\ \bibnamefont
  {Aquino}}, \bibinfo {author} {\bibfnamefont {J.~J.}\ \bibnamefont
  {Szymczak}}, \bibinfo {author} {\bibfnamefont {D.}~\bibnamefont
  {Nachtigallov{\'a}}}, \bibinfo {author} {\bibfnamefont {P.}~\bibnamefont
  {Hobza}}, \ and\ \bibinfo {author} {\bibfnamefont {H.}~\bibnamefont
  {Lischka}},\ }\href {https://doi.org/10.1073/pnas.1014982107} {\bibfield
  {journal} {\bibinfo  {journal} {Proc. Natl. Acad. Sci. U.S.A.}\ }\textbf
  {\bibinfo {volume} {107}},\ \bibinfo {pages} {21453} (\bibinfo {year}
  {2010})}\BibitemShut {NoStop}%
\bibitem [{\citenamefont {Franc{\'e}s-Monerris}\ \emph
  {et~al.}(2018)\citenamefont {Franc{\'e}s-Monerris}, \citenamefont {Gattuso},
  \citenamefont {Roca-Sanju{\'a}n}, \citenamefont {Tu{\~n}{\'o}n},
  \citenamefont {Marazzi}, \citenamefont {Dumont},\ and\ \citenamefont
  {Monari}}]{frances2018dynamics}%
  \BibitemOpen
  \bibfield  {author} {\bibinfo {author} {\bibfnamefont {A.}~\bibnamefont
  {Franc{\'e}s-Monerris}}, \bibinfo {author} {\bibfnamefont {H.}~\bibnamefont
  {Gattuso}}, \bibinfo {author} {\bibfnamefont {D.}~\bibnamefont
  {Roca-Sanju{\'a}n}}, \bibinfo {author} {\bibfnamefont {I.}~\bibnamefont
  {Tu{\~n}{\'o}n}}, \bibinfo {author} {\bibfnamefont {M.}~\bibnamefont
  {Marazzi}}, \bibinfo {author} {\bibfnamefont {E.}~\bibnamefont {Dumont}}, \
  and\ \bibinfo {author} {\bibfnamefont {A.}~\bibnamefont {Monari}},\ }\href
  {https://doi.org/10.1039/c8sc03252a} {\bibfield  {journal} {\bibinfo
  {journal} {Chem. Sci.}\ }\textbf {\bibinfo {volume} {9}},\ \bibinfo {pages}
  {7902} (\bibinfo {year} {2018})}\BibitemShut {NoStop}%
\bibitem [{\citenamefont {May}\ and\ \citenamefont
  {K{\"u}hn}(2008)}]{may2008charge}%
  \BibitemOpen
  \bibfield  {author} {\bibinfo {author} {\bibfnamefont {V.}~\bibnamefont
  {May}}\ and\ \bibinfo {author} {\bibfnamefont {O.}~\bibnamefont {K{\"u}hn}},\
  }\href@noop {} {\emph {\bibinfo {title} {Charge and energy transfer dynamics
  in molecular systems}}}\ (\bibinfo  {publisher} {John Wiley \& Sons},\
  \bibinfo {year} {2008})\BibitemShut {NoStop}%
\bibitem [{\citenamefont {Ho}\ and\ \citenamefont
  {Lasorne}(2019)}]{ho2019diabatic}%
  \BibitemOpen
  \bibfield  {author} {\bibinfo {author} {\bibfnamefont {E.~K.-L.}\
  \bibnamefont {Ho}}\ and\ \bibinfo {author} {\bibfnamefont {B.}~\bibnamefont
  {Lasorne}},\ }\href {https://doi.org/10.1016/j.comptc.2019.03.013} {\bibfield
   {journal} {\bibinfo  {journal} {Comp. Theo. Chem.}\ }\textbf {\bibinfo
  {volume} {1156}},\ \bibinfo {pages} {25} (\bibinfo {year}
  {2019})}\BibitemShut {NoStop}%
\bibitem [{\citenamefont {Gozem}\ \emph {et~al.}(2014)\citenamefont {Gozem},
  \citenamefont {Melaccio}, \citenamefont {Valentini}, \citenamefont {Filatov},
  \citenamefont {Huix-Rotllant}, \citenamefont {Ferré}, \citenamefont
  {Frutos}, \citenamefont {Angeli}, \citenamefont {Krylov}, \citenamefont
  {Granovsky} \emph {et~al.}}]{gozem2014shape}%
  \BibitemOpen
  \bibfield  {author} {\bibinfo {author} {\bibfnamefont {S.}~\bibnamefont
  {Gozem}}, \bibinfo {author} {\bibfnamefont {F.}~\bibnamefont {Melaccio}},
  \bibinfo {author} {\bibfnamefont {A.}~\bibnamefont {Valentini}}, \bibinfo
  {author} {\bibfnamefont {M.}~\bibnamefont {Filatov}}, \bibinfo {author}
  {\bibfnamefont {M.}~\bibnamefont {Huix-Rotllant}}, \bibinfo {author}
  {\bibfnamefont {N.}~\bibnamefont {Ferré}}, \bibinfo {author} {\bibfnamefont
  {L.~M.}\ \bibnamefont {Frutos}}, \bibinfo {author} {\bibfnamefont
  {C.}~\bibnamefont {Angeli}}, \bibinfo {author} {\bibfnamefont {A.~I.}\
  \bibnamefont {Krylov}}, \bibinfo {author} {\bibfnamefont {A.~A.}\
  \bibnamefont {Granovsky}},  \emph {et~al.},\ }\href
  {https://doi.org/10.1021/ct500154k} {\bibfield  {journal} {\bibinfo
  {journal} {J. Chem. Theory Comput.}\ }\textbf {\bibinfo {volume} {10}},\
  \bibinfo {pages} {3074} (\bibinfo {year} {2014})}\BibitemShut {NoStop}%
\bibitem [{\citenamefont {Siegbahn}\ \emph {et~al.}(1981)\citenamefont
  {Siegbahn}, \citenamefont {Alml{\"o}f}, \citenamefont {Heiberg},\ and\
  \citenamefont {Roos}}]{siegbahn1981complete}%
  \BibitemOpen
  \bibfield  {author} {\bibinfo {author} {\bibfnamefont {P.~E.}\ \bibnamefont
  {Siegbahn}}, \bibinfo {author} {\bibfnamefont {J.}~\bibnamefont
  {Alml{\"o}f}}, \bibinfo {author} {\bibfnamefont {A.}~\bibnamefont {Heiberg}},
  \ and\ \bibinfo {author} {\bibfnamefont {B.~O.}\ \bibnamefont {Roos}},\
  }\href {https://doi.org/10.1063/1.441359} {\bibfield  {journal} {\bibinfo
  {journal} {J. Chem. Phys.}\ }\textbf {\bibinfo {volume} {74}},\ \bibinfo
  {pages} {2384} (\bibinfo {year} {1981})}\BibitemShut {NoStop}%
\bibitem [{\citenamefont {Helgaker}\ \emph {et~al.}(2014)\citenamefont
  {Helgaker}, \citenamefont {Jorgensen},\ and\ \citenamefont
  {Olsen}}]{helgaker2014molecular}%
  \BibitemOpen
  \bibfield  {author} {\bibinfo {author} {\bibfnamefont {T.}~\bibnamefont
  {Helgaker}}, \bibinfo {author} {\bibfnamefont {P.}~\bibnamefont {Jorgensen}},
  \ and\ \bibinfo {author} {\bibfnamefont {J.}~\bibnamefont {Olsen}},\
  }\href@noop {} {\emph {\bibinfo {title} {Molecular electronic-structure
  theory}}}\ (\bibinfo  {publisher} {John Wiley \& Sons},\ \bibinfo {year}
  {2014})\BibitemShut {NoStop}%
\bibitem [{\citenamefont {Klessinger}\ and\ \citenamefont
  {Michl}(1995)}]{klessinger1995excited}%
  \BibitemOpen
  \bibfield  {author} {\bibinfo {author} {\bibfnamefont {M.}~\bibnamefont
  {Klessinger}}\ and\ \bibinfo {author} {\bibfnamefont {J.}~\bibnamefont
  {Michl}},\ }\href@noop {} {\emph {\bibinfo {title} {Excited states and
  photochemistry of organic molecules}}}\ (\bibinfo  {publisher} {VCH
  publishers},\ \bibinfo {year} {1995})\BibitemShut {NoStop}%
\bibitem [{\citenamefont {Robb}\ \emph {et~al.}(1995)\citenamefont {Robb},
  \citenamefont {Bernardi},\ and\ \citenamefont {Olivucci}}]{robb1995conical}%
  \BibitemOpen
  \bibfield  {author} {\bibinfo {author} {\bibfnamefont {M.~A.}\ \bibnamefont
  {Robb}}, \bibinfo {author} {\bibfnamefont {F.}~\bibnamefont {Bernardi}}, \
  and\ \bibinfo {author} {\bibfnamefont {M.}~\bibnamefont {Olivucci}},\ }\href
  {https://doi.org/10.1351/pac199567050783} {\bibfield  {journal} {\bibinfo
  {journal} {Pure Appl. Chem.}\ }\textbf {\bibinfo {volume} {67}},\ \bibinfo
  {pages} {783} (\bibinfo {year} {1995})}\BibitemShut {NoStop}%
\bibitem [{\citenamefont {Bernardi}\ \emph {et~al.}(1996)\citenamefont
  {Bernardi}, \citenamefont {Olivucci},\ and\ \citenamefont
  {Robb}}]{bernardi1996potential}%
  \BibitemOpen
  \bibfield  {author} {\bibinfo {author} {\bibfnamefont {F.}~\bibnamefont
  {Bernardi}}, \bibinfo {author} {\bibfnamefont {M.}~\bibnamefont {Olivucci}},
  \ and\ \bibinfo {author} {\bibfnamefont {M.~A.}\ \bibnamefont {Robb}},\
  }\href {https://doi.org/10.1039/CS9962500321} {\bibfield  {journal} {\bibinfo
   {journal} {Chem. Soc. Rev.}\ }\textbf {\bibinfo {volume} {25}},\ \bibinfo
  {pages} {321} (\bibinfo {year} {1996})}\BibitemShut {NoStop}%
\bibitem [{\citenamefont {Domcke}\ \emph {et~al.}(2004)\citenamefont {Domcke},
  \citenamefont {Yarkony} \emph {et~al.}}]{domcke2004conical}%
  \BibitemOpen
  \bibfield  {author} {\bibinfo {author} {\bibfnamefont {W.}~\bibnamefont
  {Domcke}}, \bibinfo {author} {\bibfnamefont {D.}~\bibnamefont {Yarkony}},
  \emph {et~al.},\ }\href@noop {} {\emph {\bibinfo {title} {Conical
  intersections: electronic structure, dynamics \& spectroscopy}}},\
  Vol.~\bibinfo {volume} {15}\ (\bibinfo  {publisher} {World Scientific},\
  \bibinfo {year} {2004})\BibitemShut {NoStop}%
\bibitem [{\citenamefont {Birge}(1990)}]{birge1990nature}%
  \BibitemOpen
  \bibfield  {author} {\bibinfo {author} {\bibfnamefont {R.~R.}\ \bibnamefont
  {Birge}},\ }\href {https://doi.org/10.1016/0005-2728(90)90163-X} {\bibfield
  {journal} {\bibinfo  {journal} {Biochimica et Biophysica Acta
  (BBA)-Bioenergetics}\ }\textbf {\bibinfo {volume} {1016}},\ \bibinfo {pages}
  {293} (\bibinfo {year} {1990})}\BibitemShut {NoStop}%
\bibitem [{\citenamefont {Menzel}\ \emph {et~al.}(2019)\citenamefont {Menzel},
  \citenamefont {de~Groot},\ and\ \citenamefont
  {Buda}}]{menzel2019photoinduced}%
  \BibitemOpen
  \bibfield  {author} {\bibinfo {author} {\bibfnamefont {J.~P.}\ \bibnamefont
  {Menzel}}, \bibinfo {author} {\bibfnamefont {H.~J.}\ \bibnamefont
  {de~Groot}}, \ and\ \bibinfo {author} {\bibfnamefont {F.}~\bibnamefont
  {Buda}},\ }\href {https://doi.org/10.1021/acs.jpclett.9b02408} {\bibfield
  {journal} {\bibinfo  {journal} {J. Phys. Chem. Lett.}\ }\textbf {\bibinfo
  {volume} {10}},\ \bibinfo {pages} {6504} (\bibinfo {year}
  {2019})}\BibitemShut {NoStop}%
\bibitem [{\citenamefont {Baldo}\ \emph {et~al.}(1998)\citenamefont {Baldo},
  \citenamefont {'Brien}, \citenamefont {You}, \citenamefont {Shoustiko},
  \citenamefont {Sibley}, \citenamefont {Thompson},\ and\ \citenamefont
  {Forrest}}]{Baldo98}%
  \BibitemOpen
  \bibfield  {author} {\bibinfo {author} {\bibfnamefont {M.}~\bibnamefont
  {Baldo}}, \bibinfo {author} {\bibfnamefont {F.}~\bibnamefont {'Brien}},
  \bibinfo {author} {\bibfnamefont {Y.}~\bibnamefont {You}}, \bibinfo {author}
  {\bibfnamefont {A.}~\bibnamefont {Shoustiko}}, \bibinfo {author}
  {\bibfnamefont {S.}~\bibnamefont {Sibley}}, \bibinfo {author} {\bibfnamefont
  {M.}~\bibnamefont {Thompson}}, \ and\ \bibinfo {author} {\bibfnamefont
  {S.}~\bibnamefont {Forrest}},\ }\href {https://doi.org/10.1038/25954}
  {\bibfield  {journal} {\bibinfo  {journal} {Nature}\ }\textbf {\bibinfo
  {volume} {395}},\ \bibinfo {pages} {151} (\bibinfo {year}
  {1998})}\BibitemShut {NoStop}%
\bibitem [{\citenamefont {Marian}(2012)}]{Marian2012}%
  \BibitemOpen
  \bibfield  {author} {\bibinfo {author} {\bibfnamefont {C.~M.}\ \bibnamefont
  {Marian}},\ }\href {https://doi.org/10.1002/wcms.83} {\bibfield  {journal}
  {\bibinfo  {journal} {Wiley Interdiscip. Rev. Comput. Mol. Sci.}\ }\textbf
  {\bibinfo {volume} {2}},\ \bibinfo {pages} {187} (\bibinfo {year}
  {2012})}\BibitemShut {NoStop}%
\bibitem [{\citenamefont {Gatti}\ \emph {et~al.}(2017)\citenamefont {Gatti},
  \citenamefont {Lasorne}, \citenamefont {Meyer},\ and\ \citenamefont
  {Nauts}}]{gatti2017applications}%
  \BibitemOpen
  \bibfield  {author} {\bibinfo {author} {\bibfnamefont {F.}~\bibnamefont
  {Gatti}}, \bibinfo {author} {\bibfnamefont {B.}~\bibnamefont {Lasorne}},
  \bibinfo {author} {\bibfnamefont {H.-D.}\ \bibnamefont {Meyer}}, \ and\
  \bibinfo {author} {\bibfnamefont {A.}~\bibnamefont {Nauts}},\ }\href
  {https://link.springer.com/book/10.1007\%2F978-3-319-53923-2} {\emph
  {\bibinfo {title} {Applications of quantum dynamics in chemistry}}},\
  Vol.~\bibinfo {volume} {98}\ (\bibinfo  {publisher} {Springer},\ \bibinfo
  {year} {2017})\BibitemShut {NoStop}%
\bibitem [{\citenamefont {Gonz{\'a}lez}\ and\ \citenamefont
  {Lindh}(2020)}]{gonzalez2020quantum}%
  \BibitemOpen
  \bibfield  {author} {\bibinfo {author} {\bibfnamefont {L.}~\bibnamefont
  {Gonz{\'a}lez}}\ and\ \bibinfo {author} {\bibfnamefont {R.}~\bibnamefont
  {Lindh}},\ }\href
  {https://www.wiley.com/en-fr/Quantum+Chemistry+and+Dynamics+of+Excited+States:+Methods+and+Applications-p-9781119417750}
  {\enquote {\bibinfo {title} {Quantum chemistry and dynamics of excited
  states: Methods and applications},}\ } (\bibinfo {year} {2020})\BibitemShut
  {NoStop}%
\bibitem [{\citenamefont {Casida}(1995)}]{casida1995time}%
  \BibitemOpen
  \bibfield  {author} {\bibinfo {author} {\bibfnamefont {M.~E.}\ \bibnamefont
  {Casida}},\ }in\ \href {https://doi.org/10.1142/9789812830586_0005} {\emph
  {\bibinfo {booktitle} {Recent Advances In Density Functional Methods: (Part
  I)}}}\ (\bibinfo  {publisher} {World Scientific},\ \bibinfo {year} {1995})\
  pp.\ \bibinfo {pages} {155--192}\BibitemShut {NoStop}%
\bibitem [{\citenamefont {Marques}\ and\ \citenamefont
  {Gross}(2004)}]{marques2004time}%
  \BibitemOpen
  \bibfield  {author} {\bibinfo {author} {\bibfnamefont {M.~A.}\ \bibnamefont
  {Marques}}\ and\ \bibinfo {author} {\bibfnamefont {E.~K.}\ \bibnamefont
  {Gross}},\ }\href {https://doi.org/10.1146/annurev.physchem.55.091602.094449}
  {\bibfield  {journal} {\bibinfo  {journal} {Annu. Rev. Phys. Chem.}\ }\textbf
  {\bibinfo {volume} {55}},\ \bibinfo {pages} {427} (\bibinfo {year}
  {2004})}\BibitemShut {NoStop}%
\bibitem [{\citenamefont {Ullrich}(2011)}]{ullrich2011time}%
  \BibitemOpen
  \bibfield  {author} {\bibinfo {author} {\bibfnamefont {C.~A.}\ \bibnamefont
  {Ullrich}},\ }\href@noop {} {\emph {\bibinfo {title} {Time-dependent
  density-functional theory: concepts and applications}}}\ (\bibinfo
  {publisher} {OUP Oxford},\ \bibinfo {year} {2011})\BibitemShut {NoStop}%
\bibitem [{\citenamefont {Casida}\ and\ \citenamefont
  {Huix-Rotllant}(2012)}]{casida2012progress}%
  \BibitemOpen
  \bibfield  {author} {\bibinfo {author} {\bibfnamefont {M.~E.}\ \bibnamefont
  {Casida}}\ and\ \bibinfo {author} {\bibfnamefont {M.}~\bibnamefont
  {Huix-Rotllant}},\ }\href
  {https://doi.org/10.1146/annurev-physchem-032511-143803} {\bibfield
  {journal} {\bibinfo  {journal} {Ann. Rev. Phys. Chem.}\ }\textbf {\bibinfo
  {volume} {63}},\ \bibinfo {pages} {287} (\bibinfo {year} {2012})}\BibitemShut
  {NoStop}%
\bibitem [{\citenamefont {Maitra}\ \emph {et~al.}(2004)\citenamefont {Maitra},
  \citenamefont {Zhang}, \citenamefont {Cave},\ and\ \citenamefont
  {Burke}}]{maitra2004double}%
  \BibitemOpen
  \bibfield  {author} {\bibinfo {author} {\bibfnamefont {N.~T.}\ \bibnamefont
  {Maitra}}, \bibinfo {author} {\bibfnamefont {F.}~\bibnamefont {Zhang}},
  \bibinfo {author} {\bibfnamefont {R.~J.}\ \bibnamefont {Cave}}, \ and\
  \bibinfo {author} {\bibfnamefont {K.}~\bibnamefont {Burke}},\ }\href
  {https://doi.org/10.1063/1.1651060} {\bibfield  {journal} {\bibinfo
  {journal} {J. Chem. Phys.}\ }\textbf {\bibinfo {volume} {120}},\ \bibinfo
  {pages} {5932} (\bibinfo {year} {2004})}\BibitemShut {NoStop}%
\bibitem [{\citenamefont {Fuks}\ and\ \citenamefont
  {Maitra}(2014)}]{fuks2014challenging}%
  \BibitemOpen
  \bibfield  {author} {\bibinfo {author} {\bibfnamefont {J.~I.}\ \bibnamefont
  {Fuks}}\ and\ \bibinfo {author} {\bibfnamefont {N.~T.}\ \bibnamefont
  {Maitra}},\ }\href {https://doi.org/10.1039/C4CP00118D} {\bibfield  {journal}
  {\bibinfo  {journal} {Phys. Chem. Chem. Phys.}\ }\textbf {\bibinfo {volume}
  {16}},\ \bibinfo {pages} {14504} (\bibinfo {year} {2014})}\BibitemShut
  {NoStop}%
\bibitem [{\citenamefont {Vogiatzis}\ \emph {et~al.}(2017)\citenamefont
  {Vogiatzis}, \citenamefont {Ma}, \citenamefont {Olsen}, \citenamefont
  {Gagliardi},\ and\ \citenamefont {De~Jong}}]{vogiatzis2017pushing}%
  \BibitemOpen
  \bibfield  {author} {\bibinfo {author} {\bibfnamefont {K.~D.}\ \bibnamefont
  {Vogiatzis}}, \bibinfo {author} {\bibfnamefont {D.}~\bibnamefont {Ma}},
  \bibinfo {author} {\bibfnamefont {J.}~\bibnamefont {Olsen}}, \bibinfo
  {author} {\bibfnamefont {L.}~\bibnamefont {Gagliardi}}, \ and\ \bibinfo
  {author} {\bibfnamefont {W.~A.}\ \bibnamefont {De~Jong}},\ }\href
  {https://doi.org/10.1063/1.4989858} {\bibfield  {journal} {\bibinfo
  {journal} {J. Chem. Phys.}\ }\textbf {\bibinfo {volume} {147}},\ \bibinfo
  {pages} {184111} (\bibinfo {year} {2017})}\BibitemShut {NoStop}%
\bibitem [{\citenamefont {Williams}\ \emph {et~al.}(2020)\citenamefont
  {Williams}, \citenamefont {Yao}, \citenamefont {Li}, \citenamefont {Chen},
  \citenamefont {Shi}, \citenamefont {Motta}, \citenamefont {Niu},
  \citenamefont {Ray}, \citenamefont {Guo}, \citenamefont {Anderson} \emph
  {et~al.}}]{williams2020direct}%
  \BibitemOpen
  \bibfield  {author} {\bibinfo {author} {\bibfnamefont {K.~T.}\ \bibnamefont
  {Williams}}, \bibinfo {author} {\bibfnamefont {Y.}~\bibnamefont {Yao}},
  \bibinfo {author} {\bibfnamefont {J.}~\bibnamefont {Li}}, \bibinfo {author}
  {\bibfnamefont {L.}~\bibnamefont {Chen}}, \bibinfo {author} {\bibfnamefont
  {H.}~\bibnamefont {Shi}}, \bibinfo {author} {\bibfnamefont {M.}~\bibnamefont
  {Motta}}, \bibinfo {author} {\bibfnamefont {C.}~\bibnamefont {Niu}}, \bibinfo
  {author} {\bibfnamefont {U.}~\bibnamefont {Ray}}, \bibinfo {author}
  {\bibfnamefont {S.}~\bibnamefont {Guo}}, \bibinfo {author} {\bibfnamefont
  {R.~J.}\ \bibnamefont {Anderson}},  \emph {et~al.},\ }\href
  {https://doi.org/10.1103/PhysRevX.10.011041} {\bibfield  {journal} {\bibinfo
  {journal} {Phys. Rev. X}\ }\textbf {\bibinfo {volume} {10}},\ \bibinfo
  {pages} {011041} (\bibinfo {year} {2020})}\BibitemShut {NoStop}%
\bibitem [{\citenamefont {Stair}\ and\ \citenamefont
  {Evangelista}(2020)}]{stair2020exploring}%
  \BibitemOpen
  \bibfield  {author} {\bibinfo {author} {\bibfnamefont {N.~H.}\ \bibnamefont
  {Stair}}\ and\ \bibinfo {author} {\bibfnamefont {F.~A.}\ \bibnamefont
  {Evangelista}},\ }\href {https://arxiv.org/abs/2005.11349} {\bibfield
  {journal} {\bibinfo  {journal} {arXiv:2005.11349}\ } (\bibinfo {year}
  {2020})}\BibitemShut {NoStop}%
\bibitem [{\citenamefont {Eriksen}\ \emph {et~al.}(2020)\citenamefont
  {Eriksen}, \citenamefont {Anderson}, \citenamefont {Deustua}, \citenamefont
  {Ghanem}, \citenamefont {Hait}, \citenamefont {Hoffmann}, \citenamefont
  {Lee}, \citenamefont {Levine}, \citenamefont {Magoulas}, \citenamefont {Shen}
  \emph {et~al.}}]{eriksen2020ground}%
  \BibitemOpen
  \bibfield  {author} {\bibinfo {author} {\bibfnamefont {J.~J.}\ \bibnamefont
  {Eriksen}}, \bibinfo {author} {\bibfnamefont {T.~A.}\ \bibnamefont
  {Anderson}}, \bibinfo {author} {\bibfnamefont {J.~E.}\ \bibnamefont
  {Deustua}}, \bibinfo {author} {\bibfnamefont {K.}~\bibnamefont {Ghanem}},
  \bibinfo {author} {\bibfnamefont {D.}~\bibnamefont {Hait}}, \bibinfo {author}
  {\bibfnamefont {M.~R.}\ \bibnamefont {Hoffmann}}, \bibinfo {author}
  {\bibfnamefont {S.}~\bibnamefont {Lee}}, \bibinfo {author} {\bibfnamefont
  {D.~S.}\ \bibnamefont {Levine}}, \bibinfo {author} {\bibfnamefont
  {I.}~\bibnamefont {Magoulas}}, \bibinfo {author} {\bibfnamefont
  {J.}~\bibnamefont {Shen}},  \emph {et~al.},\ }\href
  {https://arxiv.org/abs/2008.02678} {\bibfield  {journal} {\bibinfo  {journal}
  {arXiv:2008.02678}\ } (\bibinfo {year} {2020})}\BibitemShut {NoStop}%
\bibitem [{\citenamefont {Loos}\ \emph {et~al.}(2020)\citenamefont {Loos},
  \citenamefont {Damour},\ and\ \citenamefont {Scemama}}]{loos2020note}%
  \BibitemOpen
  \bibfield  {author} {\bibinfo {author} {\bibfnamefont {P.-F.}\ \bibnamefont
  {Loos}}, \bibinfo {author} {\bibfnamefont {Y.}~\bibnamefont {Damour}}, \ and\
  \bibinfo {author} {\bibfnamefont {A.}~\bibnamefont {Scemama}},\ }\href
  {https://arxiv.org/abs/2008.11145} {\bibfield  {journal} {\bibinfo  {journal}
  {arXiv:2008.11145}\ } (\bibinfo {year} {2020})}\BibitemShut {NoStop}%
\bibitem [{\citenamefont {Lloyd}(1996)}]{lloyd1996universal}%
  \BibitemOpen
  \bibfield  {author} {\bibinfo {author} {\bibfnamefont {S.}~\bibnamefont
  {Lloyd}},\ }\href {https://www.jstor.org/stable/2899535} {\bibfield
  {journal} {\bibinfo  {journal} {Science}\ ,\ \bibinfo {pages} {1073}}
  (\bibinfo {year} {1996})}\BibitemShut {NoStop}%
\bibitem [{\citenamefont {Whitfield}\ \emph {et~al.}(2011)\citenamefont
  {Whitfield}, \citenamefont {Biamonte},\ and\ \citenamefont
  {Aspuru-Guzik}}]{whitfield2011simulation}%
  \BibitemOpen
  \bibfield  {author} {\bibinfo {author} {\bibfnamefont {J.~D.}\ \bibnamefont
  {Whitfield}}, \bibinfo {author} {\bibfnamefont {J.}~\bibnamefont {Biamonte}},
  \ and\ \bibinfo {author} {\bibfnamefont {A.}~\bibnamefont {Aspuru-Guzik}},\
  }\href {https://doi.org/10.1080/00268976.2011.552441} {\bibfield  {journal}
  {\bibinfo  {journal} {Mol. Phys.}\ }\textbf {\bibinfo {volume} {109}},\
  \bibinfo {pages} {735} (\bibinfo {year} {2011})}\BibitemShut {NoStop}%
\bibitem [{\citenamefont {Berry}\ \emph {et~al.}(2019)\citenamefont {Berry},
  \citenamefont {Gidney}, \citenamefont {Motta}, \citenamefont {McClean},\ and\
  \citenamefont {Babbush}}]{berry2019qubitization}%
  \BibitemOpen
  \bibfield  {author} {\bibinfo {author} {\bibfnamefont {D.~W.}\ \bibnamefont
  {Berry}}, \bibinfo {author} {\bibfnamefont {C.}~\bibnamefont {Gidney}},
  \bibinfo {author} {\bibfnamefont {M.}~\bibnamefont {Motta}}, \bibinfo
  {author} {\bibfnamefont {J.~R.}\ \bibnamefont {McClean}}, \ and\ \bibinfo
  {author} {\bibfnamefont {R.}~\bibnamefont {Babbush}},\ }\href
  {https://doi.org/10.22331/q-2019-12-02-208} {\bibfield  {journal} {\bibinfo
  {journal} {Quantum}\ }\textbf {\bibinfo {volume} {3}},\ \bibinfo {pages}
  {208} (\bibinfo {year} {2019})}\BibitemShut {NoStop}%
\bibitem [{\citenamefont {Bauman}\ \emph
  {et~al.}(2019{\natexlab{b}})\citenamefont {Bauman}, \citenamefont {Bylaska},
  \citenamefont {Krishnamoorthy}, \citenamefont {Low}, \citenamefont {Wiebe},
  \citenamefont {Granade}, \citenamefont {Roetteler}, \citenamefont {Troyer},\
  and\ \citenamefont {Kowalski}}]{bauman2019downfolding}%
  \BibitemOpen
  \bibfield  {author} {\bibinfo {author} {\bibfnamefont {N.~P.}\ \bibnamefont
  {Bauman}}, \bibinfo {author} {\bibfnamefont {E.~J.}\ \bibnamefont {Bylaska}},
  \bibinfo {author} {\bibfnamefont {S.}~\bibnamefont {Krishnamoorthy}},
  \bibinfo {author} {\bibfnamefont {G.~H.}\ \bibnamefont {Low}}, \bibinfo
  {author} {\bibfnamefont {N.}~\bibnamefont {Wiebe}}, \bibinfo {author}
  {\bibfnamefont {C.~E.}\ \bibnamefont {Granade}}, \bibinfo {author}
  {\bibfnamefont {M.}~\bibnamefont {Roetteler}}, \bibinfo {author}
  {\bibfnamefont {M.}~\bibnamefont {Troyer}}, \ and\ \bibinfo {author}
  {\bibfnamefont {K.}~\bibnamefont {Kowalski}},\ }\href
  {https://doi.org/10.1063/1.5094643} {\bibfield  {journal} {\bibinfo
  {journal} {J. Chem. Phys.}\ }\textbf {\bibinfo {volume} {151}},\ \bibinfo
  {pages} {014107} (\bibinfo {year} {2019}{\natexlab{b}})}\BibitemShut
  {NoStop}%
\bibitem [{\citenamefont {Takeshita}\ \emph {et~al.}(2020)\citenamefont
  {Takeshita}, \citenamefont {Rubin}, \citenamefont {Jiang}, \citenamefont
  {Lee}, \citenamefont {Babbush},\ and\ \citenamefont
  {McClean}}]{takeshita2020increasing}%
  \BibitemOpen
  \bibfield  {author} {\bibinfo {author} {\bibfnamefont {T.}~\bibnamefont
  {Takeshita}}, \bibinfo {author} {\bibfnamefont {N.~C.}\ \bibnamefont
  {Rubin}}, \bibinfo {author} {\bibfnamefont {Z.}~\bibnamefont {Jiang}},
  \bibinfo {author} {\bibfnamefont {E.}~\bibnamefont {Lee}}, \bibinfo {author}
  {\bibfnamefont {R.}~\bibnamefont {Babbush}}, \ and\ \bibinfo {author}
  {\bibfnamefont {J.~R.}\ \bibnamefont {McClean}},\ }\href
  {http://dx.doi.org/10.1103/PhysRevX.10.011004} {\bibfield  {journal}
  {\bibinfo  {journal} {Phys. Rev. X}\ }\textbf {\bibinfo {volume} {10}}
  (\bibinfo {year} {2020})}\BibitemShut {NoStop}%
\bibitem [{\citenamefont {Kowalski}\ and\ \citenamefont
  {Bauman}(2020)}]{kowalski2020subsystem}%
  \BibitemOpen
  \bibfield  {author} {\bibinfo {author} {\bibfnamefont {K.}~\bibnamefont
  {Kowalski}}\ and\ \bibinfo {author} {\bibfnamefont {N.~P.}\ \bibnamefont
  {Bauman}},\ }\href {http://dx.doi.org/10.1063/5.0008436} {\bibfield
  {journal} {\bibinfo  {journal} {J. Chem. Phys.}\ }\textbf {\bibinfo {volume}
  {152}},\ \bibinfo {pages} {244127} (\bibinfo {year} {2020})}\BibitemShut
  {NoStop}%
\bibitem [{\citenamefont {Motta}\ \emph
  {et~al.}(2020{\natexlab{b}})\citenamefont {Motta}, \citenamefont {Gujarati},
  \citenamefont {Rice}, \citenamefont {Kumar}, \citenamefont {Masteran},
  \citenamefont {Latone}, \citenamefont {Lee}, \citenamefont {Valeev},\ and\
  \citenamefont {Takeshita}}]{motta2020quantum}%
  \BibitemOpen
  \bibfield  {author} {\bibinfo {author} {\bibfnamefont {M.}~\bibnamefont
  {Motta}}, \bibinfo {author} {\bibfnamefont {T.~P.}\ \bibnamefont {Gujarati}},
  \bibinfo {author} {\bibfnamefont {J.~E.}\ \bibnamefont {Rice}}, \bibinfo
  {author} {\bibfnamefont {A.}~\bibnamefont {Kumar}}, \bibinfo {author}
  {\bibfnamefont {C.}~\bibnamefont {Masteran}}, \bibinfo {author}
  {\bibfnamefont {J.~A.}\ \bibnamefont {Latone}}, \bibinfo {author}
  {\bibfnamefont {E.}~\bibnamefont {Lee}}, \bibinfo {author} {\bibfnamefont
  {E.~F.}\ \bibnamefont {Valeev}}, \ and\ \bibinfo {author} {\bibfnamefont
  {T.~Y.}\ \bibnamefont {Takeshita}},\ }\href
  {https://arxiv.org/abs/2006.02488} {\bibfield  {journal} {\bibinfo  {journal}
  {arXiv:2006.02488}\ } (\bibinfo {year} {2020}{\natexlab{b}})}\BibitemShut
  {NoStop}%
\bibitem [{\citenamefont {Metcalf}\ \emph {et~al.}(2020)\citenamefont
  {Metcalf}, \citenamefont {Bauman}, \citenamefont {Kowalski},\ and\
  \citenamefont {de~Jong}}]{metcalf2020resource}%
  \BibitemOpen
  \bibfield  {author} {\bibinfo {author} {\bibfnamefont {M.}~\bibnamefont
  {Metcalf}}, \bibinfo {author} {\bibfnamefont {N.~P.}\ \bibnamefont {Bauman}},
  \bibinfo {author} {\bibfnamefont {K.}~\bibnamefont {Kowalski}}, \ and\
  \bibinfo {author} {\bibfnamefont {W.~A.}\ \bibnamefont {de~Jong}},\ }\href
  {https://arxiv.org/abs/2004.07721} {\bibfield  {journal} {\bibinfo  {journal}
  {arXiv:2004.07721}\ } (\bibinfo {year} {2020})}\BibitemShut {NoStop}%
\bibitem [{\citenamefont {Urbanek}\ \emph {et~al.}(2020)\citenamefont
  {Urbanek}, \citenamefont {Camps}, \citenamefont {Beeumen},\ and\
  \citenamefont {de~Jong}}]{urbanek2020chemistry}%
  \BibitemOpen
  \bibfield  {author} {\bibinfo {author} {\bibfnamefont {M.}~\bibnamefont
  {Urbanek}}, \bibinfo {author} {\bibfnamefont {D.}~\bibnamefont {Camps}},
  \bibinfo {author} {\bibfnamefont {R.~V.}\ \bibnamefont {Beeumen}}, \ and\
  \bibinfo {author} {\bibfnamefont {W.~A.}\ \bibnamefont {de~Jong}},\ }\href
  {https://arxiv.org/abs/2002.12902} {\bibfield  {journal} {\bibinfo  {journal}
  {arXiv:2002.12902}\ } (\bibinfo {year} {2020})}\BibitemShut {NoStop}%
\bibitem [{\citenamefont {McArdle}\ and\ \citenamefont
  {Tew}(2020)}]{mcardle2020improving}%
  \BibitemOpen
  \bibfield  {author} {\bibinfo {author} {\bibfnamefont {S.}~\bibnamefont
  {McArdle}}\ and\ \bibinfo {author} {\bibfnamefont {D.~P.}\ \bibnamefont
  {Tew}},\ }\href {https://arxiv.org/abs/2006.11181} {\bibfield  {journal}
  {\bibinfo  {journal} {arXiv:2006.11181}\ } (\bibinfo {year}
  {2020})}\BibitemShut {NoStop}%
\bibitem [{\citenamefont {Bylaska}\ \emph {et~al.}(2020)\citenamefont
  {Bylaska}, \citenamefont {Song}, \citenamefont {Bauman}, \citenamefont
  {Kowalski}, \citenamefont {Claudino},\ and\ \citenamefont
  {Humble}}]{bylaska2020quantum}%
  \BibitemOpen
  \bibfield  {author} {\bibinfo {author} {\bibfnamefont {E.~J.}\ \bibnamefont
  {Bylaska}}, \bibinfo {author} {\bibfnamefont {D.}~\bibnamefont {Song}},
  \bibinfo {author} {\bibfnamefont {N.~P.}\ \bibnamefont {Bauman}}, \bibinfo
  {author} {\bibfnamefont {K.}~\bibnamefont {Kowalski}}, \bibinfo {author}
  {\bibfnamefont {D.}~\bibnamefont {Claudino}}, \ and\ \bibinfo {author}
  {\bibfnamefont {T.~S.}\ \bibnamefont {Humble}},\ }\href
  {https://arxiv.org/abs/2009.00080} {\bibfield  {journal} {\bibinfo  {journal}
  {arXiv:2009.00080}\ } (\bibinfo {year} {2020})}\BibitemShut {NoStop}%
\bibitem [{\citenamefont {Rossmannek}\ \emph {et~al.}(2020)\citenamefont
  {Rossmannek}, \citenamefont {Barkoutsos}, \citenamefont {Ollitrault},\ and\
  \citenamefont {Tavernelli}}]{rossmannek2020quantum}%
  \BibitemOpen
  \bibfield  {author} {\bibinfo {author} {\bibfnamefont {M.}~\bibnamefont
  {Rossmannek}}, \bibinfo {author} {\bibfnamefont {P.~K.}\ \bibnamefont
  {Barkoutsos}}, \bibinfo {author} {\bibfnamefont {P.~J.}\ \bibnamefont
  {Ollitrault}}, \ and\ \bibinfo {author} {\bibfnamefont {I.}~\bibnamefont
  {Tavernelli}},\ }\href {https://arxiv.org/abs/2009.01872} {\bibfield
  {journal} {\bibinfo  {journal} {arxiv:2009.01872}\ } (\bibinfo {year}
  {2020})}\BibitemShut {NoStop}%
\bibitem [{\citenamefont {Andersson}\ \emph {et~al.}(1992)\citenamefont
  {Andersson}, \citenamefont {Malmqvist},\ and\ \citenamefont
  {Roos}}]{andersson1992second}%
  \BibitemOpen
  \bibfield  {author} {\bibinfo {author} {\bibfnamefont {K.}~\bibnamefont
  {Andersson}}, \bibinfo {author} {\bibfnamefont {P.-{\AA}.}\ \bibnamefont
  {Malmqvist}}, \ and\ \bibinfo {author} {\bibfnamefont {B.~O.}\ \bibnamefont
  {Roos}},\ }\href {https://doi.org/10.1063/1.462209} {\bibfield  {journal}
  {\bibinfo  {journal} {J. Chem. Phys.}\ }\textbf {\bibinfo {volume} {96}},\
  \bibinfo {pages} {1218} (\bibinfo {year} {1992})}\BibitemShut {NoStop}%
\bibitem [{\citenamefont {Angeli}\ \emph {et~al.}(2001)\citenamefont {Angeli},
  \citenamefont {Cimiraglia}, \citenamefont {Evangelisti}, \citenamefont
  {Leininger},\ and\ \citenamefont {Malrieu}}]{angeli2001introduction}%
  \BibitemOpen
  \bibfield  {author} {\bibinfo {author} {\bibfnamefont {C.}~\bibnamefont
  {Angeli}}, \bibinfo {author} {\bibfnamefont {R.}~\bibnamefont {Cimiraglia}},
  \bibinfo {author} {\bibfnamefont {S.}~\bibnamefont {Evangelisti}}, \bibinfo
  {author} {\bibfnamefont {T.}~\bibnamefont {Leininger}}, \ and\ \bibinfo
  {author} {\bibfnamefont {J.-P.}\ \bibnamefont {Malrieu}},\ }\href
  {https://doi.org/10.1063/1.1361246} {\bibfield  {journal} {\bibinfo
  {journal} {J. Chem. Phys.}\ }\textbf {\bibinfo {volume} {114}},\ \bibinfo
  {pages} {10252} (\bibinfo {year} {2001})}\BibitemShut {NoStop}%
\bibitem [{\citenamefont {Bona{\v{c}}i{\'c}-Kouteck{\`y}}\ and\ \citenamefont
  {Michl}(1985)}]{bonavcic1985photochemical}%
  \BibitemOpen
  \bibfield  {author} {\bibinfo {author} {\bibfnamefont {V.}~\bibnamefont
  {Bona{\v{c}}i{\'c}-Kouteck{\`y}}}\ and\ \bibinfo {author} {\bibfnamefont
  {J.}~\bibnamefont {Michl}},\ }\href {https://doi.org/10.1007/BF00698750}
  {\bibfield  {journal} {\bibinfo  {journal} {Theoretica chimica acta}\
  }\textbf {\bibinfo {volume} {68}},\ \bibinfo {pages} {45} (\bibinfo {year}
  {1985})}\BibitemShut {NoStop}%
\bibitem [{\citenamefont {Chahre}(1985)}]{chahre1985trigger}%
  \BibitemOpen
  \bibfield  {author} {\bibinfo {author} {\bibfnamefont {M.}~\bibnamefont
  {Chahre}},\ }\href {https://doi.org/10.1146/annurev.bb.14.060185.001555}
  {\bibfield  {journal} {\bibinfo  {journal} {Annual review of biophysics and
  biophysical chemistry}\ }\textbf {\bibinfo {volume} {14}},\ \bibinfo {pages}
  {331} (\bibinfo {year} {1985})}\BibitemShut {NoStop}%
\bibitem [{\citenamefont {Bonet-Monroig}\ \emph {et~al.}(2019)\citenamefont
  {Bonet-Monroig}, \citenamefont {Babbush},\ and\ \citenamefont
  {O'Brien}}]{bonet2019nearly}%
  \BibitemOpen
  \bibfield  {author} {\bibinfo {author} {\bibfnamefont {X.}~\bibnamefont
  {Bonet-Monroig}}, \bibinfo {author} {\bibfnamefont {R.}~\bibnamefont
  {Babbush}}, \ and\ \bibinfo {author} {\bibfnamefont {T.~E.}\ \bibnamefont
  {O'Brien}},\ }\href {https://arxiv.org/abs/1908.05628} {\bibfield  {journal}
  {\bibinfo  {journal} {arXiv:1908.05628}\ } (\bibinfo {year}
  {2019})}\BibitemShut {NoStop}%
\bibitem [{\citenamefont {Gross}\ \emph {et~al.}(1988)\citenamefont {Gross},
  \citenamefont {Oliveira},\ and\ \citenamefont {Kohn}}]{gross1988rayleigh}%
  \BibitemOpen
  \bibfield  {author} {\bibinfo {author} {\bibfnamefont {E.~K.}\ \bibnamefont
  {Gross}}, \bibinfo {author} {\bibfnamefont {L.~N.}\ \bibnamefont {Oliveira}},
  \ and\ \bibinfo {author} {\bibfnamefont {W.}~\bibnamefont {Kohn}},\ }\href
  {https://doi.org/10.1103/PhysRevA.37.2805} {\bibfield  {journal} {\bibinfo
  {journal} {Phys. Rev. A}\ }\textbf {\bibinfo {volume} {37}},\ \bibinfo
  {pages} {2805} (\bibinfo {year} {1988})}\BibitemShut {NoStop}%
\bibitem [{\citenamefont {Fletcher}(2013)}]{fletcher2013practical}%
  \BibitemOpen
  \bibfield  {author} {\bibinfo {author} {\bibfnamefont {R.}~\bibnamefont
  {Fletcher}},\ }\href@noop {} {\emph {\bibinfo {title} {Practical methods of
  optimization}}}\ (\bibinfo  {publisher} {John Wiley \& Sons},\ \bibinfo
  {year} {2013})\BibitemShut {NoStop}%
\bibitem [{\citenamefont {Yarkony}(1995)}]{yarkony1995modern}%
  \BibitemOpen
  \bibfield  {author} {\bibinfo {author} {\bibfnamefont {D.}~\bibnamefont
  {Yarkony}},\ }\href@noop {} {\emph {\bibinfo {title} {Modern electronic
  structure theory}}},\ Vol.~\bibinfo {volume} {2}\ (\bibinfo  {publisher}
  {World Scientific},\ \bibinfo {year} {1995})\BibitemShut {NoStop}%
\bibitem [{\citenamefont {Grimsley}\ \emph {et~al.}(2019)\citenamefont
  {Grimsley}, \citenamefont {Economou}, \citenamefont {Barnes},\ and\
  \citenamefont {Mayhall}}]{grimsley2019adaptive}%
  \BibitemOpen
  \bibfield  {author} {\bibinfo {author} {\bibfnamefont {H.~R.}\ \bibnamefont
  {Grimsley}}, \bibinfo {author} {\bibfnamefont {S.~E.}\ \bibnamefont
  {Economou}}, \bibinfo {author} {\bibfnamefont {E.}~\bibnamefont {Barnes}}, \
  and\ \bibinfo {author} {\bibfnamefont {N.~J.}\ \bibnamefont {Mayhall}},\
  }\href {https://doi.org/10.1038/s41467-019-10988-2} {\bibfield  {journal}
  {\bibinfo  {journal} {Nature comm.}\ }\textbf {\bibinfo {volume} {10}},\
  \bibinfo {pages} {1} (\bibinfo {year} {2019})}\BibitemShut {NoStop}%
\bibitem [{\citenamefont {Tang}\ \emph {et~al.}(2019)\citenamefont {Tang},
  \citenamefont {Barnes}, \citenamefont {Grimsley}, \citenamefont {Mayhall},\
  and\ \citenamefont {Economou}}]{tang2019qubit}%
  \BibitemOpen
  \bibfield  {author} {\bibinfo {author} {\bibfnamefont {H.~L.}\ \bibnamefont
  {Tang}}, \bibinfo {author} {\bibfnamefont {E.}~\bibnamefont {Barnes}},
  \bibinfo {author} {\bibfnamefont {H.~R.}\ \bibnamefont {Grimsley}}, \bibinfo
  {author} {\bibfnamefont {N.~J.}\ \bibnamefont {Mayhall}}, \ and\ \bibinfo
  {author} {\bibfnamefont {S.~E.}\ \bibnamefont {Economou}},\ }\href
  {https://arxiv.org/abs/1911.10205} {\bibfield  {journal} {\bibinfo  {journal}
  {arXiv:1911.10205}\ } (\bibinfo {year} {2019})}\BibitemShut {NoStop}%
\bibitem [{\citenamefont {McClean}\ \emph {et~al.}(2020)\citenamefont
  {McClean}, \citenamefont {Rubin}, \citenamefont {Sung}, \citenamefont
  {Kivlichan}, \citenamefont {Bonet-Monroig}, \citenamefont {Cao},
  \citenamefont {Dai}, \citenamefont {Fried}, \citenamefont {Gidney},
  \citenamefont {Gimby}, \citenamefont {Gokhale}, \citenamefont {H\"aner},
  \citenamefont {Hardikar}, \citenamefont {Havl{\'{\i}}{\v{c}}ek},
  \citenamefont {Higgott}, \citenamefont {Huang}, \citenamefont {Izaac},
  \citenamefont {Jiang}, \citenamefont {Liu}, \citenamefont {McArdle},
  \citenamefont {Neeley}, \citenamefont {O'Brien}, \citenamefont {O'Gorman},
  \citenamefont {Ozfidan}, \citenamefont {Radin}, \citenamefont {Romero},
  \citenamefont {Sawaya}, \citenamefont {Senjean}, \citenamefont {Setia},
  \citenamefont {Sim}, \citenamefont {Steiger}, \citenamefont {Steudtner},
  \citenamefont {Sun}, \citenamefont {Sun}, \citenamefont {Wang}, \citenamefont
  {Zhang},\ and\ \citenamefont {Babbush}}]{mcclean2020openfermion}%
  \BibitemOpen
  \bibfield  {author} {\bibinfo {author} {\bibfnamefont {J.~R.}\ \bibnamefont
  {McClean}}, \bibinfo {author} {\bibfnamefont {N.~C.}\ \bibnamefont {Rubin}},
  \bibinfo {author} {\bibfnamefont {K.~J.}\ \bibnamefont {Sung}}, \bibinfo
  {author} {\bibfnamefont {I.~D.}\ \bibnamefont {Kivlichan}}, \bibinfo {author}
  {\bibfnamefont {X.}~\bibnamefont {Bonet-Monroig}}, \bibinfo {author}
  {\bibfnamefont {Y.}~\bibnamefont {Cao}}, \bibinfo {author} {\bibfnamefont
  {C.}~\bibnamefont {Dai}}, \bibinfo {author} {\bibfnamefont {E.~S.}\
  \bibnamefont {Fried}}, \bibinfo {author} {\bibfnamefont {C.}~\bibnamefont
  {Gidney}}, \bibinfo {author} {\bibfnamefont {B.}~\bibnamefont {Gimby}},
  \bibinfo {author} {\bibfnamefont {P.}~\bibnamefont {Gokhale}}, \bibinfo
  {author} {\bibfnamefont {T.}~\bibnamefont {H\"aner}}, \bibinfo {author}
  {\bibfnamefont {T.}~\bibnamefont {Hardikar}}, \bibinfo {author}
  {\bibfnamefont {V.}~\bibnamefont {Havl{\'{\i}}{\v{c}}ek}}, \bibinfo {author}
  {\bibfnamefont {O.}~\bibnamefont {Higgott}}, \bibinfo {author} {\bibfnamefont
  {C.}~\bibnamefont {Huang}}, \bibinfo {author} {\bibfnamefont
  {J.}~\bibnamefont {Izaac}}, \bibinfo {author} {\bibfnamefont
  {Z.}~\bibnamefont {Jiang}}, \bibinfo {author} {\bibfnamefont
  {X.}~\bibnamefont {Liu}}, \bibinfo {author} {\bibfnamefont {S.}~\bibnamefont
  {McArdle}}, \bibinfo {author} {\bibfnamefont {M.}~\bibnamefont {Neeley}},
  \bibinfo {author} {\bibfnamefont {T.}~\bibnamefont {O'Brien}}, \bibinfo
  {author} {\bibfnamefont {B.}~\bibnamefont {O'Gorman}}, \bibinfo {author}
  {\bibfnamefont {I.}~\bibnamefont {Ozfidan}}, \bibinfo {author} {\bibfnamefont
  {M.~D.}\ \bibnamefont {Radin}}, \bibinfo {author} {\bibfnamefont
  {J.}~\bibnamefont {Romero}}, \bibinfo {author} {\bibfnamefont {N.~P.~D.}\
  \bibnamefont {Sawaya}}, \bibinfo {author} {\bibfnamefont {B.}~\bibnamefont
  {Senjean}}, \bibinfo {author} {\bibfnamefont {K.}~\bibnamefont {Setia}},
  \bibinfo {author} {\bibfnamefont {S.}~\bibnamefont {Sim}}, \bibinfo {author}
  {\bibfnamefont {D.~S.}\ \bibnamefont {Steiger}}, \bibinfo {author}
  {\bibfnamefont {M.}~\bibnamefont {Steudtner}}, \bibinfo {author}
  {\bibfnamefont {Q.}~\bibnamefont {Sun}}, \bibinfo {author} {\bibfnamefont
  {W.}~\bibnamefont {Sun}}, \bibinfo {author} {\bibfnamefont {D.}~\bibnamefont
  {Wang}}, \bibinfo {author} {\bibfnamefont {F.}~\bibnamefont {Zhang}}, \ and\
  \bibinfo {author} {\bibfnamefont {R.}~\bibnamefont {Babbush}},\ }\href
  {https://doi.org/10.1088/2058-9565/ab8ebc} {\bibfield  {journal} {\bibinfo
  {journal} {Quantum Sci. Technol.}\ } (\bibinfo {year} {2020})}\BibitemShut
  {NoStop}%
\bibitem [{cir()}]{cirq}%
  \BibitemOpen
  \href@noop {} {\ }\bibinfo {note}
  {\url{https://github.com/quantumlib/Cirq}}\BibitemShut {NoStop}%
\bibitem [{\citenamefont {Smith}\ \emph {et~al.}(2020)\citenamefont {Smith},
  \citenamefont {Burns}, \citenamefont {Simmonett}, \citenamefont {Parrish},
  \citenamefont {Schieber}, \citenamefont {Galvelis}, \citenamefont {Kraus},
  \citenamefont {Kruse}, \citenamefont {Di~Remigio}, \citenamefont {Alenaizan}
  \emph {et~al.}}]{smith2020psi4}%
  \BibitemOpen
  \bibfield  {author} {\bibinfo {author} {\bibfnamefont {D.~G.}\ \bibnamefont
  {Smith}}, \bibinfo {author} {\bibfnamefont {L.~A.}\ \bibnamefont {Burns}},
  \bibinfo {author} {\bibfnamefont {A.~C.}\ \bibnamefont {Simmonett}}, \bibinfo
  {author} {\bibfnamefont {R.~M.}\ \bibnamefont {Parrish}}, \bibinfo {author}
  {\bibfnamefont {M.~C.}\ \bibnamefont {Schieber}}, \bibinfo {author}
  {\bibfnamefont {R.}~\bibnamefont {Galvelis}}, \bibinfo {author}
  {\bibfnamefont {P.}~\bibnamefont {Kraus}}, \bibinfo {author} {\bibfnamefont
  {H.}~\bibnamefont {Kruse}}, \bibinfo {author} {\bibfnamefont
  {R.}~\bibnamefont {Di~Remigio}}, \bibinfo {author} {\bibfnamefont
  {A.}~\bibnamefont {Alenaizan}},  \emph {et~al.},\ }\href
  {https://doi.org/10.1063/5.0006002} {\bibfield  {journal} {\bibinfo
  {journal} {The Journal of Chemical Physics}\ }\textbf {\bibinfo {volume}
  {152}},\ \bibinfo {pages} {184108} (\bibinfo {year} {2020})}\BibitemShut
  {NoStop}%
\bibitem [{\citenamefont {Baer}(2006)}]{baer2006beyond}%
  \BibitemOpen
  \bibfield  {author} {\bibinfo {author} {\bibfnamefont {M.}~\bibnamefont
  {Baer}},\ }\href@noop {} {\emph {\bibinfo {title} {Beyond Born-Oppenheimer:
  electronic nonadiabatic coupling terms and conical intersections}}}\
  (\bibinfo  {publisher} {John Wiley \& Sons},\ \bibinfo {year}
  {2006})\BibitemShut {NoStop}%
\bibitem [{\citenamefont {Zhang}\ \emph
  {et~al.}(2020{\natexlab{d}})\citenamefont {Zhang}, \citenamefont {Su},
  \citenamefont {Lasorne}, \citenamefont {Bra{\"\i}da},\ and\ \citenamefont
  {Wu}}]{zhang2020novel}%
  \BibitemOpen
  \bibfield  {author} {\bibinfo {author} {\bibfnamefont {Y.}~\bibnamefont
  {Zhang}}, \bibinfo {author} {\bibfnamefont {P.}~\bibnamefont {Su}}, \bibinfo
  {author} {\bibfnamefont {B.}~\bibnamefont {Lasorne}}, \bibinfo {author}
  {\bibfnamefont {B.}~\bibnamefont {Bra{\"\i}da}}, \ and\ \bibinfo {author}
  {\bibfnamefont {W.}~\bibnamefont {Wu}},\ }\href
  {https://doi.org/10.1021/acs.jpclett.0c01466} {\bibfield  {journal} {\bibinfo
   {journal} {J. Phys. Chem. Lett.}\ } (\bibinfo {year}
  {2020}{\natexlab{d}})}\BibitemShut {NoStop}%
\bibitem [{\citenamefont {Plasser}\ \emph {et~al.}(2016)\citenamefont
  {Plasser}, \citenamefont {Ruckenbauer}, \citenamefont {Mai}, \citenamefont
  {Oppel}, \citenamefont {Marquetand},\ and\ \citenamefont
  {Gonz{\'a}lez}}]{plasser2016efficient}%
  \BibitemOpen
  \bibfield  {author} {\bibinfo {author} {\bibfnamefont {F.}~\bibnamefont
  {Plasser}}, \bibinfo {author} {\bibfnamefont {M.}~\bibnamefont
  {Ruckenbauer}}, \bibinfo {author} {\bibfnamefont {S.}~\bibnamefont {Mai}},
  \bibinfo {author} {\bibfnamefont {M.}~\bibnamefont {Oppel}}, \bibinfo
  {author} {\bibfnamefont {P.}~\bibnamefont {Marquetand}}, \ and\ \bibinfo
  {author} {\bibfnamefont {L.}~\bibnamefont {Gonz{\'a}lez}},\ }\href
  {https://doi.org/10.1021/acs.jctc.5b01148} {\bibfield  {journal} {\bibinfo
  {journal} {J. Chem. Theory Comput.}\ }\textbf {\bibinfo {volume} {12}},\
  \bibinfo {pages} {1207} (\bibinfo {year} {2016})}\BibitemShut {NoStop}%
\bibitem [{\citenamefont {L{\"o}wdin}(1955)}]{lowdin1955quantum}%
  \BibitemOpen
  \bibfield  {author} {\bibinfo {author} {\bibfnamefont {P.-O.}\ \bibnamefont
  {L{\"o}wdin}},\ }\href {https://doi.org/10.1103/PhysRev.97.1474} {\bibfield
  {journal} {\bibinfo  {journal} {Phys. Rev.}\ }\textbf {\bibinfo {volume}
  {97}},\ \bibinfo {pages} {1474} (\bibinfo {year} {1955})}\BibitemShut
  {NoStop}%
\bibitem [{\citenamefont {Malmqvist}\ \emph {et~al.}(2002)\citenamefont
  {Malmqvist}, \citenamefont {Roos},\ and\ \citenamefont
  {Schimmelpfennig}}]{malmqvist2002restricted}%
  \BibitemOpen
  \bibfield  {author} {\bibinfo {author} {\bibfnamefont {P.~{\AA}.}\
  \bibnamefont {Malmqvist}}, \bibinfo {author} {\bibfnamefont {B.~O.}\
  \bibnamefont {Roos}}, \ and\ \bibinfo {author} {\bibfnamefont
  {B.}~\bibnamefont {Schimmelpfennig}},\ }\href
  {https://doi.org/10.1016/S0009-2614(02)00498-0} {\bibfield  {journal}
  {\bibinfo  {journal} {Chem. Phys. Lett.}\ }\textbf {\bibinfo {volume}
  {357}},\ \bibinfo {pages} {230} (\bibinfo {year} {2002})}\BibitemShut
  {NoStop}%
\bibitem [{\citenamefont {Shepard}\ and\ \citenamefont
  {Brozell}(2019)}]{shepard2019all}%
  \BibitemOpen
  \bibfield  {author} {\bibinfo {author} {\bibfnamefont {R.}~\bibnamefont
  {Shepard}}\ and\ \bibinfo {author} {\bibfnamefont {S.~R.}\ \bibnamefont
  {Brozell}},\ }\href {https://doi.org/10.1080/00268976.2019.1635275}
  {\bibfield  {journal} {\bibinfo  {journal} {Mol. Phys}\ }\textbf {\bibinfo
  {volume} {117}},\ \bibinfo {pages} {2374} (\bibinfo {year}
  {2019})}\BibitemShut {NoStop}%
\bibitem [{\citenamefont {Kassal}\ and\ \citenamefont
  {Aspuru-Guzik}(2009)}]{kassal2009quantum}%
  \BibitemOpen
  \bibfield  {author} {\bibinfo {author} {\bibfnamefont {I.}~\bibnamefont
  {Kassal}}\ and\ \bibinfo {author} {\bibfnamefont {A.}~\bibnamefont
  {Aspuru-Guzik}},\ }\href {https://doi.org/10.1063/1.3266959} {\bibfield
  {journal} {\bibinfo  {journal} {J. Chem. Phys.}\ }\textbf {\bibinfo {volume}
  {131}},\ \bibinfo {pages} {224102} (\bibinfo {year} {2009})}\BibitemShut
  {NoStop}%
\bibitem [{\citenamefont {O’Brien}\ \emph {et~al.}(2019)\citenamefont
  {O’Brien}, \citenamefont {Senjean}, \citenamefont {Sagastizabal},
  \citenamefont {Bonet-Monroig}, \citenamefont {Dutkiewicz}, \citenamefont
  {Buda}, \citenamefont {DiCarlo},\ and\ \citenamefont
  {Visscher}}]{obrien2019calculating}%
  \BibitemOpen
  \bibfield  {author} {\bibinfo {author} {\bibfnamefont {T.~E.}\ \bibnamefont
  {O’Brien}}, \bibinfo {author} {\bibfnamefont {B.}~\bibnamefont {Senjean}},
  \bibinfo {author} {\bibfnamefont {R.}~\bibnamefont {Sagastizabal}}, \bibinfo
  {author} {\bibfnamefont {X.}~\bibnamefont {Bonet-Monroig}}, \bibinfo {author}
  {\bibfnamefont {A.}~\bibnamefont {Dutkiewicz}}, \bibinfo {author}
  {\bibfnamefont {F.}~\bibnamefont {Buda}}, \bibinfo {author} {\bibfnamefont
  {L.}~\bibnamefont {DiCarlo}}, \ and\ \bibinfo {author} {\bibfnamefont
  {L.}~\bibnamefont {Visscher}},\ }\href
  {https://doi.org/10.1038/s41534-019-0213-4} {\bibfield  {journal} {\bibinfo
  {journal} {npj Quantum Inf.}\ }\textbf {\bibinfo {volume} {5}},\ \bibinfo
  {pages} {1} (\bibinfo {year} {2019})}\BibitemShut {NoStop}%
\bibitem [{\citenamefont {Mitarai}\ \emph {et~al.}(2020)\citenamefont
  {Mitarai}, \citenamefont {Nakagawa},\ and\ \citenamefont
  {Mizukami}}]{mitarai2020theory}%
  \BibitemOpen
  \bibfield  {author} {\bibinfo {author} {\bibfnamefont {K.}~\bibnamefont
  {Mitarai}}, \bibinfo {author} {\bibfnamefont {Y.~O.}\ \bibnamefont
  {Nakagawa}}, \ and\ \bibinfo {author} {\bibfnamefont {W.}~\bibnamefont
  {Mizukami}},\ }\href {https://doi.org/10.1103/PhysRevResearch.2.013129}
  {\bibfield  {journal} {\bibinfo  {journal} {Phys. Rev. Res.}\ }\textbf
  {\bibinfo {volume} {2}},\ \bibinfo {pages} {013129} (\bibinfo {year}
  {2020})}\BibitemShut {NoStop}%
\bibitem [{\citenamefont {Sokolov}\ \emph
  {et~al.}(2020{\natexlab{b}})\citenamefont {Sokolov}, \citenamefont
  {Barkoutsos}, \citenamefont {Moeller}, \citenamefont {Suchsland},
  \citenamefont {Mazzola},\ and\ \citenamefont
  {Tavernelli}}]{sokolov2020microcanonical}%
  \BibitemOpen
  \bibfield  {author} {\bibinfo {author} {\bibfnamefont {I.~O.}\ \bibnamefont
  {Sokolov}}, \bibinfo {author} {\bibfnamefont {P.~K.}\ \bibnamefont
  {Barkoutsos}}, \bibinfo {author} {\bibfnamefont {L.}~\bibnamefont {Moeller}},
  \bibinfo {author} {\bibfnamefont {P.}~\bibnamefont {Suchsland}}, \bibinfo
  {author} {\bibfnamefont {G.}~\bibnamefont {Mazzola}}, \ and\ \bibinfo
  {author} {\bibfnamefont {I.}~\bibnamefont {Tavernelli}},\ }\href
  {https://arxiv.org/abs/2008.08144} {\bibfield  {journal} {\bibinfo  {journal}
  {arXiv:2008.08144}\ } (\bibinfo {year} {2020}{\natexlab{b}})}\BibitemShut
  {NoStop}%
\bibitem [{\citenamefont {Paulus}\ \emph {et~al.}(2020)\citenamefont {Paulus},
  \citenamefont {Adelman}, \citenamefont {Jamula},\ and\ \citenamefont
  {McCusker}}]{paulus2020leveraging}%
  \BibitemOpen
  \bibfield  {author} {\bibinfo {author} {\bibfnamefont {B.~C.}\ \bibnamefont
  {Paulus}}, \bibinfo {author} {\bibfnamefont {S.~L.}\ \bibnamefont {Adelman}},
  \bibinfo {author} {\bibfnamefont {L.~L.}\ \bibnamefont {Jamula}}, \ and\
  \bibinfo {author} {\bibfnamefont {J.~K.}\ \bibnamefont {McCusker}},\ }\href
  {https://doi.org/10.1038/s41586-020-2353-2} {\bibfield  {journal} {\bibinfo
  {journal} {Nature}\ }\textbf {\bibinfo {volume} {582}},\ \bibinfo {pages}
  {214} (\bibinfo {year} {2020})}\BibitemShut {NoStop}%
\bibitem [{\citenamefont {Ligthart}\ \emph {et~al.}(2018)\citenamefont
  {Ligthart}, \citenamefont {de~Vries}, \citenamefont {Zhang}, \citenamefont
  {Pols}, \citenamefont {Bobbert}, \citenamefont {van Eersel},\ and\
  \citenamefont {Coehoorn}}]{Lighthart2018}%
  \BibitemOpen
  \bibfield  {author} {\bibinfo {author} {\bibfnamefont {A.}~\bibnamefont
  {Ligthart}}, \bibinfo {author} {\bibfnamefont {X.}~\bibnamefont {de~Vries}},
  \bibinfo {author} {\bibfnamefont {L.}~\bibnamefont {Zhang}}, \bibinfo
  {author} {\bibfnamefont {M.~C. W.~M.}\ \bibnamefont {Pols}}, \bibinfo
  {author} {\bibfnamefont {P.~A.}\ \bibnamefont {Bobbert}}, \bibinfo {author}
  {\bibfnamefont {H.}~\bibnamefont {van Eersel}}, \ and\ \bibinfo {author}
  {\bibfnamefont {R.}~\bibnamefont {Coehoorn}},\ }\href
  {https://doi.org/10.1002/adfm.201804618} {\bibfield  {journal} {\bibinfo
  {journal} {{Adv. Funct. Mater}}\ }\textbf {\bibinfo {volume} {{28}}}
  (\bibinfo {year} {{2018}})}\BibitemShut {NoStop}%
\bibitem [{\citenamefont {Ostroumov}\ \emph {et~al.}(2020)\citenamefont
  {Ostroumov}, \citenamefont {Götze}, \citenamefont {Reus}, \citenamefont
  {Lambrev},\ and\ \citenamefont {Holzwarth}}]{Ostroumov.Holzwarth.2020}%
  \BibitemOpen
  \bibfield  {author} {\bibinfo {author} {\bibfnamefont {E.~E.}\ \bibnamefont
  {Ostroumov}}, \bibinfo {author} {\bibfnamefont {J.~P.}\ \bibnamefont
  {Götze}}, \bibinfo {author} {\bibfnamefont {M.}~\bibnamefont {Reus}},
  \bibinfo {author} {\bibfnamefont {P.~H.}\ \bibnamefont {Lambrev}}, \ and\
  \bibinfo {author} {\bibfnamefont {A.~R.}\ \bibnamefont {Holzwarth}},\ }\href
  {https://doi.org/10.1007/s11120-020-00745-8} {\bibfield  {journal} {\bibinfo
  {journal} {Photosynth. Res.}\ }\textbf {\bibinfo {volume} {144}},\ \bibinfo
  {pages} {171} (\bibinfo {year} {2020})}\BibitemShut {NoStop}%
\bibitem [{\citenamefont {Markovitsi}(2016)}]{Markovitsi2016}%
  \BibitemOpen
  \bibfield  {author} {\bibinfo {author} {\bibfnamefont {D.}~\bibnamefont
  {Markovitsi}},\ }\href {https://doi.org/10.1111/php.12533} {\bibfield
  {journal} {\bibinfo  {journal} {Photochem. Photobiol.}\ }\textbf {\bibinfo
  {volume} {92}},\ \bibinfo {pages} {45} (\bibinfo {year} {2016})}\BibitemShut
  {NoStop}%
\bibitem [{\citenamefont {M}(1973)}]{yoshimine1973}%
  \BibitemOpen
  \bibfield  {author} {\bibinfo {author} {\bibfnamefont {Y.}~\bibnamefont
  {M}},\ }\href@noop {} {\bibfield  {journal} {\bibinfo  {journal} {J. Comp.
  Phys.}\ }\textbf {\bibinfo {volume} {11}},\ \bibinfo {pages} {449} (\bibinfo
  {year} {1973})}\BibitemShut {NoStop}%
\bibitem [{\citenamefont {Feyereisen}\ \emph {et~al.}(1993)\citenamefont
  {Feyereisen}, \citenamefont {Fitzgerald},\ and\ \citenamefont
  {Komornicki}}]{feyereisen1993use}%
  \BibitemOpen
  \bibfield  {author} {\bibinfo {author} {\bibfnamefont {M.}~\bibnamefont
  {Feyereisen}}, \bibinfo {author} {\bibfnamefont {G.}~\bibnamefont
  {Fitzgerald}}, \ and\ \bibinfo {author} {\bibfnamefont {A.}~\bibnamefont
  {Komornicki}},\ }\href@noop {} {\bibfield  {journal} {\bibinfo  {journal}
  {Chemical physics letters}\ }\textbf {\bibinfo {volume} {208}},\ \bibinfo
  {pages} {359} (\bibinfo {year} {1993})}\BibitemShut {NoStop}%
\bibitem [{\citenamefont {Kreplin}\ \emph {et~al.}(2020)\citenamefont
  {Kreplin}, \citenamefont {Knowles},\ and\ \citenamefont
  {Werner}}]{kreplin2020mcscf}%
  \BibitemOpen
  \bibfield  {author} {\bibinfo {author} {\bibfnamefont {D.~A.}\ \bibnamefont
  {Kreplin}}, \bibinfo {author} {\bibfnamefont {P.~J.}\ \bibnamefont
  {Knowles}}, \ and\ \bibinfo {author} {\bibfnamefont {H.-J.}\ \bibnamefont
  {Werner}},\ }\href@noop {} {\bibfield  {journal} {\bibinfo  {journal} {The
  Journal of Chemical Physics}\ }\textbf {\bibinfo {volume} {152}},\ \bibinfo
  {pages} {074102} (\bibinfo {year} {2020})}\BibitemShut {NoStop}%
\bibitem [{\citenamefont {Kreplin}\ \emph {et~al.}(2019)\citenamefont
  {Kreplin}, \citenamefont {Knowles},\ and\ \citenamefont
  {Werner}}]{kreplin2019second}%
  \BibitemOpen
  \bibfield  {author} {\bibinfo {author} {\bibfnamefont {D.~A.}\ \bibnamefont
  {Kreplin}}, \bibinfo {author} {\bibfnamefont {P.~J.}\ \bibnamefont
  {Knowles}}, \ and\ \bibinfo {author} {\bibfnamefont {H.-J.}\ \bibnamefont
  {Werner}},\ }\href@noop {} {\bibfield  {journal} {\bibinfo  {journal} {The
  Journal of chemical physics}\ }\textbf {\bibinfo {volume} {150}},\ \bibinfo
  {pages} {194106} (\bibinfo {year} {2019})}\BibitemShut {NoStop}%
\bibitem [{\citenamefont {Seeley}\ \emph {et~al.}(2012)\citenamefont {Seeley},
  \citenamefont {Richard},\ and\ \citenamefont {Love}}]{seeley2012bravyi}%
  \BibitemOpen
  \bibfield  {author} {\bibinfo {author} {\bibfnamefont {J.~T.}\ \bibnamefont
  {Seeley}}, \bibinfo {author} {\bibfnamefont {M.~J.}\ \bibnamefont {Richard}},
  \ and\ \bibinfo {author} {\bibfnamefont {P.~J.}\ \bibnamefont {Love}},\
  }\href@noop {} {\bibfield  {journal} {\bibinfo  {journal} {The Journal of
  chemical physics}\ }\textbf {\bibinfo {volume} {137}},\ \bibinfo {pages}
  {224109} (\bibinfo {year} {2012})}\BibitemShut {NoStop}%
\bibitem [{\citenamefont {Parrish}\ and\ \citenamefont
  {McMahon}(2019)}]{parrish2019quantumfilter}%
  \BibitemOpen
  \bibfield  {author} {\bibinfo {author} {\bibfnamefont {R.~M.}\ \bibnamefont
  {Parrish}}\ and\ \bibinfo {author} {\bibfnamefont {P.~L.}\ \bibnamefont
  {McMahon}},\ }\href {https://arxiv.org/abs/1909.08925} {\bibfield  {journal}
  {\bibinfo  {journal} {arXiv:1909.08925}\ } (\bibinfo {year}
  {2019})}\BibitemShut {NoStop}%
\bibitem [{\citenamefont {Bespalova}\ and\ \citenamefont
  {Kyriienko}(2020)}]{bespalova2020hamiltonian}%
  \BibitemOpen
  \bibfield  {author} {\bibinfo {author} {\bibfnamefont {T.~A.}\ \bibnamefont
  {Bespalova}}\ and\ \bibinfo {author} {\bibfnamefont {O.}~\bibnamefont
  {Kyriienko}},\ }\href {https://arxiv.org/abs/2009.03351} {\bibfield
  {journal} {\bibinfo  {journal} {arXiv:2009.03351}\ } (\bibinfo {year}
  {2020})}\BibitemShut {NoStop}%
\bibitem [{\citenamefont {Santagati}\ \emph {et~al.}(2018)\citenamefont
  {Santagati}, \citenamefont {Wang}, \citenamefont {Gentile}, \citenamefont
  {Paesani}, \citenamefont {Wiebe}, \citenamefont {McClean}, \citenamefont
  {Morley-Short}, \citenamefont {Shadbolt}, \citenamefont {Bonneau},
  \citenamefont {Silverstone} \emph {et~al.}}]{santagati2018witnessing}%
  \BibitemOpen
  \bibfield  {author} {\bibinfo {author} {\bibfnamefont {R.}~\bibnamefont
  {Santagati}}, \bibinfo {author} {\bibfnamefont {J.}~\bibnamefont {Wang}},
  \bibinfo {author} {\bibfnamefont {A.~A.}\ \bibnamefont {Gentile}}, \bibinfo
  {author} {\bibfnamefont {S.}~\bibnamefont {Paesani}}, \bibinfo {author}
  {\bibfnamefont {N.}~\bibnamefont {Wiebe}}, \bibinfo {author} {\bibfnamefont
  {J.~R.}\ \bibnamefont {McClean}}, \bibinfo {author} {\bibfnamefont
  {S.}~\bibnamefont {Morley-Short}}, \bibinfo {author} {\bibfnamefont {P.~J.}\
  \bibnamefont {Shadbolt}}, \bibinfo {author} {\bibfnamefont {D.}~\bibnamefont
  {Bonneau}}, \bibinfo {author} {\bibfnamefont {J.~W.}\ \bibnamefont
  {Silverstone}},  \emph {et~al.},\ }\href
  {https://doi.org/10.1126/sciadv.aap9646} {\bibfield  {journal} {\bibinfo
  {journal} {Sci. Adv.}\ }\textbf {\bibinfo {volume} {4}},\ \bibinfo {pages}
  {eaap9646} (\bibinfo {year} {2018})}\BibitemShut {NoStop}%
\end{thebibliography}

\newcommand{\Aa}[0]{Aa}
%

%----------------------------------------------------------------------------------------

\end{document}